\documentclass[12pt,a4paper,twoside]{report}
\usepackage[applemac]{inputenc}
\usepackage{amssymb}
\usepackage{amsmath}
\usepackage{amsopn}
\include{introduction.tex}
\newcommand{\LL}{\mathbb{L}}
\newcommand{\CC}{\mathbb{C}}

\newcommand{\BB}{\mathbb{B}}

\newcommand{\KK}{\mathbb{K}}
\newcommand{\MM}{\mathbb{M}}
\newcommand{\NN}{\mathbb{N}}

\newcommand{\PP}{\mathbb{P}}
\newcommand{\QQ}{\mathbb{Q}}
\newcommand{\RR}{\mathbb{R}}
\newcommand{\ZZ}{\mathbb{Z}}
\newcommand{\frN}{\mathfrak{N}}
\newcommand{\frA}{\mathfrak{A}}

\newcommand{\frAr}{\mathfrak{A}(\rr)}
\newcommand{\frc}{\mathfrak{C}}

\newcommand{\frb}{\mathfrak{B}}

\newcommand{\frp}{\mathfrak{p}}

\newcommand{\frm}{\mathfrak{m}}
\newcommand{\frR}{\mathfrak{R}}

\newcommand{\frU}{\mathfrak{U}}

\newcommand{\kaa}{\mathcal{A}}
\newcommand{\kbb}{\mathcal{B}}
\newcommand{\kbm}{\mathcal{B}(M)}
\newcommand{\kbr}{\mathcal{B}(\RR)}
\newcommand{\ktm}{\mathcal{T}(M)}
\newcommand{\kcc}{\mathcal{C}}

\newcommand{\cor}{\mathcal{C}\mathcal{O}\mathcal{N}(\rr)}
\newcommand{\kD}{\mathcal{D}}
\newcommand{\kE}{\mathcal{E}}
\newcommand{\kf}{\mathcal{F}}
\newcommand{\kg}{\mathcal{G}}
\newcommand{\kh}{\mathcal{H}}
\newcommand{\kI}{\mathcal{I}}
\newcommand{\kj}{\mathcal{J}}
\newcommand{\kK}{\mathcal{K}}
\newcommand{\kL}{\mathcal{L}}
\newcommand{\mm}{\mathcal{M}}
\newcommand{\nn}{\mathcal{N}}
\newcommand{\kO}{\mathcal{O}}
\newcommand{\kP}{\mathcal{P}}
\newcommand{\kQ}{\mathcal{Q}}
\newcommand{\kR}{\mathcal{R}}
\newcommand{\kS}{\mathcal{S}}

\newcommand{\kT}{\mathcal{T}}
\newcommand{\kV}{\mathcal{V}}
\newcommand{\ga}{\alpha}
\newcommand{\gb}{\beta}
\newcommand{\gd}{\delta}
\newcommand{\eps}{\varepsilon}
\newcommand{\gga}{\gamma}
\newcommand{\gG}{\Gamma}

\newcommand{\gl}{\lambda}
\newcommand{\go}{\omega}
\newcommand{\gO}{\Omega}
\newcommand{\gf}{\varphi}

\newcommand{\gF}{\Phi}
\newcommand{\gr}{\varrho}
\newcommand{\gs}{\sigma}

\newcommand{\gt}{\tau}
\newcommand{\gtt}{\theta}
\newcommand{\gT}{\Theta}
\newcommand{\gz}{\zeta}
\newcommand{\tm}{\subseteq}

\newcommand{\∞}{\infty}
\newcommand{\ten}{\otimes}
\newcommand{\Ds}{\bigoplus}
\newtheorem{definition}{Definition}[chapter]
\newtheorem{proposition}{Proposition}[chapter]
\newtheorem{theorem}{Theorem}[chapter]
\newtheorem{lemma}{Lemma}[chapter]
\newtheorem{corollary}{Corollary}[chapter]
\newtheorem{remark}{Remark}[chapter]
\newtheorem{example}{Example}[chapter]

\newcommand{\bg}{\begin}
\newcommand{\por}{\kP_{0}(\kR)}
\newcommand{\poa}{\kP_{0}(\kaa)}
\newcommand{\pob}{\kP_{0}(\kbb)}

\newcommand{\pr}{\kP(\kR)}
\newcommand{\ph}{\kP(\lh)}
\newcommand{\pph}{\PP\kh}

\newcommand{\pa}{\kP(\kaa)}

\newcommand{\pmm}{\kP(\mm)}

\newcommand{\qr}{\mathcal{Q}(\mathcal{R})}
\newcommand{\qm}{\kQ(\kT(M))}
\newcommand{\qpr}{\mathcal{Q}_{P}(\mathcal{R})}
\newcommand{\qpa}{\mathcal{Q}_{P}(\mathcal{A})}

\newcommand{\qqka}{\mathcal{Q}_{Q_{k}}(\mathcal{A})}
\newcommand{\qpja}{\mathcal{Q}_{P_{j}}(\mathcal{A})}

\newcommand{\qh}{\mathcal{Q}(\mathcal{H})} 
\newcommand{\qa}{\mathcal{Q}(\mathcal{A})}
\newcommand{\qb}{\mathcal{Q}(\mathcal{B})}
\newcommand{\qab}{\mathcal{Q}_{a}(\mathcal{B})}
\newcommand{\qc}{\mathcal{Q}(\mathcal{C})}

\newcommand{\qA}{\mathcal{Q}(\frA)}
\newcommand{\qbm}{\mathcal{Q}(\mathcal{B}(M))}

\newcommand{\ql}{\mathcal{Q}(\LL)}

\newcommand{\bpl}{\mathcal{Q}^b(\LL)}
\newcommand{\qal}{\mathcal{Q}_{a}(\LL)}
\newcommand{\qbl}{\kQ_{b}(\LL)}
\newcommand{\lh}{\mathcal{L}(\mathcal{H})}

\newcommand{\llh}{\LL(\kh)}

\newcommand{\all}{\forall}
\newcommand{\ex}{\exists}
\newcommand{\rr}{\kR}

\newcommand{\hr}{\kR_{sa}}
\newcommand{\el}{E_{\gl}}

\newcommand{\emm}{E_{\mu}}

\newcommand{\eal}{E^{A}_{\gl}}

\newcommand{\eamu}{E^{A}_{\mu}}

\newcommand{\ea}{E^{A}}

\newcommand{\fa}{F^{A}}
\newcommand{\fal}{F^{A}_{\gl}}
\newcommand{\famu}{F^{A}_{\mu}}
\newcommand{\ssl}{\gs(\gl)}

\newcommand{\we}{\wedge}
\newcommand{\We}{\bigwedge}
\newcommand{\Ve}{\bigvee}

\newcommand{\tto}{\mapsto}
\newcommand{\lra}{\Longrightarrow}
\newcommand{\llra}{\Longleftrightarrow}
\newcommand{\smm}{\setminus}
\newcommand{\dl}{\kD(\LL)}
\newcommand{\dprl}{\kD_{pr}(\LL)}

\newcommand{\dr}{\kD(\rr)}

\newcommand{\kba}{\gr^{\kbb}_{\kaa}}
\newcommand{\pp}{\perp}
\newcommand{\irr}{\in \RR}

\newcommand{\lir}{\gl \in \RR}

\newcommand{\inn}{\in \NN}
\newcommand{\nin}{n \in \NN}
\newcommand{\ikk}{\in \KK}
\newcommand{\kik}{k \in \KK}
\newcommand{\oco}{\kO\kcc(\gO)}
\newcommand{\ocod}{\kO\kcc(\gO_{d})}
\newcommand{\urb}{\overset{-1}}
\newcommand{\ffi}{\kf(\kI)} 
\newcommand{\qAm}{\kQ(\frA(M))}
\begin{document}

\title{\Huge{Observables \\ I : Stone Spectra}}

\author{Hans F.\ de Groote\footnote{degroote@math.uni-frankfurt.de;
FB Mathematik, J.W.Goethe-Universität Frankfurt a.\ M.}}

\titlepage
\maketitle
\begin{abstract}
    In this work we discuss the notion of observable - both quantum
    and classical - from a new point of view. In classical mechanics, 
    an observable is represented as a function (measurable, continuous
    or smooth), whereas in (von Neumann's approach to) quantum
    physics, an observable is represented as a bonded selfadjoint
    operator on Hilbert space. We will show in part II of this work
    that there is a common structure behind these two different
    concepts. If $\mathcal{R}$ is a von Neumann algebra, a selfadjoint
    element $A \in \mathcal{R}$ induces a continuous function $f_{A} :
    \mathcal{Q}(\mathcal{P(R)}) \to \mathbb{R}$ defined on the
    \emph{Stone spectrum} $\mathcal{Q}(\mathcal{P(R)})$ of the lattice
    $\mathcal{P(R)}$ of projections in $\mathcal{R}$. The Stone spectrum
    $\mathcal{Q}(\mathbb{L})$ of a general lattice $\mathbb{L}$ is the set
    of maximal dual ideals in $\mathbb{L}$, equipped with a canonical
    topology. $\mathcal{Q}(\mathbb{L})$ coincides with Stone's
    construction if $\mathbb{L}$ is a Boolean algebra (thereby
    ``Stone'') and is homeomorphic to the Gelfand spectrum of an
    abelian von Neumann algebra $\mathcal{R}$ in case of $\mathbb{L} =
    \mathcal{P(R)}$ (thereby ``spectrum''). Moreover,
    $\mathcal{Q}(\mathbb{L})$ appears quite naturally in the construction
    of the sheafification of presheaves on a lattice $\mathbb{L}$. On 
    the other hand, measurable or continuous functions can be
    described by spectral families and, therefore, as functions on
    appropriate Stone spectra. In this first part of our work, we
    investigate general properties of Stone spectra and, in more
    detail, Stone spectra of two specific classes of lattices:
    $\sigma$-algebras and projection lattices $\mathcal{P(R)}$ of von 
    Neumann algebras $\mathcal{R}$.  
\end{abstract}

\begin{center}
    {\large Für Karin}
\end{center}

\tableofcontents

\chapter{Introduction and Overview}
\label{In}
\pagestyle{myheadings}
\markboth{Introduction and Overview}{Introduction and Overview}

\begin{quote}
    \emph{Man hat viel erreicht, wenn einen sein Leben an ein volles Fass
    erinnert und nicht an einen leeren Eimer. \\
    (Hildegunst von Mythenmetz \cite{moers})}
\end{quote}

In this work I shall develop an unusual view of the notion of
\emph{observable}, both in quantum and in classical physics. \\
Following Araki \cite{ara}, an observable is an equivalence class of
measuring instruments, two measuring instruments being equivalent if
in any ``state'' of the ``physical system'' they lead upon a ``large
number of measurements'' to the same distribution of (relative
frequencies of) results. From this concept one can ``derive'' that
\begin{itemize}
    \item [(i)] in (von Neumann's axiomatic approach to) quantum physics,
    an observable is represented by a bounded selfadjoint operator $A$
    acting on a Hilbert space $\kh$, and  

    \item [(ii)] in classical mechanics, an observable is represented by a
    real valued (smooth or continuous or measurable) function on an
    appropriate phase space.
\end{itemize}
Here a natural question arises: is the structural difference
between classical and quantum observables fundamental, or is there
some background structure, showing that classical and quantum observables
are on the same footing? Indeed, such a background structure exists,
and I shall describe some of its features and consequences. \\
~\\
This work consists of three parts. In part I we introduce and study
the \emph{Stone spectrum of a lattice} $\LL$. This is a zero dimensional
Hausdorff space that is of twofold importance: it is the base space for
the etale space in the sheafification of a presheaf on $\LL$ and it is
a generalization of Stone's representation of Boolean algebras as
Boolean algebras of sets. If, in particular, $\LL$ is the lattice of
projections in an abelian von Neumann algebra $\kaa$, then the Stone
spectrum of $\LL$ is homeomorphic to the Gelfand spectrum of $\kaa$.   
Furthermore, we review in part I some basic definitions and results from
lattice theory and the theory of operator algebras.\\
In part II we show that the selfadjoint operators $A$ in a von Neumann
algebra $\rr$ can be represented by bounded continuous functions
$f_{A} : \qr \to \RR$ on the Stone spectrum $\qr$ of the projection
lattice $\pr$ of $\rr$. The mapping $A \tto f_{A}$ from $\hr$ to
$C_{b}(\qr, \RR)$ is injective, but it is surjective if and only if
$\rr$ is abelian. In this case, $f_{A}$ is the Gelfand transform of
$A$. The main result of part II is an abstract characterization of observable
functions. In the second chapter of part II we show that continuous
real valued functions on a Hausdorff space $M$ (``classical observables'')
can be characterized by certain spectral families in the lattice of
open subsets of $M$. Similar results are proved for measurable functions.\\
In part III we come back to the presheaf perspective and use the
abstract characterization of observable functions to define the
restriction of selfadjoint elements of a von Neumann algebra $\rr$ to 
a von Neumann subalgebra $\mm$ of $\rr$. This leads to the notion of
\emph{contextual observables} as global sections of a presheaf on the 
semi-lattice of abelian von Neumann subalgebras of $\rr$.\\
~\\
Now we describe the content in more detail.\\    
~\\
A continuous classical observable is a continuous function $f : M \to 
\RR$ on a (locally compact) Hausdorff space $M$. Equivalently, $f$ can
be considered as a global section of the presheaf $\kcc_{M}$ of all 
real valued continuous functions that are defined on some nonempty
open subsets of $M$. This situation leads to a natural generalization.
The set $\ktm$ of all open subsets of $M$ can be seen as a
\emph{complete lattice}\footnote{In the English language the word
``lattice'' has two different meanings. Either it is a subgroup of the
additive group $\ZZ^d$ for some $d \in \NN$ (this is called ``Gitter''
in German) or it means a partially ordered set with certain additional
properties. This is called ``Verband'' in German. We always use 
``lattice'' in this second meaning.} (definition \ref{P1})\footnote
{definition (n.k) refers to the k-th definition in chapter n. The same
system of internal reference is used for propositons, theorems etc.},
the lattice operations being defined by
\[
    \Ve_{k \ikk}U_{k} := \bigcup_{k \ikk}U_{k}, \qquad \We_{k \ikk}U_{k}
    := int (\bigcap_{k \ikk}U_{k}). 
\] 
It is straightforward to define presheaves and complete presheaves on 
an arbitrary complete lattice. (As is well known from topos theory
(\cite{mm}), the theory of presheaves can be built on an arbitrary
category.) It turns out, however, that on some important lattices,
like the lattice $\llh$ of all closed subspaces of a Hilbert space
$\kh$, there are no nontrivial complete presheaves. \\
It is well known that one can associate to each preasheaf $\kS_{M}$ on a
topological space $M$ a sheaf on $M$ in the following way: \\
If $\kS$ is a presheaf on a topological space $M$, 
i.e. on the lattice $\kT(M)$, then the corresponding \emph{etale
space} $\kE(\kS)$ of $\kS$ is the disjoint union of the \emph{stalks}
of $\kS$ at points in $M$:
\[  \kE(\kS) = \coprod_{x \in M}\kS_{x} \]
where
\[  \kS_{x} = \lim_{\overset{\longrightarrow}{U \in \frU}}\kS(U), \]
the \emph{inductive limit} of the family $(\kS(U))_{U \in \frU(x)}$
(here $\frU(x)$ denotes the set of all open neighbourhoods of $x$),
is the stalk in $x \in M$. The stalk $\kS_{x}$ consists of the
\emph{germs} in $x$ of elements $f \in \kS(U), \ U \in \frU(x)$. Germs
are defined quite analogously to the case of ordinary functions. Let
$\pi : \kE(\kS) \to M$ be the mapping that sends a germ in $x$ to its 
basepoint $x$. $\kE(\kS)$ can be given a topology for which $\pi$
is a local homeomorphism. It is easy to see that the local sections of
$\pi$ form a complete presheaf on $M$. If $\kS$ was already complete, 
then this presheaf of local sections of $\pi$ is isomorphic to $\kS$.\\
A first attempt to generalize this construction to the situation of a 
presheaf on a general lattice $\LL$ is to define a suitable notion of
``point in a lattice''. This can be done in a quite natural manner,
and it turns out that, for \emph{regular} topological spaces $M$,
the points in $\ktm$ are of the form $\frU(x)$, hence correspond to
the elements of $M$. But it also turns out that some important
lattices, like $\llh$, do not have points at all (proposition \ref{SS6})! \\  

For the definition of an inductive limit, however, we do not need a 
point, like $\frU(x)$, but only a partially ordered set $I$ with the property
\[  \forall \ \alpha, \beta \in I\ \  \exists \gamma \in I: \gamma \leq 
\alpha \ \  \mbox{and}  \ \ \gamma \leq \beta. \]
In other words: a \emph{filter base} $B$ in a lattice $\LL$ is 
sufficient. It is obvious how to define a filter base in an 
arbitrary lattice $\LL$ (definition \ref{SS7}). The set of all filter 
bases in $\LL$ is of course a rather unstructered object. 
Therefore it is reasonable to consider \emph{maximal} filter bases in 
$\LL$. (By Zorn's lemma, every filter base is contained in a maximal 
filter base in $\LL$.) This leads to the following
\pagebreak
\begin{definition}\label{in1}
    A nonempty subset $\frb$ of a lattice $\LL$ is 
    called a {\bf quasipoint} in $\LL$ if and only if it is a maximal
    subset of $\LL$ with the properties 
\begin{enumerate}
    \item  [(i)] $0 \notin \frb$,

    \item  [(ii)] $\all \ a, b \in \frb \ \ex \ c \in \frb : \ c ≤ a, \ 
    c ≤ b.$ 
\end{enumerate} 
\end{definition}
It is easy to see that a quasipoint is nothing else but a \emph{maximal
dual ideal} in $\LL$. By the way, it is rather obvious how to
generalize this definition to small categories. \\
~\\
In 1936 M.H.Stone (\cite{stone}) showed that the set $\kQ(\kbb)$ of 
quasipoints in a Boolean algebra $\kbb$ can be given a topology such 
that $\kQ(\kbb)$ is a \emph{compact zero dimensional} Hausdorff space 
and that the Boolean algebra $\kbb$ is isomorphic to the Boolean 
algebra of all \emph{closed open} subsets of $\kQ(\kbb)$. A basis for 
this topology is simply given by the sets
\[ \kQ_{a}(\kbb) := \{ \frb \in \kQ(\kbb) \mid a \in \frb \} \]
where $a$ is an arbitrary element of $\kbb$. \\
~\\
Of course we can generalize this construction to an arbitrary 
lattice $\LL$. For $a \in \LL$ let
\[ \kQ_{a}(\LL) := \{ \frb \in \kQ(\LL) \mid a \in \frb \}. \]
It is quite obvious from the definition of a quasipoint that
\[ \kQ_{a \wedge b}(\LL) = \kQ_{a}(\LL) \cap \kQ_{b}(\LL), \]
\[\kQ_{0}(\LL) = \emptyset \quad \text{and} \quad \kQ_{I}(\LL)
    = \kQ(\LL)  \]
hold. Hence $\{ \kQ_{a}(\LL) \mid a \in \LL \}$ is a basis for a 
topology on $\kQ(\LL)$. It is easy to see, using the 
maximality of quasipoints, that in this topology the sets 
$\kQ_{a}(\LL)$ are open and closed. Moreover, this topology is
Hausdorff, zero-dimensional, and therefore also completely regular. \\

\begin{definition}\label{in2}
    $\kQ(\LL)$, together with the topology defined by the basis $\{
    \qal \ | \ a \in \LL \}$, is called the {\bf Stone spectrum
    of the lattice $\LL$.}
\end{definition}

Then we can mimic the construction of the etale space of a presheaf on
a topological space $M$ to obtain from a presheaf $\kS$ on a lattice
$\LL$ an etale space $\kE(\kS)$ \emph{over the Stone spectrum $\ql$}
and a local homeomorphism $\pi_{\kS} : \kE(\kS) \to \ql$.
From the etale space $\kE(\kS)$ over $\kQ(\LL)$ we obtain 
a complete presheaf $\kS^{\kQ}$ on the topological space $\kQ(\LL)$ by
  \[ \kS^{\kQ}(\kV) := \gG(\kV,\ \kE(\kS)) \]
where $\kV \subseteq \kQ(\LL)$ is an open set and $\gG(\kV,\ 
\kE(\kS))$ is the set of {\bf sections of $\pi_{\kS}$ 
over $\kV$}, i.e. of all (necessarily continuous) mappings $s_{\kV} : \kV \to 
\kE(\kS)$ such that $\pi_{\kS} \circ s_{\kV} = id_{\kV}$. If $\kS$ 
is a presheaf of modules, then $\gG(\kV, \kE(\kS))$ is a module,
too.\\

\begin{definition}\label{in3}
   The complete presheaf $\kS^{\kQ}$ on the Stone spectrum $\kQ(\LL)$ 
   is called the {\bf sheaf associated to the presheaf $\kS$ on $\LL$}.
\end{definition}

Of course, Stone had quite another motivation for introducing the
space $\qb$ of a Boolean algebra $\kbb$, namely to represent $\kbb$ as
a Boolean algebra of sets. The remarkable fact is that we arrive at a
generalization of Stone's concept from a completely different point of
view.\\
~\\
In chapter \ref{SS} we will study properties of Stone spectra in
general and of some specific types of lattices. In particular, it is
shown that the Stone spectrum of a $\gs$-algebra $\frA$ of subsets of a
nonempty set $M$ is homeomorphic to the Gelfand spectrum of the
$C^\ast$-algebra $\kf_{\frA}(M, \CC)$ of all bounded $\frA$-measurable
functions $M \to \CC$. Quite analogously, the Stone spectrum of the 
projection lattice of an abelian von Neumann algebra $\kaa$ is homeomorphic
to its Gelfand spectrum. \\
Therefore, the Stone spectrum of an arbitrary von Neumann algebra is a
noncommutative generalization of the Gelfand spectrum of an abelian
von Neumann algebra. \\
But the real meaning of the Stone spectrum $\qr$ of a von Neumann
algebra $\rr$ is, that any selfadjoint element $A \in \rr$ can be
represented as a \emph{continuous} function $f_{A} : \qr \to \RR$.
Because selfadjoint operators are the (mathematical description of)
observables in Quantum Theory, we have coined the name
\emph{observable function of $A$} for $f_{A}$. If $A \in \hr$, and if 
$(\el)_{\lir}$ is the spectral resolution of $A$, then $f_{A}$ is
defined by
\[
    \all \ \frb \in \qr : \ f_{A}(\frb) := \inf \{ \lir \mid \el \in
    \frb \}.
\]  
As the Stone spectrum is a generalization of the Gelfand spectrum, the
mapping $A \tto f_{A}$ will be proved to be a generalization of the Gelfand
transformation. We motivate the definition of $f_{A}$ in Part II
using the presheaf of bounded spectral families in the lattice $\ph$ of
projections in $\lh$.\\
~\\
In Part II we shall study the properties of observable
functions for general von Neumann algebras $\rr$. It will be shown
that observable functions are continuous and that the range of $f_{A}$
is precisely the spectrum of the operator $A$. But the mapping $A \tto
f_{A}$ from $\hr$, the real vectorspace of selfadjoint operators in
$\rr$, to $C_{b}(\qr, \RR)$, the real vectorspace of real valued
bounded continuous functions on $\qr$, is linear if and only if $\rr$ 
is abelian. This may appear as a shortcoming of the theory, because
linear structures are indispensable in the theory of operator algebras.
From the physical point of view, however, the possibility of adding 
two given observables to obtain a new one, is merely a mathematical
reflex: what is the meaning of the sum of the position and the
momentum operator or the sum of two different spin operators? Perhaps 
a similar question appears with the completion $\RR$ of the rationals 
$\QQ$: it is indispensable for analysis, but in the light of quantum theory
it is worth to debate whether the continuum is of physical
significance or not (\cite{ish4}).      \\
~\\
For the case $\rr = \lh$, we give an abstract characterization of
observable functions, considered as functions on projective Hilbert space
$\pph$. We generalize this characterization for arbitrary von Neumann
algebras. In order to achieve this, we extend the domain of definition of
$f_{A}$ from the Stone spectrum$\qr$ to the space $\dr$ of \emph{all}
dual ideals in $\pr$ in an obvious manner:
\[
    \all \ \kj \in \dr : \ f_{A}(\kj) := \inf \{ \lir \mid \el \in \kj
    \}. 
\]
The space $\dr$ can be equipped with a topology in the very same way
as $\qr$. It is not difficult to show that observable functions
$f_{A}$, considered as functions on $\dr$, have the following
properties:
\begin{enumerate}
    \item  [(i)] Let $(\kj_{j})_{j \in J}$ be a family in $\dr$. Then
    \[
	f_{A}(\bigcap_{j \in J}\kj_{j}) = \sup_{j \in J}f_{A}(\kj_{j}).
    \]

    \item  [(ii)] $f_{A} : \dr \to \RR$ is upper semicontinuous.
\end{enumerate}
Giving $(i)$ and $(ii)$ the status of \emph{defining} properties, we
get the notion of an \emph{abstract observable function}.

\begin{definition}\label{in4}
    A function $f : \dr \to \RR$ is called an {\bf abstract observable
    function} if it is \emph{upper semicontinuous} and satisfies the
    \emph{intersection condition}
    \[
	f(\bigcap_{j \in J}\kj_{j}) = \sup_{j \in J}f(\kj_{j}) 
    \]
    \nopagebreak
    for all families $(\kj_{j})_{j \in J}$ in $\dr$.
\end{definition}

The intersection condition implies that an abstract observable
function is decreasing. Let 
\[
    H_{P} := \{ Q \in \pr \mid Q ≥ P \}
\]
be the \emph{principle dual ideal} in $\pr$, defined by $P \in \por$ 
($\por := \pr \smm \{0\}$). Then the definition of abstract
observable functions can be reformulated so that it does not refer to 
the topology of $\dr$:

\begin{remark}\label{in5}
    $f : \dr \to \RR$ is an observable function if and only if the
    following two properties hold for $f$:
    \begin{enumerate}
       \item  [(i)]  $\all \ \kj \in \dr  : \ f(\kj) = \inf \{ f(H_{P}) |
       \ P \in \kj \}$,
       
       \item  [(ii)] $f(\bigcap_{j \in J}\kj_{j}) = \sup_{j \in J}f(\kj_{j})$
	for all families $(\kj_{j})_{j \in J}$ in $\dr$.
    \end{enumerate}
\end{remark}

\noindent{The} central result in Part II, is the following

\begin{theorem}\label{in6}
    Let $f : \dr \to \RR$ be an abstract observable function. Then
    there is a unique $A \in \hr$ such that $f = f_{A}$.
\end{theorem}

\noindent{In} fact, this is a theorem about an abstract characterization
of spectral families, and an inspection of its proof shows that it
also holds for spectral families in any complete orthomodular lattice.\\
~\\
The set of non-zero elements $a$ of a lattice $\LL$ is in one-to-one
correspondence to the set $\dprl$ of principal dual ideals $H_{a}$ in 
$\LL$. Hence any bounded function
$r : \LL \smm \{0\} \to \RR$ induces by 
\[
    \all \ \kj \in \dr  : \ f(\kj) := \inf \{ r(a) | \ a \in \kj \}
\]
a function $f : \dr \to \RR$. Of course $r$ must satisfy some
condition so that $f$ becomes an (abstract) observable function. In a 
complete lattice we have
\[
    H_{\Ve_{\kik}a_{k}} = \bigcap_{\kik}H_{a_{k}}.
\]
A necessary condition for $f$ to be an observable function is
therefore
\[
    f(H_{\Ve_{\kik}a_{k}}) = \sup_{\kik}f(H_{a_{k}}).
\] 
If $\LL$ is a complete lattice, this requirement leads to the condition 
that
\[
     r(\Ve_{k \in \KK}a_{k}) = \sup_{k \in \KK}r(a_{k})
\]
must be satisfied for every family $(a_{k})_{k \in \KK}$ in
$\LL \smm \{0\}$. In this case $r$ is called \emph{completely
increasing}. If this latter condition is fulfilled, then
\[
    f(H_{a}) = r(a)
\]
for all $a \in \LL \smm \{0\}$ and $f$ is an observable function.
Conversely, if $f : \dl \to \RR$ is an observable function, then
$r_{f}(a) := f(H_{a})$ defines a completely increasing function $r_{f}
: \LL \smm \{0\} \to \RR$. This gives a bijection $f \tto r_{f}$
between observable functions and completely increasing functions.\\
~\\
If $M$ is a nonempty set and $\frA$ is a $\gs$-algebra of subsets
of $M$, then every $\frA$-measurable function $g : M \to \RR$ defines 
a spectral family $\gs_{g}$ in $\frA$ by
\[
    \all \ \lir : \gs_{g}(\gl) := \urb{g}(]-\∞, \gl]).
\] 
Conversely, any spectral family $\gs$ in $\frA$ induces a function
$g_{\gs} : M \to \RR$ by 
\[
    \all \ x \in M : \ g_{\gs}(x) := \inf \{ \lir \mid x \in \gs(\gl) 
    \}.
\]
$g_{\gs}$ is $\frA$-measurable because 
\[
    \all \ \lir : \ \urb{g_{\gs}}(]-\∞, \gl]) = \gs(\gl).
\]
Moreover, we will show that these constructions are inverse to each
other, i.e. 
\[
    g_{\gs_{g}} = g \quad \text{and} \quad \gs_{g_{\gs}} = \gs
\]
for all $\frA$-measurable functions $g : M \to \RR$ and all spectral
families $\gs$ in $\frA$. On the other hand, every $\frA$-measurable
function $g : M \to \RR$ induces a function $f_{g} : \qA \to \RR$ on the
Stone spectrum $\qA$ of $\frA$, defined by
\[
    f_{g}(\frb) := \inf \{ \lir \mid \gs_{g}(\gl) \in \frb \}.
\]
We will show that $f_{g}$ is the Gelfand transformation of $g$:

\begin{theorem}\label{in7}
    Let $\frA(M)$ be a $\gs$-algebra of subsets of a nonempty set $M$ and
    let $\kf_{\frA(M)}(M, \CC)$ be the $C^\ast$-algebra of all bounded
    $\frA(M)$-measurable functions $g : M \to \CC$. Then the Gelfand
    spectrum $\gO(\kf_{\frA(M)}(M, \CC))$ of $\kf_{\frA(M)}(M, \CC)$ is
    homeomorphic to the Stone spectrum $\qAm$ of $\frA(M)$ and the
    restriction of the Gelfand transformation to $\kf_{\frA(M)}(M, \RR)$ is
    given, up to the homeomorphism $\qAm \cong 
    \gO(\kf_{\frA(M)}(M, \CC))$, by $g \tto f_{g}$, where 
    \[
	f_{g}(\frb) = \inf \{ \lir \mid \urb{g}(]-\∞, \gl]) \in \frb \}  
    \]
    for all $\frb \in \qAm$.
\end{theorem}
Moreover, we will generalize this theorem to $\gs$-algebras of the
form $\frA(M) / \kI$, where $\kI$ is a $\gs$-ideal in $\frA(M)$, which
is, by a theorem of Loomis and Sikorski (\cite{Sik}), up to isomorphy 
the general form of $\gs$-algebras. 
~\\
If $M$ is a Hausdorff space, then the interplay between
\emph{continuous} functions $f : M \to \RR$ and spectral families
$\gs$ in the lattice $\ktm$ of all open subsets of $M$ is not as simple
as in the measurable case. This is due to the fact that for a family
$(U_{k})_{\kik}$ the infimum $\We_{\kik}U_{k}$ may the empty set but
$\bigcap_{\kik}U_{k}$ is not empty.\\
Every continuous function $f : M \to \RR$ defines a spectral family
$\gs_{f}$ in $\ktm$ by
\[
    \gs_{f}(\gl) := int \urb{f}(]-\∞, \gl]).
\] 
Conversely, if a spectral family $\gs$ in $\ktm$ is given, then 
\[
     f_{\gs}(x) := \inf \{ \lir \mid x \in \gs(\gl) \}
\] 
is not necessarily defined for all $x \in M$. This leads to the
following

\begin{definition}\label{in8}
  Let $\gs : \RR \to \kT(M)$ be a spectral family in $\kT(M)$. Then
  \begin{displaymath}
	    \kD(\gs) := \{x \in M \mid \exists \ \gl \in \RR : \ x \notin 
		    \gs(\gl) \}
  \end{displaymath}
  is called the {\bf admissible domain of $\gs$}.
\end{definition}
It is easy to see that $\kD(\gs)$ is dense in $M$.

\begin{definition}\label{in9}
  Let $\gs : \RR \to \kT(M)$ be a spectral family with admissible 
  domain $\kD(\gs)$. Then the function $f_{\gs} : \kD(\gs) \to 
  \RR$, defined by
  \begin{displaymath}
      \forall \ x \in \kD(\gs) : \ f_{\gs}(x) := \inf \{\gl \in \RR 
		    \mid x \in \gs(\gl) \},
  \end{displaymath}
  is called the {\bf function induced by $\gs$}.
\end{definition}
If $\gs = \gs_{f}$ for a \emph{continuous} function $f : M \to \RR$,
then $\gs$ is regular in the following sense:

\begin{definition}\label{in10}
A spectral family $\gs : \RR \to \kT(M)$ is called {\bf regular} if
\begin{displaymath}
	  \forall \ \gl < \mu : \ \overline{\gs(\gl)} \tm \gs(\mu)
\end{displaymath}
holds.
\end{definition}      
If $\gs$ is a regular spectral family in $\ktm$, then each $\ssl$ is a
regular open set, i.e. it is the interior of its closure. Thus a regular
spectral family has values in the complete Boolean algebra
$\kT_{r}(M)$ of regular open subsets of $M$. 

\begin{theorem}\label{in11}
Let $M$ be a Hausdorff space. Then every continuous function $f : 
M \to \RR$ induces a regular spectral family $\gs_{f} : \RR 
\to \kT(M)$ by
\begin{displaymath}
  \forall \ \gl \in \RR : \ \gs_{f}(\gl) := int(\overset{-1}{f}(]-\infty,
  \gl])).
\end{displaymath}
The admissible domain $\kD(\gs_{f})$ equals $M$ and the function 
$f_{\gs_{f}} : M \to \RR$ induced by $\gs_{f}$ is $f$. 
Conversely, if $\gs : \RR \to \kT(M)$ is a regular spectral 
family, then the admissible domain of $\gs$ is open and dense in $M$, 
the function
\begin{displaymath}
		  f_{\gs} : \kD(\gs) \to \RR
\end{displaymath}
induced by $\gs$ is continuous and the induced spectral family 
$\gs_{f_{\gs}}$ in $\kT(\kD(\gs))$ is the restriction of $\gs$ to 
the admissible domain $\kD(\gs)$:
\begin{displaymath}
  \forall \gl \in \RR : \ \gs_{f_{\gs}}(\gl) = \gs(\gl) \cap \kD(\gs). 
\end{displaymath}
\end{theorem}
One may wonder why we have defined the function, that is induced by a
spectral family $\gs$, on $M$ and not on the Stone spectrum $\qm$. A
quasipoint $\frb \in \qm$ is called \emph{finite} if $\bigcap_{U \in
\frb}\overline{U} \ne \emptyset$. If $\frb$ is finite, then this
intersection consists of a single element $x_{\frb} \in M$, and we call
$\frb$ a quasipoint \emph{over $x_{\frb}$}. Note that for a compact space
$M$, all quasipoints are finite. Moreover, one can show that for
compact $M$, the mapping $pt : \frb \tto x_{\frb}$ from $\qm$ onto $M$
is continuous and \emph{identifying}.

\begin{remark}\label{in12}
    Let $\gs : \RR \to \kT(M)$ be a regular spectral family and let
    $x \in \kD(\gs)$. Then for all quasipoints $\frb_{x} \in \qm$ over
    $x$ we have
    \[
	f_{\gs}(\frb_{x}) = f_{\gs}(x).
    \]
\end{remark}
Therefore, if $M$ is compact, it makes no difference whether we define
$f_{\gs}$ in $M$ or in $\qm$.  \\
~\\
In part III we come back to the presheaf perspective. Basic to that is
the \emph{semilattice} $\frAr$ of all \emph{abelian} von Neumann
subalgebras of a von Neumann algebra $\rr$. It can be seen also as the
set of objects of a (small) \emph{category} $\cor$ whose morphisms are
simply the inclusion maps. It is called the \emph{context category of 
the von Neumann algebra $\rr$.} Let $\kaa, \kbb \in \frAr$ such that $\kaa 
\tm \kbb$, we can define, using the results of part II, a restriction 
map
\[
    \kba : \kbb \to \kaa
\]
in the following way. Identify $B \in \kbb_{sa}$ with the corresponding
completely increasing function $r_{B} : \pob \to \RR$. Then $r_{\kba B}
:= {r_{B}}_{\mid_{\poa}} : \poa \to \RR$ is completely increasing and,
therefore, corresponds to a unique $\kba B \in \kaa_{sa}$. For an
arbitrary $B \in \kbb$ we define $\kba B := \kba B_{1} + i \kba
B_{2}$, where $B = B_{1} + i B_{2}$ is the decomposition of $B$ into
selfadjoint parts. Obviously, the abelian von Neumann subalgebras of
$\rr$, together with the restriction maps $\kba$, form a presheaf
\[
    \gT_{\rr} := (\gT(\kaa), \kba)_{\kaa \tm \kbb} := (\kaa,
    \kba)_{\kaa \tm \kbb}
\] 
on $\cor$. We call $\gT_{\rr}$ the \emph{tautological presheaf} on
$\cor$. The presheaf
\[
    \gT_{\hr} := (\kaa_{sa}, \kba)_{\kaa \tm \kbb}
\]
on $\cor$ is a sub-presheaf of $\gT_{\rr}$, which is called the
\emph{real tautological presheaf} on $\cor$ because
\[
    \gT_{\rr} = \gT_{\hr} \oplus i \gT_{\hr}.
\] 
Of course the definition of restricting an operator to a von Neumann
subalgebra works for any pair $(\mm, \nn)$ of von Neumann algebras
such that $\mm \tm \nn$. We will interpret the restriction
$\gr^\rr_{\mm}A$ of $A \in \rr$ to a von Neumann subalgebra $\mm$ of
$\rr$ as a \emph{coarse graining} of $A$. This can already be seen in 
the following example: if $P \in \rr$ is a projection, then
$\gr^\rr_{\mm}P = s_{\mm}(P)$, where $s_{\mm}(P) := \We \{ Q \in \pmm
\mid Q ≥ P \}$ is the $\mm$-support of $P$.\\  
If $A$ is an observable, i.e. $A \in \hr$, then the family
$(A_{\kaa})_{\kaa \in \frAr}$, defined by $A_{\kaa} :=
\gr^\rr_{\mm}A$, is a \emph{global section} of the presheaf
$\gT_{\hr}$. This means that the family $(A_{\kaa})_{\kaa \in \frAr}$ 
satisfies the conditions
\[
    A_{\kaa} = \kba A_{\kbb} \quad \text{if} \quad \kaa \tm \kbb.
\]
Here the question arises whether every global section of $\gT_{\hr}$
is induced by an operator $A \in \hr$. This is trivially the case if
$\rr$ is abelian, but not for $\rr = \kL(\CC^2)$. However, it is not
only the notorious type $I_{2}$ exception: we will present a
generalizable example for $\kL(\CC^3)$. The reason for that phenomenon
lies in the following result:

\begin{proposition}\label{in13}
    Let $\rr$ be a von Neumann algebra. There is a one-to-one
    correspondence between global sections of the real tautological presheaf
    $\gT_{\hr}$ and functions $f : \por \to \RR$ that satisfy
    \begin{enumerate}
	\item  [(i)] $f(\Ve_{k \ikk}P_{k}) = \sup_{k \ikk}f(P_{k})$ for all
	{\bf commuting} families $(P_{k})_{k \ikk}$ in $\por$,
    
	\item  [(ii)] $f_{|_{\por \cap \kaa}}$ is bounded for all
	$\kaa \in \frAr$.
    \end{enumerate}
\end{proposition}
Therefore, if one takes contextuality in quantum physics serious, it
is natural to generalize the notion of quantum observable:

\begin{definition}\label{in14}
    Let $\rr$ be a von Neumann algebra. The global sections of the
    real tautological presheaf $\gT_{\hr}$ are called {\bf contextual
    observables}.
\end{definition}
Moreover, we will discuss in part III some applications of our theory 
to positive operator valued measures.\\
~\\
To finish this introduction, I like to stress that the development of 
the theory presented here was \emph{motivated} by conceptual notions of
physics, it has been \emph{guided}, however, by mathematical naturalness.
This is not only due to the fact that I am a mathematician (although
with strong inclination to physics), but mainly to my belief that the 
ultimate theory of physics will be in good mathematical shape.

\chapter{Preliminaries}
\label{P}

In this chapter, we present the basic definitions and results from
lattice theory and the theory of operator algebras that we shall use
throughout this work. We omit proofs for most of the presented results
because they can be found in the standard literature. An exception is 
section \ref{ol} which contains complete proofs.    

\section{Lattices}
\label{latt}
\pagestyle{myheadings}
\markboth{Preliminaries}{Lattices}

\begin{definition}\label{P1}
A \emph{lattice} is a partially ordered set $(\LL, \leq)$  such 
that any two elements $a,b \in\LL$ possess a \emph{maximum} $a\vee b 
\in\LL$ and a \emph{minimum} $a\wedge b \in\LL$.\\
Let $\frm$ be an infinite cardinal number.\\
The lattice $\LL$ is called $\mathfrak{m}$-complete, if every family 
$(a_{i})_{i\in I}$ has a supremum $\bigvee_{i \in I}a_{i}$ and an 
infimum $\bigwedge_{i\in I}a_{i}$ in $\LL$, provided that $\# I\leq 
\mathfrak{m}$ holds.
A lattice $\LL$ is simply called complete, if every family 
$(a_{i})_{i \in I}$ in $\LL$ (without any restriction of the 
cardinality of $I$) has a supremum and an infimum in $\LL$.\\
$\LL$ is said to be boundedly complete if every bounded family in 
$\LL$ has a supremum and an infimum. \\
If a lattice has a \emph{zero element} $0$ ( i.e. $\forall a \in \LL: 
0 \leq a$) and a \emph{unit element} $1$ (i.e. $\forall a\in \LL : a\leq 
1$), completeness and bounded completeness are the same.\\
A lattice $\LL$ is called \emph{distributive} if the two distributive 
laws
\begin{eqnarray*}
       a \wedge (b \vee c) & = & (a \wedge b) \vee (a \wedge c)  \\
       a \vee (b \wedge c) & = & (a \vee b) \wedge (a \vee c)
\end{eqnarray*}
hold for all elements $a, b, c \in \LL$. 
\end{definition}

In fact it is an easy exercise to show that if one of these
distributive laws is satisfied for all $a, b, c \in \LL$, so is the
other. \\

$\bigvee_{i \in I}a_{i}$ is characterized by the following universal 
property:
\begin{enumerate}
       \item [(i)]  $\forall j\in I :\quad a_{j}\leq \bigvee_{i \in I}a_{i}$

       \item [(ii)]  $\forall c\in \LL :\quad ((\forall i\in I : a_{i} \leq c) 
       \Rightarrow \bigvee_{i}a_{i} \leq c ).$
\end{enumerate}
An analogous universal property characterizes the infimum 
$\bigwedge_{i}a_{i}$.

Note that if  $\LL$ is a distributive complete lattice, then in 
general
\begin{displaymath}
       a \wedge (\bigvee_{i \in I}b_{i}) \ne \bigvee_{i \in I}(a \wedge 
       b_{i}),
\end{displaymath}
so completeness and distributivity together do not imply \emph{complete 
distributivity}! \\

Let us give some important examples.
\begin{example}\label{P2}
Let $M$ be a topological space and $\kT(M)$ the topology of $M$, i.e. 
the set of all open subsets of $M$. $\kT(M)$ is a completely distributive
lattice. The supremum of a family $(U_{i})_{i \in I}$ of open 
subsets $U_{i}$ of $M$ is given by
\begin{displaymath}
	\bigvee_{i \in I}U_{i} = \bigcup_{i \in I}U_{i},
\end{displaymath}
the infimum, however, is given by
\begin{displaymath}
	\bigwedge_{i \in I}U_{i} = int(\bigcap_{i \in I}U_{i}),
\end{displaymath}
where $int N$ denotes the interior of a subset $N$ of $M$.
\end{example}

\begin{example}\label{P3}
If $U \in \kT(M)$, then always
\begin{displaymath}
	U \tm int \bar{U},
\end{displaymath}
but $U \ne int \bar{U}$ in general. $U$ fails to be the interior of 
its adherence $\bar{U}$, if for example $U$ has a ``crack'' or is 
obtained from an open set $V$ by deleting some points of $V$.\\
We call $U$ a {\bf regular open set}, if $U = int \bar{U}$. Each $U 
\in \kT(M)$ has a {\bf pseudocomplement}, defined by
\begin{displaymath}
	U^{c} := M \setminus \bar{U},
\end{displaymath}
and together with the operation of pseudocomplementation $\kT(M)$ is 
a \mbox{{\bf Heyting algebra}}:
\begin{displaymath}
	\forall \ U \in \kT(M) : \ U^{ccc} = U^{c}.
\end{displaymath}
$U \in \kT(M)$ is regular if and only if $U = U^{cc}$. Let 
$\kT_{r}(M)$ be the set of regular open subsets of $M$. If $U, V \in 
\kT_{r}(M)$, then also $U \cap V \in \kT_{r}(M)$. The union of two 
regular open sets, however, is not regular in general. Therefore one 
is forced to define the maximum of two elements $U, V \in \kT_{r}(M)$ 
as
\begin{displaymath}
	U \vee V := (U \cup V)^{cc}.
\end{displaymath}
It is then easy to see that $\kT_{r}(M)$ is a distributive complete 
lattice with the lattice operations
\begin{displaymath}
	U \wedge V := U \cap V, \quad U \vee V := (U \cup V)^{cc}.
\end{displaymath}
The pseudocomplement on $\kT(M)$, restricted to $\kT_{r}(M)$, gives 
an {\bf orthocomplement} $U \mapsto U^{c}$ on $\kT_{r}(M)$:
\begin{displaymath}
	U^{cc} = U, \ U^{c} \vee U = M, \ U^{c} \wedge U = \emptyset, \ (U 
	\wedge V)^{c} = U^{c} \vee V^{c}
\end{displaymath}
for all $U, V \in \kT_{r}(M)$. Thus $\kT_{r}(M)$ is a {\bf complete 
Boolean lattice} i.e. a complete Boolean algebra.
\end{example}

\begin{example}\label{P4}
Let $M$ be a topological space and $\kbb(M)$ the set of Borel subsets 
of $M$. $\kbb(M)$ together with the usual set theoretic operations is a 
distributive $\aleph_{0}$-complete Boolean lattice, usually called 
the $\gs$-algebra of Borel subsets of $M$.
\end{example}

\begin{example}\label{P5}
Let $\kh$ be a (complex) Hilbert space and $\LL(\kh)$ the set of all 
closed subspaces of $\kh$. $\LL(\kh)$ is a complete lattice with 
lattice operations defined by
\begin{eqnarray*}
	U \wedge V & := & U \cap V  \\
	U \vee V & := & (U + V)^{-}  \\
	U^{\perp} & := & \text{orthogonal complement of $U$ in $\kh$.}
\end{eqnarray*}
Contrary to the foregoing examples, $\LL(\kh)$ is highly {\bf 
non-distributive}!\\
Of course $\LL(\kh)$ is isomorphic to the lattice $\ph := 
\{P_{U} \mid U \in \LL(\kh) \}$ of all orthogonal projections in the 
algebra $\kL(\kh)$ of bounded linear operators of $\kh$. The 
non-distributivity of $\LL(\kh)$ is equivalent to the fact that two 
projections $P_{U}, P_{V} \in \ph$ do not commute in 
general.\\
$\LL(\kh)$ is the basic lattice of quantum mechanics (\cite{Jauch}). 
It represents ``quantum logic'' in contrast to classical ``Boolean logic''. \\
\end{example}

\section{Orthomodular Lattices}
\label{ol}
\pagestyle{myheadings}
\markboth{Preliminaries}{Orthomodular Lattices}

The most prominent examples of \emph{orthomodular lattices} are
Boolean algebras and the lattice of projections in a von Neumann
algebra. A less popular example is the lattice of \emph{causally
closed subsets} of a spacetime (\cite{cas}). For the sake of
completeness we give here the necessary definitions and prove the
results we will use. Of course neither the results nor, probably,
the presented proofs are new. Our general references are \cite{birk}
and \cite{kal}.

\begin{definition}\label{ol1}
    Let $\LL$ be a lattice with a minimal element $0$ and a maximal
    element $1$. An orthocomplement for $\LL$ is a mapping 
    $\perp : \LL \to \LL, \ \ a \tto a^\perp$ with the following
    properties:
    \begin{itemize}
       \item  [(i)] $a \we a^\perp = 0, \ \ a \vee a^\perp = 1$
       for all $a \in \LL$, 
    
       \item  [(ii)] $(a \we b)^\perp = a^\perp \vee b^\perp, \ \  (a
       \vee b)^\perp = a^\perp \we b^\perp$ for all $a, b \in \LL$,
    
       \item  [(iii)] $a^{\perp\perp} = a$ for all $a \in \LL$.
    \end{itemize}
    $\LL$ together with an orthocomplement $\perp$ is called an
    orthocomplemented lattice (or an ortholattice for short). 
\end{definition}
Immediate consequences of these definitions are

\begin{remark}\label{ol2}
    \begin{itemize}
       \item  [(1)] $0^\perp = 1$,

       \item  [(2)] $1^\perp = 0$,

       \item  [(3)] $a ≤ b \ \llra \ b^\perp ≤ a^\perp$, 
       and, if $\LL$ is complete, 
    
       \item  [(4)] $(\We_{k \in \KK}a_{k})^\perp = \Ve_{k \in
       \KK}a_{k}^\perp$ and $(\Ve_{k \in \KK}a_{k})^\perp = \We_{k \in
       \KK}a_{k}^\perp$ for all families $(a_{k})_{k \in \KK}$ in $\LL$.
    \end{itemize}
\end{remark}
\emph{Proof:} (1) and (2) follow from $a = (a^\perp \vee 0)^\perp = 
a \we 0^\perp$ and $a = (a^\perp \we 1)^\perp = a \vee 1^\perp$,
(3) follows from $a ≤ b \ \ \llra \ \ a \we b = a \ \ \llra a \vee b =
b$, and (4) from the universal property of meet and join:
\begin{eqnarray*}
    \Ve_{i}a_{i}^\perp ≥ a_{k}^\perp & \lra &
    (\Ve_{i}a_{i}^\perp)^\perp ≤ a_{k}  \\
     & \lra & (\Ve_{i}a_{i}^\perp)^\perp ≤ \We_{k}a_{k}  \\
     & \lra & \Ve_{i}a_{i}^\perp ≥ (\We_{k}a_{k})^\perp
\end{eqnarray*}
and
\begin{eqnarray*}
    \We_{i}a_{i} ≤ a_{k} & \lra & (\We_{i}a_{i})^\perp ≥ a_{k}^\perp  \\
     & \lra & (\We_{i}a_{i})^\perp ≥ \Ve_{k}a_{k}^\perp. \ \ \Box
\end{eqnarray*}

\begin{definition}\label{ol3}
    A lattice $\LL$ is called modular if 
    \[
	\all \ a, b, c \in \LL : \ (b ≤ a \ \ \lra \ \ a \we (b \vee
	c) = (a \we b) \vee (a \we c))
    \]
    holds.
\end{definition}

The projection lattice $\pr$ of a von Neumann algebra $\rr$ is modular
if $\rr$ is \emph{finite} (\cite{tak1}). In general $\pr$ is only
\emph{orthomodular}:

\begin{definition}\label{ol4}
    An ortholattice $\LL$ is called orthomodular if
    \[
	\all \ a, b, c \in \LL : \ (b ≤ a \ \ \text{and} \ \  c ≤ a^\pp \ \ \lra 
	\ \ a \we (b \vee c) = (a \we b) \vee (a \we c))
    \]
    holds.
\end{definition}
It is easy to see that the orthomodular law is equivalent to 
\[
    \all \ a, b \in \LL : \ (b ≤ a \ \ \lra \ \ b = a \we (a^\pp \vee 
    b)).
\]
An important consequence of orthomodularity is that an element $d \in 
\LL$ has at most one decomposition $d = b \vee c$ with $b ≤ a$ and
$c ≤ a^\pp$. \\

One can define commutativity of an element $a$ with an element $b$
in an arbitrary ortholattice $\LL$ by the relation
\[
    a \kcc b \ \ :\llra \ \ a = (a \we b) \vee (a \we b^\pp). 
\]
However, the relation $\kcc$ is \emph{not symmetric} in general. In
fact the symmetry of $\kcc$ is equivalent to the orthomodularity of
$\LL$:
\begin{proposition}(Nakamura)\label{ol5}
    The commutativity relation $\kcc$ in $\LL$ is symmetric if and
    only if $\LL$ is orthomodular.
\end{proposition}
\emph{Proof:} If $\LL$ is orthomodular and $(a, b) \in
\kcc$, i.e. $a = (a \we b) \vee (a \we b^\pp)$, then $a^\pp = (a \we
b)^\pp \we (a^\pp \vee b)$, hence 
\[
     a^\pp \we b = (a \we b)^\pp \we (a^\pp \vee b) \we b = (a \we
     b)^\pp \we b,
\]  
i.e. 
\[
    ((a \we b) \vee (a^\pp \we b))^\pp = (a \we b)^\pp \we ((a \we b) 
    \vee b^\pp).
\]
$a \we b ≤ b$ implies $b^\pp ≤ (a \we b)^\pp$. Then we obtain, using
the orthomodularity of $\LL$,
\[
     (a \we b)^\pp \we ((a \we b) \vee b^\pp) = b^\pp,
\] 
hence 
\[
    b = (a \we b) \vee (a^\pp \we b),
\]
i.e. $(b, a) \in \kcc $. \\
Conversely, let $\kcc$ be symmetric. If $b ≤ a$ then $(b, a) \in \kcc$
and therefore $(a, b) \in \kcc$. Hence $(a, b^\pp)$ and so $(b^\pp, a)$
and $(b^\pp, a^\pp)$ also belong to $\kcc$. We obtain 
\[
    (a \we (a^\pp \vee b))^\pp = a^\pp \vee (a \we b^\pp) = (a^\pp \we
    b^\pp) \vee (a \we b^\pp) = b^\pp.
\]
This shows that $\LL$ is orthomodular. \ \ $\Box$ \\

\begin{definition}\label{ol6}
    Let $\LL$ be an orthomodular lattice and let $M$ be a (nonvoid)
    subset of $\LL$. Then 
    \[
	M^\kcc := \{ b \in \LL \ | \ \all \ a \in M : \ a \kcc b \}
    \]
    is called the commutant of $M$ in $\LL$. If $M = \{a\}$, we simply
    write $a^\kcc$ instead of $\{a\}^\kcc$. 
\end{definition}

\begin{proposition}\label{ol7}
    Let $M$ be a subset of an orthomodular lattice $\LL$. Then
    $M^\kcc$ is a lattice. If $\LL$ is complete then $M^\kcc$ is
    complete, too. 
\end{proposition}
\emph{Proof:} Without loss of generality we can assume that $M = \{a\}$.
Obviously $0, 1 \in a^\kcc$ and $b^\pp \in a^\kcc$ for $b \in
\kcc$. Let $(b_{k})_{k \in \KK}$ be an arbitrary family in $a^\kcc$,
where we assume that $\KK$ is finite if $\LL$ is not complete. Then 
\begin{eqnarray*}
    \Ve_{k}b_{k} & = & \Ve_{k}((b_{k} \we a) \vee (b_{k} \we a^\pp))  \\
     & = & (\Ve_{k}(b_{k} \we a)) \vee (\Ve_{k}(b_{k} \we a^\pp))  \\
     & ≤ & ((\Ve_{k}b_{k}) \we a) \vee ((\Ve_{k}b_{k}) \we a^\pp)  \\
     & ≤ & \Ve_{k}b_{k}.
\end{eqnarray*}
Hence
\[
    ((\Ve_{k}b_{k}) \we a) \vee ((\Ve_{k}b_{k}) \we a^\pp) = \Ve_{k}b_{k}
\]
and, by orthomodularity,
\[
    (\Ve_{k}b_{k}) \we a = \Ve_{k}(b_{k} \we a).
\]
The case of the meet $\We_{k}b_{k}$ reduces to that of the join
because of $\We_{k}b_{k} = (\Ve_{k}b_{k}^\pp)^\pp$. \ \ $\Box$ \\

\begin{remark}\label{ol8}
    In the course of the foregoing proof we have also shown that $M$
    distributes over $M^\kcc$, i.e.
    \[
	\all \ a \in M, (b_{k})_{k \in \KK} \ \text{in} \ M^\kcc : \ a
	\we (\Ve_{k}b_{k}) = \Ve_{k}(a \we b_{k}).
    \]
\end{remark}

\begin{lemma}\label{ol9}
    Let $M$ and $N$ be subsets of an orthomodular lattice $\LL$. Then 
    \begin{enumerate}
       \item  [(i)] $M \tm N \ \lra \ N^\kcc \tm M^\kcc$,

       \item  [(ii)] $M \tm M^{\kcc\kcc}$,

       \item  [(iii)] $M^\kcc = M^{\kcc\kcc\kcc}$.
    \end{enumerate}
\end{lemma}
\emph{Proof:} Property $(i)$ is obvious from the definition of $\kcc$.
Properties $(ii)$ and $(iii)$ are essentially consequences of the
symmetry of $\kcc$: If $a \in M$ and $b \in M^\kcc$, then $b \kcc a$, 
hence $a \kcc b$ and therefore $a \in M^{\kcc\kcc}$. This proves
$(ii)$. From $(i)$ and $(ii)$ we obtain $M^{\kcc\kcc\kcc} \tm M^\kcc$ 
and $(ii)$ implies the opposite inclusion. \ \ $\Box$ \\

\begin{proposition}\label{ol10}
    Let $\LL$ be a complete orthomodular lattice and $\MM \tm \LL$ a
    sublattice. Then 
    \begin{enumerate}
       \item  [(i)] $\MM$ is distributive if and only if $\MM \tm
       \MM^\kcc$,
    
       \item  [(ii)] $\MM$  is maximal distributive if and only if
       $\MM = \MM^\kcc$,
    
       \item  [(iii)] A maximal distributive sublattice is complete.
    \end{enumerate}
\end{proposition}
\emph{Proof:} $(i)$ is obvious from the definition of $\kcc$. Let
$\MM$ be maximal distributive and $a \in \MM^\kcc$. Then $\{a\} \cup
\MM \tm (\{a\} \cup \MM)^\kcc$ and therefore, by proposition \ref{ol7},
$<\{a\} \cup \MM> \tm (\{a\} \cup \MM)^\kcc$. (If $M$ is a subset of
$\LL$ then $<M>$ denotes the sublattice generated by $M$.) Hence $a
\in \MM$ since $\MM$ is a maximal distributive sublattice. Conversely,
let $\MM = \MM^\kcc$ and let $\MM_{max}$ be a maximal distributive
sublattice of $\LL$ that contains $\MM$. Then $\MM_{max} =
\MM_{max}^\kcc$ and therefore $\MM_{max} = \MM_{max}^\kcc \tm
\MM^\kcc = \MM$, i.e. $\MM$ is maximal. $(iii)$ follows from $(ii)$
and proposition \ref{ol7}. \ \ $\Box$ \\

\begin{definition}\label{ol11}
    A maximal distributive sublattice of a complete orthomodular lattice
    $\LL$ is called a \emph{Boolean sector of $\LL$}.
\end{definition}
Usually a maximal distributive sublattice of a complete orthomodular lattice
$\LL$ is called a \emph{block}. It will become clear in the next
chapter why we deviate from common use.

\section{Operator Algebras}
\label{oa}
  
\pagestyle{myheadings}
\markboth{Preliminaries}{Operator Algebras}

We do not intend to give a real introduction into the subject of
operator algebras here. We only want to fix our notations and to
present some of the basic definitions and results in order to make
this work more self-contained. Moreover, we restrict our discussion to
operator algebras that are contained in $\lh$, the algebra of all
bounded linear operators of some (complex) Hilbert space $\kh$. By the
GNS-construction (\cite{kr1}), this is no real loss of generality. 
Our standard references are \cite{kr1, kr2, kr3, kr4, tak1}.\\

\noindent{In} what follows, $\kh$ denotes an arbitrary Hilbert space. 

\begin{definition}\label{oa1}
    An operator algebra is a subalgebra $\rr$ of the algebra $\lh$ of all
    bounded linear operators $\kh \to \kh$ such that $T^\ast \in \rr$ 
    whenever $T \in \rr$. $\rr$ is called a {\bf $C^\ast$-algebra} if it
    closed in the norm topology, and a {\bf von Neumann algebra} if it is
    closed in the weak operator topology of $\lh$. 
\end{definition}

\begin{remark}\label{oa1a}
    More generally, an involutory algebra is an algebra $\frA$ over
    $\CC$ that possesses an involution, i.e. a conjugate-linear
    mapping $\ast : \frA \to \frA, \ a \tto a^\ast$ which satisfies
    \begin{enumerate}
	\item  [(i)] $a^{\ast \ast} = a$ and
    
	\item  [(ii)] $(ab)^\ast = b^\ast a^\ast$
    \end{enumerate}
    for all $a, b \in \frA$. A Banach algebra $\frA$ with an
    involution $\ast$ is called an abstract $C^\ast$- algebra if the
    norm of $\frA$ satisfies
    \[
	\all \ a \in \frA : \ |a^\ast a| = |a|^2.
    \]
    The involution $\ast$ of an abstract $C^\ast$- algebra is
    necessarily isometric.\\
    A homomorphism $\gF : \frA \to \frb$ between involutary algebras
    $\frA$ and $\frb$ is called a $\ast$- homomorphism if $\gF(a^\ast)
    = \gF(a)^\ast$ holds for every $a \in \frA$. A $\ast$- homomorphism
    between $C^\ast$- algebras is continuous with norm less or equal
    to $1$ and a $\ast$- isomorphism is necessarily isometric. 
    An abstract $C^\ast$- algebra is, by the Gelfand-Neumark theorem
    (\cite{kr1}), $\ast$- isomorphic to a (concrete) $C^\ast$- algebra
    $\rr$ in $\lh$ for a suitable Hilbert space $\kh$. \\
    Also von Neumann algebras have an abstract description: 
    A $W^\ast$- algebra is a $C^\ast$- algebra that is the topological
    dual of a Banach space. $\lh$, for example, is a $W^\ast$-
    algebra, since $\lh$ is isomorphic to the topological dual of
    $\kL_{1}(\kh)$, the Banach space of trace-class operators in $\lh$.
    It can be shown (\cite{kr2}) that any $W^\ast$- algebra $\frc$ is 
    $\ast$- isomorphic to a von Neumann algebra $\rr$ in $\lh$ for
    some Hilbert space $\kh$.
\end{remark}

In contrast to general $C^\ast$-algebras, a von Neumann algebra always
contains enough projections to be generated by them. We denote by
$\pr$ the set of projections in $\rr$. $\pr$ is a complete lattice,
hence there is a unique maximal projection $P_{I}$ in $\rr$. It is an 
immediate consequence of the \emph{spectral theorem} that
\[
    AP_{I} = P_{I}A = A
\] 
holds for all $A \in \rr$. This means that $P_{I}$ is the unit element
of $\rr$ and that the closed subspace $\kh_{I} := P_{I}\kh$ of $\kh$
is $\rr$- invariant. Therefore, we can always assume that a von Neumann
algebra $\rr \tm \lh$ contains the identity operator $I := id_{\kh}$. 
\\
~\\
~\\
The {\bf spectral theorem} is a fundamental result that is used
ubiquitously in the theory of operator algebras. It generalizes the
diagonalization of hermitean matrices $A \in \MM_{n}(\CC)$ to
selfadjoint operators defined in $\kh$.\\
Let $\gl_{1}, \ldots , \gl_{m}$ be the distinct eigenvalues of the
hermitean matrix $A \in \MM_{n}(\CC)$, numbered in ascending
order: $\gl_{1} < \cdots < \gl_{m}$. Moreover, let $P_{\gl_{k}}$ be
the orthogonal projection onto the eigenspace $\kE_{\gl_{k}}$ for the 
eigenvalue $\gl_{k}$ of $A$. Since the distinct eigenspaces of $A$ are
pairwise orthogonal, the family $\ea = (\eal)_{\lir}$, defined by
\[
    \eal := \sum \{ P_{\gl_{k}} \mid \gl_{k} ≤ \gl \},
\]
has the following properties:
\begin{enumerate}
    \item [(i)] $\eal ≤ \eamu$ for $\gl ≤ \mu$,

    \item [(ii)] $\eal = \We_{\mu > \gl}\eamu$,            

    \item [(iii)] $\eal = 0$ for $\gl < \gl_{1}$ and $\eal = I$ for $\gl ≥
    \gl_{m}$,

    \item [(iv)] $A = \sum_{k = 1}^{m}\gl_{k}(\ea_{\gl_{k}} - 
    \ea_{\gl_{k - 1}})$, where $\gl_{0} < \gl_{1}$.
\end{enumerate}
Property $(ii)$ expresses that $\ea$ is continuous from the right.
Equally well we could define
\[
    \fal := \sum \{ P_{\gl_{k}} \mid \gl_{k} < \gl \}.
\] 
Then the family $\fa = (\fal)_{\lir}$ is continuous from the left, i.e.
\[
    \fal = \Ve_{\mu < \gl}\famu,
\]
and $A$ can be represented as
\[
    A = \sum_{k = 2}^{m + 1}\gl_{k - 1}(\fa_{\gl_{k}} - \fa_{\gl_{k -
    1}}),
\]
where $\gl_{m + 1} > \gl_{m}$.\\
~\\
This situation is generalized to arbitrary selfadjoint operators $A
\in \lh$ in the following couple of theorems.

\begin{theorem}\label{oa2}(\cite{kr1}, Theorem 5.2.2)
    If $A \in \lh$ is a selfadjoint operator and $\kaa$ is an abelian 
    von Neumann algebra containing $A$, there is a family
    $(\eal)_{\lir}$ of projections in $\kaa$, called the spectral
    resolution of $A$, such that
    \begin{enumerate}
	\item  [(i)] $\eal = 0$ if $\gl < - |A|$, and $\eal = I$ if $\gl
	≥ |A|$; 
    
	\item  [(ii)] $\eal ≤ \eamu$ if $\gl ≤ \mu$;
    
	\item  [(iii)] $\eal = \We_{\mu > \gl}\eamu$;
    
	\item  [(iv)] $A\eal ≤ \gl \eal$ and $\gl (I - \eal) ≤ A (I -
	\eal)$ for each $\lir$;
    
	\item  [(v)] $A = \int_{- |A|}^{|A|}\gl d\eal$ in the sense of 
	norm convergence of approximating Riemann sums;
	and $A$ is the norm limit of finite linear combinations with
	coefficients in $sp(A)$ of orthogonal projections
	$\eamu - \eal$.
     \end{enumerate}
\end{theorem}
This theorem is proved using the Gelfand representation of $\kaa$. As 
is shown in \cite{na}, $\eal$ can be described quite explicitly: it is 
the projection onto the kernel of $(A - \gl I)^{+}$.\\
The spectral resolution of $A \in \lh_{sa}$ is a (bounded) \emph{spectral
family}, a notion which we need throughout this work not only in the
projection lattice of a von Neumann algebra, but also in other
complete lattices.
 
\begin{definition}\label{oa3}
    Let $\LL$ be a complete lattice. A (right-continuous) spectral family
    is a family $E = (\el)_{\lir}$ in $\LL$ satisfying
    \begin{enumerate}
	\item  [(i)] $\el ≤ \emm$ for $\gl ≤ \mu$,    
    
	\item  [(ii)] $\el = \We_{\mu > \gl}\emm$,
    
	\item  [(iii)] $\We_{\lir}\el = 0$  and $\Ve_{\lir}\el = 1$.
    \end{enumerate}          
    $E$ is called bounded if there are $a_{0}, a_{1} \irr$ such that
    $\el = 0$ for $\gl < a_{0}$ and $\el = 1$ for $\gl > a_{1}$.
\end{definition}
The converse of theorem \ref{oa2} are

\begin{theorem}\label{oa4}(\cite{kr1}, Theorem 5.2.3)
    If $(\el)_{\lir}$ is a spectral family and $A \in \lh$ is a
    selfadjoint operator such that $A \el ≤ \gl \el$ and $\gl(I - \el)
    ≤ A(I - \el)$ for each $\lir$, or if $A = \int_{- a}^{a}\gl d\el$ 
    for each $a$ exceeding some $b \irr$, then $(\el)_{\lir}$ is the
    spectral resolution of $A$ in $\kaa_{0}$, the abelian von Neumann 
    algebra generated by $A$ and $I$.
\end{theorem}
and

\begin{theorem}\label{oa5}(\cite{kr1}, Theorem 5.2.4)
    If $(\el)_{\lir}$ is a bounded spectral family in $\ph$, then
    $\int_{-a}^{a}\gl d\el$ converges to a selfadjoint operator $A$ on
    $\kh$ such that $|A| ≤ a$ and for which $(\el)_{\lir}$ is the
    spectral resolution, where $\el = 0$ if $\gl ≤ -a$ and $\el = I$
    if $\gl ≥ a$.
\end{theorem}
If $\LL$ is an \emph{orthomodular} complete lattice, then the notion
of a spectral family can be generalized to that of a \emph{spectral
measure}: 

\begin{definition}\label{oa6}
    A spectral measure in a complete orthomodular lattice $\LL$ is
    a mapping $\kE : \kbr \to \LL$, where $\kbr$ denotes the 
    $\gs$- algebra of all Borel subsets of $\RR$, such that the
    following two properties 
    \begin{enumerate}
	   \item  [(i)] $\kE(\bigcup_{\nin}M_{n}) = \Ve^{\perp}_{\nin}
	  \kE(M_{n})$ for all pairwise disjoint sequences $(M_{n})_{\nin}$
	  in $\kbr$, where $\Ve^\pp$ indicates that $(\kE(M_{n}))_{\nin}$ is 
	  a sequence of pairwise orthogonal elements in $\LL$.

	 \item  [(ii)] $\kE(\RR) = 1$.
    \end{enumerate}
    are satisfied.
\end{definition}
These two properties imply that a spectral measure has the following
properties, too:
\begin{enumerate}
    \item  [(iii)] $\kE(M \smm N) = \kE(M) \we \kE(N)^\pp$ for all $M, N
    \in \kbr, \ N \tm M$.
    
    \item  [(iv)] $\kE(\bigcup_{\nin}M_{n}) = \Ve_{\nin}
    \kE(M_{n})$ for all sequences $(M_{n})_{\nin}$
    in $\kbr$.
    
    \item  [(v)] $\kE(\bigcap_{\nin}M_{n}) = \We_{\nin}
    \kE(M_{n})$ for all sequences $(M_{n})_{\nin}$
    in $\kbr$. 
\end{enumerate}

A spectral family $E$ in an \emph{orthomodular} complete lattice $\LL$
induces a \emph{spectral measure} $\kE : \kbr \to \LL$ by
\[
    \all \ a, b \irr, \ a < b : \ \kE(]a, b]) := E_{b} \we
    E_{a}^\perp, 
\]
and, consistent with this definition,
\[
    \all \ a \irr : \ \kE(]-\∞, a]) := E_{a}.   
\]
This can be seen using standard measure theoretic techniques, since
the complete orthomodular lattice generated by $\{ \el \mid \lir \}$
is completely distributive (by remark \ref{ol8}).\\
~\\
Conversely, if we start from a spectral measure, i.e. from a map $\kE 
: \kbr \to \LL$ with the properties $(i)$ and $(ii)$, we obtain two
spectral families:
\begin{enumerate}
    \item  $E = (\el)_{\lir}$, defined by $\el := \kE(]-\∞, \gl])$,
    which is right-continuous, i.e. $\el = \We_{\mu > \gl}\emm$ for
    all $\lir$, and

    \item  $F = (F_{\gl})_{\lir}$, defined by $F_{\gl} := \kE(]-\∞,
    \gl[)$, which is left-continuous, i.e. $F_{\gl} = \Ve_{\mu <
    \gl}F_{\mu}$ for all $\lir$.
\end{enumerate} 

Of course there is an analogous formulation of the spectral theorem
for left-continuous spectral families (see, for example, appendix C in
\cite{hr}). Continuity from one side is necessary in order to make the
spectral resolution of a selfadjoint operator unique. Any choice
between left and right continuity brings, in principle, an asymmetry into
the theory. In operator theory, this asymmetry does not play any
r\^{o}le. But we shall see in part III of this work that it
becomes manifest in some important constructions. Our preferred choice
are right-continuous spectral families. \\
~\\
\emph{Abelian von Neumann algebras} will play a significant r\^{o}le
in our work. The fundamental theorem for abelian operator algebras is
the {\bf Gelfand representation theorem}. \\
Let $\kaa$ be an abelian $C^\ast$-algebra and assume for simplicity
that $\kaa$ has a unit element. Let $\gO(\kaa)$ be the set of all
non-zero \emph{multiplicative} linear functionals $\gf : \kaa \to \CC$
that are \emph{positive} in the following sense:
\[
    \all \ A \in \kaa : \ \gf(A^\ast A) ≥ 0.
\] 
The elements of $\gO(\kaa)$ are called \emph{characters} of $\kaa$,
and the set $\gO(\kaa)$ itself is called the \emph{Gelfand spectrum}
of $\kaa$. The sets
\[
    N_{A, \eps}(\gf_{0}) := \{ \gf \in \gO(\kaa) \mid |\gf(A) - \gf_{0}(A)|
    < \eps \} \quad (A \in \kaa, \ \eps > 0)
\]
form a subbasis of neighbourhoods of $\gf_{0} \in \gO(\kaa)$ in a
topology for $\gO(\kaa)$. This topology on $\gO(\kaa)$ is induced by
the \emph{weak*- topology} on $\kaa^\prime$, the topological dual of
the Banach space $\kaa$. Since $\gO(\kaa)$ is closed in the unit ball
$\{ \psi \in \kaa^\prime \mid |\psi| ≤ 1 \}$ of $\kaa^\prime$ 
with respect to the weak*- topology, and the latter is compact,
$\gO(\kaa)$ becomes a compact Hausdorff space. The space
$C(\gO(\kaa))$ of all continuous functions $f : \gO(\kaa) \to \CC$
with pointwise defined algebraic operations, norm $|f|_{\∞} :=
\sup_{\gf \in \gO(\kaa)}|f(\gf)|$ and ``adjoint'' $f^\ast(\gf) :=
\overline{f(\gf)}$ is an abelian $C^\ast$- algebra, whose characters
are the evaluation functionals
\[
    f \tto f(\gf) \quad (\gf \in \gO(\kaa)).
\] 
The Gelfand spectrum of $C(\gO(\kaa))$ is therefore homeomorphic to
$\gO(\kaa)$. 

\begin{theorem}\label{oa7}(Gelfand representation theorem; \cite{kr1},
    theorem 4.4.3)\\
    Let $\kaa$ be an abelian $C^\ast$- algebra with unit element. Then
    the mapping \\ $\kaa \to C(\gO(\kaa)), \ A \tto \hat{A},$ defined
    by
    \[
	\all \ \gf \in \gO(\kaa) : \ \hat{A}(\gf) := \gf(A),
    \]
    is an isometric $\ast$- isomorphism from $\kaa$ onto
    $C(\gO(\kaa))$.
\end{theorem}
The mapping $A \tto \hat{A}$ is called the \emph{Gelfand
transformation}, the function $\hat{A} \in C(\gO(\kaa))$ the 
\emph{Gelfand transform of $A$}. \\
If $A \in \lh$ is a normal operator, i.e. $A^\ast A = A A^\ast$, then 
the Gelfand spectrum of the abelian $C^\ast$- algebra $C^\ast(I, A,
A^\ast)$, generated by $I$, $A$ and $A^\ast$, is homeomorphic to
the spectrum $sp(A)$ of $A$ (\cite{kr1}, theorem 4.4.5). In contrast
to that, the Gelfand spectrum of an infinite dimensional abelian von
Neumann algebra is always of a monstrous size. We demonstrate this at 
a very simple example. Let $l^\∞(\NN)$ be the algebra of bounded
sequences in $\CC$ with norm $|(a_{n})_{\nin}|_{\∞} :=
\sup_{\nin}|a_{n}|$. $l^\∞(\NN)$ is an abelian von Neumann algebra
acting on the Hilbert space $l^2(\NN)$ by multiplication operators. 
It is not difficult to show that the Gelfand spectrum of $l^\∞(\NN)$
is homeomorphic to the \emph{Stone-\v{C}ech compactification}
$\check{\NN}$ of $\NN$. $\check{\NN}$, although separable and
compact, is a very large space: its cardinality is $2^{2^{\aleph_{0}}}$. \\
The Gelfand spectrum of an abelian infinite dimensional von Neumann
algebra is not only very large, but its topology is also rather
bizarre: every open set has open closure. Such topological spaces are 
called \emph{extremely disconnected}. However, not every extremely
disconnected compact Hausdorff space is the Gelfand spectrum of an 
abelian von Neumann algebra. For an extensive discussion of this
question we refer to \cite{tak1}.

~\\
The {\bf double commutant theorem} shows that the weak (and strong)
closure of an operator algebra $\frA$ in $\lh$ can be expressed in
purely algebraic terms. \\
Let $\kf$ be a subset of $\lh$. Then 
\[
    \kf^\kcc := \{ C \in \lh \mid \all \ A \in \kf : \ AC = CA \}
\]
is called the \emph{commutant} of $\kf$. It is easy to see that the
commutant has the following properties:
\begin{enumerate}
    \item  [(i)] If $\kf \tm \kg \tm \lh$, then $\kg^\kcc \tm
    \kf^\kcc$.

    \item  [(ii)] $\kf \tm \kf^{\kcc \kcc}$.

    \item  [(iii)] $\kf^\kcc$ is weakly closed. If $\kf$ is
    selfadjoint, i.e. $A^\ast \in \kf$ whenever $A \in \kf$, then
    $\kf^\kcc$ and $\kf^{\kcc \kcc}$ are von Neumann algebras.  
\end{enumerate}

\begin{theorem}\label{oa8}(Double commutant theorem; \cite{kr1},
    theorem 5.3.1)\\
    If $\frA$ is an operator algebra in $\lh$ containing the identity 
    operator, then the weak and the strong closure of $\frA$ coincide 
    with $\frA^{\kcc \kcc}$.
\end{theorem}
A direct consequence of this theorem is the following

\begin{corollary}\label{oa9}
    The double commutant $\kf^{\kcc \kcc}$ of a selfadjoint subset
    $\kf$ of $\lh$ containing the identity operator is the von Neumann
    algebra generated by $\kf$.
\end{corollary}
~\\
In the sequel, we need some structure theory of von Neumann algebras. 
This rests on the notion of \emph{equivalence of projections}. The
equivalence theory of projections can be seen as an adaption of naive 
set theory to the projection lattice of a von Neumann algebra.

\begin{definition}\label{oa10}
    Let $\rr$ be a von Neumann algebra and let $P, Q \in \pr$. $P$ is 
    called equivalent to $Q$ (in $\rr$), written $P \sim Q$, if there is a
    partial isometry $\gtt \in \rr$ such that $\gtt^\ast \gtt = P$ and
    $\gtt \gtt^\ast = Q$.  
\end{definition}
Recall that a partial isometry $\gtt$ is an operator in $\lh$ that is 
isometric on the orthogonal complement of its kernel. So, if $\gtt^\ast
\gtt = P$ and $\gtt \gtt^\ast = Q$, then $\gtt$ is an isometry from $P
\kh$ onto $Q\kh$ and $\gtt^\ast$ is an isometry from $Q\kh$ onto
$P\kh$. Note that the definition of equivalence requires that the
partial isometry $\gtt$, joining $P$ with $Q$, belongs to $\rr$.
Hence, if $\rr$ is abelian, $P \sim Q$ if and only if $P = Q$.

\begin{definition}\label{oa11}
    Let $P, Q \in \pr$. $P$ is called weaker than $Q$, written $P
    \precsim Q$, if there is a projection $Q_{0} \in \pr$ such that
    $Q_{0} ≤ Q$ and $P \sim Q_{0}$.
\end{definition}
One can prove that $\precsim$ is a partial ordering of the classes of 
equivalent projections (see \cite{kr2}, chapter 6). The fundamental
result of the comparison theory of projections is

\begin{theorem}\label{oa12}(Comparison Theorem; \cite{kr2}, 6.2.7)\\
    If $E$ and $F$ are projections in a von Neumann algebra $\rr$,
    there are unique orthogonal central projections $P$ and $Q$
    maximal with respect to the properties $QE \sim QF$, and, if
    $P_{0}$ is a non-zero central subprojection of $P$, then $P_{0}E
    \prec P_{0}F$. If $R_{0}$ is a non-zero central subprojection of
    $I - P - Q$, then $R_{0}F \prec R_{0}E$.    
\end{theorem}
In set theory, two sets are called equivalent if they can be mapped
bijectively onto each other. A set is defined to be finite, if it is
not equivalent to any of its proper subsets. Thus the following 
definition is natural.

\begin{definition}\label{oa13}(\cite{kr2}, 6.3.1)\\
    A projection $E$ in a von Neumann algebra $\rr$ is said to be
    infinite relative to $\rr$ when $E \sim E_{0} < E$ for some $E_{0}
    \in \pr$. Otherwise, $E$ is said to be finite relative to $\rr$.
    If $E$ is infinite and $PE$ is either $0$ or infinite for each
    central projection $P$, $E$ is said to be properly infinite. $\rr$
    is a finite or properly infinite von Neumann algebra when $I$ is, 
    respectively, finite or properly infinite.
\end{definition}
Also the following important result comes from set theory.

\begin{proposition}\label{oa14}(Halving Lemma; \cite{kr2}, 6.3.3)\\
    If $E$ is a properly infinite projection in a von Neumann algebra 
    $\rr$, there is a projection $F$ in $\rr$ such that $F ≤ E$ and $F
    \sim E - F \sim E$.
\end{proposition}

\begin{definition}\label{oa15}
    A projection $E$ in a von Neumann algebra $\rr$ is said to be
    abelian in $\rr$ when $E \rr E$ is abelian.
\end{definition}
The basic properties of abelian projections are summarized in

\begin{proposition}\label{oa16}(\cite{kr2}, 6.4.2)\\
    Each subprojection of an abelian projection in a von Neumann algebra
    $\rr$ is the product of the abelian projection and a central projection.
    A projection in $\rr$ is abelian if and only if it is minimal in
    the set of projections in $\rr$ with the same central carrier.
    Each abelian projection in $\rr$ is finite. If $\kcc$ is the
    center of $\rr$ and $E$ is an abelian projection in $\rr$, then $E
    \rr E = \kcc E$.
\end{proposition}
If $E$ is a non-zero projection in a von Neumann algebra $\rr$, then
\begin{center}
$E$ is abelian \ \ - \ \ $E$ is finite \ \ - \ \ $E$ is infinite \\
\end{center}
is a chain of properties with from left to right ascending complexity.
This leads to the following definition:

\begin{definition}\label{oa17}(\cite{kr2}, 6.5.1)\\
    A von Neumann algebra $\rr$ is said to be of type $I$ if it has an
    abelian projection with central carrier $I$ - of type $I_{n}$ if $I$
    is the sum of $n$ equivalent abelian projections. If $\rr$ has no 
    non-zero abelian projections but has a finite projection with
    central carrier $I$, then $\rr$ is said to be of type $II$ - of
    type $II_{1}$ if $I$ is finite - of type $II_{\∞}$ if $I$ is
    properly infinite. If $\rr$ has no non-zero finite projections,
    $\rr$ is said to be of type $III$. 
\end{definition}
A first insight into the structure of a von Neumann algebra $\rr$ is
given by the following theorem, which says that $\rr$ can be decomposed
into von Neumann subalgebras of different types.

\begin{theorem}\label{oa18}(Type Decomposition; \cite{kr2}, 6.5.2)\\
    If $\rr$ is a von Neumann algebra acting on a Hilbert space $\kh$,
    there are mutually orthogonal central projections $P_{n}$, $n$ not
    exceeding $dim \kh$, $P_{c_{1}}$, $P_{c_{\∞}}$, and $P_{\∞}$, with
    sum $I$, maximal with respect to the properties that $\rr P_{n}$
    is of type $I_{n}$ or $P_{n} = 0$, $\rr P_{c_{1}}$ is of type
    $II_{1}$ or $P_{c_{1}} = 0$, $\rr P_{c_{\∞}}$ is of type $II_{\∞}$
    or $P_{c_{\∞}} = 0$, and $\rr P_{\∞}$ is of type $III$ or $P_{\∞} 
    = 0$.
\end{theorem}

    

\chapter{The Stone Spectrum of a Lattice}
\label{SS}
\section{Presheaves and their Sheafification}
\label{SSS}

\pagestyle{myheadings}
\markboth{The Stone Spectrum of a Lattice}{Presheaves and their
Sheafification}

Traditionally, the notions of a {\bf presheaf} and a {\bf complete 
presheaf} (complete presheaves are usually called ``sheaves'') are 
defined for the lattice $\kT(M)$ of a topological space $M$. The very 
definition of presheaves and sheaves, however, can be formulated also 
for an arbitrary lattice:
\begin{definition}\label{SS1}
A {\bf presheaf} of sets ($R$-modules) on a lattice $\LL$ assigns to 
every element $a \in \LL$ a set ($R$-module) $\kS(a)$ and to every 
pair $(a, b) \in \LL \times \LL$ with $a \leq b$ a mapping 
($R$-module homomorphism)
\begin{displaymath}
	\rho_{a}^{b} : \kS(b) \to \kS(a)
\end{displaymath}
such that the following two properties hold:
\begin{enumerate}
    \item  [(1)] $\rho_{a}^{a} = id_{\kS(a)}$ for all $a \in \LL$,

    \item  [(2)] $\rho_{a}^{b} \circ \rho_{b}^{c} = \rho_{a}^{c}$ for 
    all $a, b, c \in \LL$ such that $a \leq b \leq c$.
\end{enumerate}
The presheaf $(\kS(a), \rho_{a}^{b})_{a \leq b}$ is called a {\bf 
complete presheaf} (or a {\bf sheaf} for short) if it has the 
additional property
\begin{enumerate}
    \item  [(3)] If $a = \bigvee_{i \in I}a_{i}$ in $\LL$ and if $f_{i} 
    \in \kS(a_{i}) \ \ (i \in I)$ are given such that
    \begin{displaymath}
	\forall \ i, j \in I : \ (a_{i} \wedge a_{j} \ne 0 \quad 
	\Longrightarrow \quad \rho_{a_{i}\wedge a_{j}}^{a_{i}}(f_{i}) = 
	\rho_{a_{i} \wedge a_{j}}^{a_{j}}(f_{j}),
    \end{displaymath}
    then there is exactly one $f \in \kS(a)$ such that
    \begin{displaymath}
	\forall \ i \in I : \ \rho_{a_{i}}^{a}(f) = f_{i}.
    \end{displaymath}
\end{enumerate}
\end{definition}
The mappings $\rho_{a}^{b} : \kS(b) \to \kS(a)$ are called {\bf 
restriction maps}.\\
~\\
One of the most elementary and at the same time instructive examples 
is the sheaf of locally defined continuous complex valued functions on 
a topological space $M$: $\kS(U)$ is the space of continuous functions 
on the open set $U \tm M$ and for $U, V \in \kT(M)$ with $U \tm V$
\begin{displaymath}
	\rho_{U}^{V} : \kS(V) \to \kS(U)
\end{displaymath}
is the restriction map $f \mapsto f|_{U}$. Property $(3)$ in 
definition \ref{SS1} expresses the elementary fact that one can glue 
together a family of locally defined continuous functions $f_{i} : 
U_{i} \to \CC$ which agree on the non-empty overlaps $U_{i} \cap 
U_{j}$ to a continuous function $f$ on $\bigcup_{i \in I}U_{i}$ which 
coincides with $f_{i}$ on $U_{i}$ for each $i \in I$.\\
~\\
Are there interesting new examples for sheaves on a lattice other 
than $\kT(M)$, in particular on the quantum lattice $\LL(\kh)$?
The story begins with a disappointing answer:

\begin{proposition}\label{SS2}
Let $(\kS(U), \rho_{U}^{V})_{U \tm V}$ be a complete presheaf of
nonempty sets on the quantum lattice $\LL(\kh)$. Then
\begin{displaymath}
	\#\kS(U) = 1
\end{displaymath}
for all $U \in \LL(\kh)$.\\
Thus complete presheaves on $\LL(\kh)$ are completely trivial!
\end{proposition}
\emph{Proof:} Each $U \in \LL(\kh)$ can be written as
\[
    U = \bigvee_{\CC x \tm U}\CC x.
\]
Because of $\CC x \cap \CC y = 0$ for $\CC x \neq \CC y$, the family
$(\kS(\CC x))_{\CC x \tm U}$ satisfies in a trivial manner the
compatibility conditions. Therefore to each family $(s_{\CC x})_{\CC x
\tm U}$ of elements $s_{\CC x} \in \kS(\CC x)$ there is a unique $s_{U}
\in \kS(U)$ such that $\rho_{\CC x}^{U}(s_{U}) = s_{\CC x}$ for all
$\CC x \tm U$. Hence there is a bijection
\[
    \kS(U) \cong \prod_{\CC x \tm U}\kS(\CC x).             
\]
Consequently, it suffices to prove that each $\kS(\CC x) \ \ (x \neq 0)$
consists of a single element. \\
Let $\CC e_{1}, \CC e_{2}$ be different lines in $\kh$, $U = \CC e_{1}
+ \CC e_{2}$ and $0 \neq \CC x \tm U$ such that $\CC x \notin \{ \CC
e_{1}, \CC e_{2} \}$. Then
\[
    U = \CC e_{1} \vee \CC e_{2} = \CC x \vee \CC e_{1} \vee \CC e_{2}
\]    
and therefore
\[
    \kS(U) \cong \kS(\CC e_{1}) \times \kS(\CC e_{2}) \cong \kS(\CC 
    x) \times \kS(\CC e_{1}) \times \kS(\CC e_{2}).  
\]
Let $s_{x}, t_{x} \in \kS(\CC x)$ and fix elements $s_{e_{k}} \in \kS(\CC
e_{k}), \ \ (k = 1, 2)$. Then there are unique $s, t \in \kS(U)$ such
that 
\[
    \rho_{\CC x}^{U}(s) = s_{x}, \ \rho_{\CC e_{k}}(s) = s_{e_{k}} \
    \ (k =1, 2), \\
    \rho_{\CC x}^{U}(t) = t_{x}, \ \rho_{\CC e_{k}}(t) = s_{e_{k}} \
    \ (k =1, 2). 
\]
$U = \CC e_{1} \vee \CC e_{2}$ implies $s = t$, hence $s_{x} =  t_{x}$.
This shows $\# \kS(\CC x) = 1$ for all lines in $\kh$ and therefore $\# 
\kS(U) = 1$ for all $U \in \LL(\kh)$.  $\Box$ \\ 
~\\
There are, however, non-trivial presheaves on $\LL(\kh)$ and one of 
them, which we shall study in part II, turns out to be quite 
fruitful for quantum mechanics and the theory of operator algebras.\\
~\\
Moreover, there is also another perspective of sheaves: the etale 
space of a presheaf. Classically, for a topological space $M$, a 
presheaf $\kS$ on $\kT(M)$ induces a sheaf of local sections of the 
etale space of $\kS$. This sheaf on $\kT(M)$ is called the 
``sheafification of the presheaf $\kS$''.\\
~\\
In what follows we shall show that to each presheaf on a (complete) 
lattice $\LL$ one can assign a sheaf on a certain topological space
derived from the lattice $\LL$, the \emph{Stone spectrum $\kQ(\LL)$}
of $\LL$. The construction is quite similar to the well-known construction 
called \emph{``sheafification of a presheaf''}.\\
If $\kS$ is a presheaf, say, of modules on a topological space $M$, 
i.e. on the lattice $\kT(M)$, then the corresponding \emph{etale
space} $\kE(\kS)$ of $\kS$ is the disjoint union of the \emph{stalks}
of $\kS$ at points in $M$:
\[  \kE(\kS) = \coprod_{x \in M}\kS_{x} \]
where
\[  
     \kS_{x} = \lim_{\overset{\longrightarrow}{U \in \frU(x)}}\kS(U),
\]
the \emph{inductive limit} of the family $(\kS(U))_{U \in \frU}$, is the
stalk in $x \in M$. \\
~\\
A first attempt to generalize stalks to the situation of lattices is
to develop a general notion of a ``point in a lattice''. This can be
done in a quite satisfactory manner. The essential hint comes from the
topological context. \\
~\\
Let $M$ and $N$ be topological spaces. The elements of $N$ are in 
one-to-one correspondence to the \emph{constant} mappings $f: M \to 
N$. These constant mappings correspond via the inverse image 
morphisms
\[ V \mapsto \overset{-1}{f}(V) \quad (V \in \kT(N))\]
to the left continuous lattice morphisms
\[ \gF : \kT(N) \to \kT(M) \]
with the property
\[ \forall \ V \in \kT(N) : \ \gF(V) \in \{\emptyset, M \}. \]
Here a lattice morphism $\gF : \LL_{1} \to \LL_{2}$ is called left
continuous if $\gF(\Ve_{k}a_{k}) = \Ve_{k}\gF(a_{k})$ holds for all
families $(a_{k})_{k \in \KK}$ in $\LL_{1}$. Analogously $\gF$ is
called right continuous if $\gF(\We_{k}a_{k}) = \We_{k}\gF(a_{k})$. 
$\gF$ is called continuous if it is both left and right continuous.
The inverse image morphism is not right continuous in general. \\
~\\
It is immediate that the set 
\[ \frp := \{ V \in \kT(N) \ |\ \gF(V) = M \} \]
has the following properties:
\begin{enumerate}
 \item[(1)] $\emptyset \notin \frp$ .

 \item[(2)] If $V,W \in \frp$, then $V\cap W \in \frp$.

 \item[(3)] If $V \in \frp$ and $W\supseteq V$ in $\kT(N)$, then $W\in 
	\frp$.
	
 \item[(4)] If $(V_{\iota})_{\iota \in I}$ is a family in $\kT(N)$ 
	and $\bigcup_{\iota \in I}V_{\iota} \in \frp$, then there is at least 
	one $\iota_{0} \in I$ such that $V_{\iota_{0}} \in \frp$.
\end{enumerate}
Now these properties make perfectly  sense in an arbitrary 
$\frm$-complete lattice, so we can use them to define \emph{points in 
a lattice}:

\begin{definition}\label{SS3}
    Let $\LL$ be an $\frm$-complete lattice. A non-empty subset
    $\frp \subseteq \LL$ is called a {\bf point in $\LL$} if the 
    following properties hold:
\begin{enumerate}
    \item  [(1)] $0 \notin \frp$. 

    \item  [(2)] $a,b \in \frp \Rightarrow a \wedge b \in \frp$.

    \item  [(3)] $a \in \frp, b \in \LL, a \leq b \Rightarrow b \in \frp$.

    \item  [(4)] Let $(a_{\iota})_{\iota \in I}$ be a family in $\LL$ 
	such that $\#I  \leq \frm$ and $\bigvee_{\iota \in I}a_{\iota} \in 
	\frp$ then $a_{\iota} \in \frp$ for at least one $\iota \in I$.
\end{enumerate}
\end{definition}

\begin{example}\label{SS4}
    Let $M$ be a non-empty set and $\LL \subseteq pot(M)$ an
    $\frm$-complete lattice such that
    \begin{eqnarray*}
	0_{\LL} & = & \emptyset  \\
	1_{\LL} & = & M  \\
	\bigvee_{\iota \in I}U_{\iota} & = & \bigcup_{\iota \in I}U_{\iota} 
	\qquad (\# I \leq \frm).
    \end{eqnarray*}
    Then for each $x \in M$
    \[  \frp_{x} := \{ U \in \LL \ | \ x \in U \}  \]
    is a point in $\LL$.
\end{example}
Conversely, if $\LL$ is the lattice $\kT(M)$ of open sets of a 
regular topological space $M$, we have

\begin{proposition}\label{SS5} Let $M$ be a regular topological space. A 
non-empty subset $\frp \subseteq \kT(M)$ is a point in the lattice 
$\kT(M)$ if and only if $\frp$ is the set of open neighbourhoods of an 
element $x \in M$. $x$ is uniquely determined by $\frp$.
\end{proposition}
\emph{Proof:} Let $\frp$ be a point in $\kT(M)$. Then $\bigcap_{U \in 
\frp}\overline{U} \ne \emptyset$, for otherwise $\bigcup_{U \in
\frp}\overline{U}^\prime = M \in \frp$ and therefore
$\overline{U}^\prime \in \frp$ for some $U \in \frp$, a contradiction.
Assume that $\bigcap_{U \in \frp}\overline{U}$ contains two different 
elements $x, y$. Since $M$ is regular, there are open neighbourhoods
$V, W$ of $x$ and $y$, respectively, such that $\overline{V} \cap
\overline{W} = \emptyset$. Then $\overline{V}^\prime \in \frp$ or
$\overline{W}^\prime \in \frp$, and we may assume that
$\overline{V}^\prime \in \frp$. But then $x \in
\overline{\overline{V}^\prime}$, and therefore $V \cap
\overline{V}^\prime \ne \emptyset$, a contradiction. Hence there is a 
unique $x_{\frp} \in M$ such that 
\[
    \{x_{\frp}\} = \bigcap_{U \in \frp}\overline{U}.
\] 
Assume that there is an open neighbourhood $V$ of $x_{\frp}$ that does
not belong to $\frp$. Take an open neighbourhood $W$ of $x_{\frp}$
such that $\overline{W} \tm V$. Then $V \cup \overline{W}^\prime = M \in
\frp$, so $\overline{W}^\prime \in \frp$, and from $x_{\frp} \in
\overline{\overline{W}^\prime}$ we get the contradiction $W \cap
\overline{W}^\prime \ne \emptyset$. Therefore, all open neighbourhoods
of $x_{\frp}$ belong to $\frp$. \\
Finally, let $U \in \frp$, but assume that $x_{\frp} \notin U$. Let
\[
     \frU := \{ V \in \kT(M) \ | \ \overline{V} \tm U \}.
\]
Since $M$ is regular, we have $U = \bigcup_{V \in \frU}V$, so $V \in
\frp$ for some $V \in \frU$. Therefore, $x_{\frp} \notin \overline{V}$
by assumption, but $x_{\frp} \in \overline{V}$ by construction. \ \
$\Box$ \\
~\\
Unfortunately, there are important lattices that do not possess any 
points!\\
There are plenty of points in $\kT(M)$ and $\kbb(M)$; $\kT_{r}(M)$
(for suitable topological spaces $M$)
and $\LL(\kh)$ possess no points at all. We will show this 
here only for the lattice $\LL(\kh)$ of closed subspaces of the 
Hilbert space $\kh$.

\begin{proposition}\label{SS6} If $dim \kh > 1$, there are no points in 
$\LL(\kh)$.
\end{proposition}
\emph{Proof:} Let $\frp \subseteq \LL(\kh)$ be a point. If 
$(e_{\alpha})_{\alpha \in A}$ is an orthonormal basis of $\kh$ then
\[  \bigvee_{\alpha \in A}\CC e_{\alpha} = \kh \in \frp,  \]
so $\CC e_{\alpha_{0}} \in \frp $ for some $\alpha_{0} \in A$. It 
follows that each $U \in \frp$ must contain the line $\CC 
e_{\alpha_{0}}$. Now choose $U\in \LL(\kh)$ such that neither $U$ nor 
$U^{\perp}$ contains $\CC e_{\alpha_{0}}$. Then $U, U^{\perp} \notin 
\frp$ but $U \vee U^{\perp} = \kh \in \frp$ which is a contradiction 
to property $(4)$ in the definition of a point in a lattice.
Therefore, there are no points in $\LL(\kh)$. $\Box$\\
~\\
Let $\kS = (\kS(U),\rho^{U} _{V})_{V\leq U}$ be a presheaf on the 
topological space $M$. The \emph{stalk} of $\kS$ at $x \in M$ is the 
direct limit
\[  \kS_{x} := \lim_{\overset{\longrightarrow}{U \in \frU(x)}}\kS(U) \]
where $\frU(x)$ denotes the set of open neighbourhoods of $x$ in $M$, 
i.e. the point in $\kT(M)$ corresponding to $x$.\\
For the definition of the direct limit (see below), however, we do not need the 
point $\frU(x)$, but only a partially ordered set $I$ with the property
\[  \forall \ \alpha, \beta \in I\ \  \exists \gamma \in I: \gamma \leq 
\alpha \ \  \mbox{and}  \ \ \gamma \leq \beta. \]
In other words: a \emph{filter base} $B$ in a lattice $\LL$ is 
sufficient. It is obvious how to define a filter base in an 
arbitrary lattice $\LL$:

\begin{definition}\label{SS7} A filter base $B$ in a lattice $\LL$ is a 
non-empty subset $B \subseteq \LL$ such that
\begin{enumerate}
    \item  [(1)] $0 \notin B$,

    \item  [(2)] $\forall \ a,b \in B \ \exists \ c \in B : \ c \leq a 
	\wedge b$.
\end{enumerate}
\end{definition}
The set of all filter bases in a lattice $\LL$ is of course a vast object. 
So it is reasonable to consider \emph{maximal} filter bases in 
$\LL$. (By Zorn's lemma, every filter base is contained in a maximal 
filter base in $\LL$.) This leads to the following

\begin{definition}\label{SS8} A nonempty subset $\frb$ of a lattice $\LL$ is 
called a {\bf quasipoint} in $\LL$ if and only if
\begin{enumerate}
    \item  [(1)] $0 \notin \frb$,

    \item  [(2)] $\forall \ a,b \in \frb \ \exists \ c \in \frb : \ c 
	\leq a \wedge b$,

    \item  [(3)] $\frb$ is a maximal subset having the properties $(1)$ 
	and $(2)$.
\end{enumerate}
\end{definition}

\begin{proposition}\label{SS9} Let $\frb$ be a quasipoint in the lattice 
$\LL$. Then 
\[ \forall \ a \in \frb \ \forall \ b \in \LL : \ (a \leq b 
    \Longrightarrow b \in \frb ). \]
In particular
\[ \forall \ a, b \in \frb : \ a \wedge b \in \frb. \]
\end{proposition}
\emph{Proof:} Let $c \in \frb$. Then $a \wedge c \leq b \wedge c$ and 
from $a, c \in \frb$ we obtain a $d \in \frb$ such that
\[  d \leq a \wedge c \leq b \wedge c. \]
Therefore $\frb \cup \{ b\}$ is a filter base in $\LL$ containing 
$\frb$. Hence $\frb = \frb \cup \{b\}$ by the maximality of $\frb$, 
i.e. $b \in \frb$. $\Box$ \\
~\\
This proposition shows that a quasipoint in $\LL$ is nothing else but 
a \emph{maximal dual ideal} in the lattice $\LL$ (\cite{birk}).
The set of quasipoints in $\LL$ is denoted by $\ql$.\\
~\\
In 1936, M.H.Stone (\cite{stone}) showed that the set $\kQ(\kbb)$ of 
quasipoints in a Boolean algebra $\kbb$ can be given a topology such 
that $\kQ(\kbb)$ is a \emph{compact zero dimensional} Hausdorff space 
and that the Boolean algebra $\kbb$ is isomorphic to the Boolean 
algebra of all \emph{closed open} subsets of $\kQ(\kbb)$. A base for 
this topology is simply given by the sets
\[ \kQ_{U}(\kbb) := \{ \frb \in \kQ(\kbb) \mid U \in \frb \} \]
where $U$ is an arbitrary element of $\kbb$. Of course we can generalize
this construction to an arbitrary lattice $\LL$. \\
~\\
For $a \in \LL$ let
\[ \kQ_{a}(\LL) := \{ \frb \in \kQ(\LL) \mid a \in \frb \}. \]
It is quite obvious from the definition of a quasipoint that
\[ \kQ_{a \wedge b}(\LL) = \kQ_{a}(\LL) \cap \kQ_{b}(\LL), \]
\[\kQ_{0}(\LL) = \emptyset \quad \text{and} \quad \kQ_{I}(\LL)
    = \kQ(\LL)  \]
hold. Hence $\{ \kQ_{a}(\LL) \mid a \in \LL \}$ is a base for a 
topology on $\kQ(\LL)$. It is easy to see, using the 
maximality of quasipoints, that in this topology the sets 
$\kQ_{a}(\LL)$ are open and closed:
By definition, $\qal$ is an open set. Let $\frb \in \ql \smm \qal$.
Then $a \notin \frb$, so there is some $b \in \frb$ such that $a \we b
= 0$ and this implies $\qal \cap \qbl =\emptyset$, hence $\qal$ is also
closed. Therefore, the topology defined by 
the basic sets $\kQ_{a}(\LL)$ is \emph{zero dimensional} and, using
the same argument, we see that it is also Hausdorff. Moreover, as the 
basic sets $\qal$ are open and closed, this topology is completely
regular. \\

\begin{definition}\label{SS10}
    $\kQ(\LL)$, together with the topology defined by the base $\{
    \qal \ | \ a \in \LL \}$, is called the {\bf Stone spectrum
    of the lattice $\LL$.}
\end{definition}
We have chosen this terminology because we will
see in section \ref{SSVN} that the Stone spectrum is a generalization of the
Gelfand spectrum of an abelian von Neumann algebra. \\
~\\
We will prove some general properties of Stone spectra for certain
classes of lattices in the next sections. \\
~\\
A general lattice has no points. Our most important 
example for this situation is the {\bf quantum lattice $\LL(\kh)$} of 
closed subspaces of the Hilbert space $\kh$. However, putting aside
some very special examples, we always have plenty of quasipoints, and we
can define the stalk of a presheaf $\kP$ on a lattice $\LL$ over a
quasipoint $\frb \in \kQ(\LL)$ in the very same manner as in the
topological situation.\\
~\\
Let $\kS = (\kS(U),\ \rho_{V}^{U})_{V \leq U}$ be a presheaf on the 
(complete) lattice $\LL$.

\begin{definition}\label{SS11}
$f \in \kS(U)$ is called equivalent to $g \in \kS(V)$ at the 
quasipoint $\frb \in \kQ_{U \wedge V}(\LL)$ if and only if
\[ \exists \ W \in \frb : \ W \leq U \wedge V \ \mbox{and} \ 
\rho_{W}^{U}(f) = \rho_{W}^{V}(g). \]
\end{definition}
If $f$ and $g$ are equivalent at the quasipoint $\frb$ we write $f 
\sim_{\frb} g$.\\
It is easy to see that $\sim_{\frb}$ is an equivalence relation. The 
equivalence class of $f \in \kS(U)$ at the quasipoint $\frb \in 
\kQ(\LL)$ is denoted by $[f]_{\frb}$. It is called the \emph{germ of 
$f$ at $\frb$}. Note that this only makes sense if $\frb \in 
\kQ_{U}(\LL)$. Let $\frb \in \kQ_{U}(\LL)$. Then we obtain a canonical 
mapping
\[  \rho_{\frb}^{U} : \kS(U) \to \kS_{\frb} \]
of $\kS(U)$ onto the set $\kS_{\frb}$ of germs at the quasipoint 
$\frb$, defined by the composition
\[ \kS(U) \overset{i_{U}}{\hookrightarrow} \coprod_{V \in \frb}\kS(V) 
\overset{\pi_{\frb}}{\to} (\coprod_{V \in \frb}\kS(V))/ \sim_{\frb}   \]
where $i_{U}$ is the canonical injection and $\pi_{\frb}$ the 
canonical projection of the equivalence relation $\sim_{\frb}$. 
($\kS_{\frb} := (\coprod_{V \in \frb}\kS(V))/\sim_{\frb}$ is nothing 
else but the direct limit $\varinjlim_{V \in \frb}\kS(V)$ 
(\cite{CondeG}) and $\rho_{\frb}^{U}(f)$ is just another notation for 
the germ $[f]_{\frb}$ of $f \in \kS(U)$.) \\
~\\
Let $\kS$ be a presheaf on the lattice $\LL$ and 
\[  \kE(\kS) := \coprod_{\frb \in \kQ(\LL)}\kS_{\frb} . \]
Moreover, let
\[ \pi_{\kS} : \kE(\kS) \to \kQ(\LL)  \]
be the projection defined by
\[  \pi_{\kS}(\kS_{\frb}) := \{\frb \}. \]
We will define a toplogy on $\kE(\kS)$ such that $\pi_{\kS}$ is a local 
homeomorphism.\\
For $U \in \LL$ and $f \in \kS(U)$ let
\[ \kO_{f,U} := \{ \rho_{\frb}^{U}(f) \mid \frb \in \kQ_{U}(\LL) \}. \]
It is quite easy to see that $\{ \kO_{f,U} \mid f \in \kS(U),\ U \in 
\LL \}$ is a base for a topology on $\kE(\kS)$. Together with this 
topology, $\kE(\kS)$ is called the {\bf etale space of $\kS$ over 
$\kQ(\LL)$.} By the very definition of this topology the projection 
$\pi_{\kS}$ is a local homeomorphism, for $\kO_{f,U}$ is mapped 
bijectively onto $\kQ_{U}(\LL)$.\\
~\\
If $\kS$ is a presheaf of modules or algebras, the algebraic 
operations can be transferred fibrewise to the etale space 
$\kE(\kS)$.\\
Addition, for example, gives a mapping from 
\[ \kE(\kS) \circ \kE(\kS) := \{ (a,b) \in \kE(\kS) \times \kE(\kS) 
\mid \pi_{\kS}(a) = \pi_{\kS}(b) \} \]
to $\kE(\kS)$, defined as follows:\\
Let $f \in \kS(U),\ g \in \kS(V)$ be such that 
\[ a = \rho_{\pi_{\kS}(a)}^{U}(f), \ \ b = \rho_{\pi_{\kS}(b)}(g) \]
and let $ W \in \pi_{\kS}(a)$ be some element such that $W \leq U 
\wedge V$. Then
\[ a + b := \rho_{\pi_{\kS}(a)}^{W}( \rho_{W}^{U}(f) + 
\rho_{W}^{V}(g)) \]
is a well defined element of $\kE(\kS)$.\\
By standard techniques one can prove that the algebraic operations
 \begin{eqnarray*}
 \kE(\kS) \circ \kE(\kS)	 &\to  & \kE(\kS)    \\
	(a,b) & \mapsto & a - b
 \end{eqnarray*}
 (and $(a,b) \mapsto ab$, if $\kS$ is a presheaf of algebras) and
 \begin{eqnarray*}
	\kE(\kS) & \to & \kE(\kS)  \\
	a & \mapsto & \ga a
  \end{eqnarray*}
  (scalar multiplication with $\ga$) are continuous.\\
  ~\\
  From the etale space $\kE(\kS)$ over $\kQ(\LL)$ we obtain - as in 
  ordinary sheaf theory -  a complete presheaf $\kS^{\kQ}$ on the 
  topological space $\kQ(\LL)$ by
  \[ \kS^{\kQ}(\kV) := \gG(\kV,\ \kE(\kS)) \]
  where $\kV \subseteq \kQ(\LL)$ is an open set and $\gG(\kV,\ 
  \kE(\kS))$ is the set of {\bf continuous sections of $\pi_{\kS}$ 
  over $\kV$}, i.e. of all continuous mappings $s_{\kV} : \kV \to 
  \kE(\kS)$ such that $\pi_{\kS} \circ s_{\kV} = id_{\kV}$. If $\kS$ 
  is a presheaf of modules, then $\gG(\kV, \kE(\kS))$ is a module, too.
  
  \begin{definition}\label{SS12}
  The complete presheaf $\kS^{\kQ}$ on the Stone spectrum $\kQ(\LL)$ 
  is called the {\bf sheaf associated to the presheaf $\kS$ on $\LL$}.
  \end{definition}

\section{General Properties of Stone Spectra}
\label{SSGP}
\pagestyle{myheadings}
\markboth{The Stone Spectrum of a Lattice}{General Properties of Stone Spectra}
    
In the following, let $\LL$ be a lattice (with minimal element $0$ and
maximal element $1$) and $\ql$ the Stone spectrum of $\LL$.   \\
~\\
We have seen that $\kQ_{a}(\LL) \cap \kQ_{b}(\LL) = \kQ_{a \we b}(\LL)$
holds for all $a, b \in \LL$. Clearly $a ≤ b$ implies $\kQ_{a}(\LL)
\tm \kQ_{b}(\LL)$, so  
\[
     \kQ_{a}(\LL) \cup \kQ_{b}(\LL) \tm \kQ_{a \vee b}(\LL).
\] 
In an arbitrary lattice however, this inclusion may be proper. 

\begin{remark}\label{SS13}
    If $\LL$ is a distributive lattice then 
    \begin{equation}
	\kQ_{a}(\LL) \cup \kQ_{b}(\LL) = \kQ_{a \vee b}(\LL)
	\label{eq:GP1}
    \end{equation}
for all $a, b \in \LL$.
\end{remark}
\emph{Proof:} Assume that there is some $\frb \in \kQ_{a \vee b}(\LL)
\smm (\kQ_{a}(\LL) \cup \kQ_{b}(\LL))$. Then, by the maximality of
quasipoints, we can choose $d, e \in \frb$ such that $d \we a = e \we b = 0$.
Because of $d \we e, a \vee b \in \frb$ we obtain the contradiction $0
\ne d \we e \we (a \vee b) = (d \we e \we a) \vee (d \we e \we b) = 0$.
\ \ $\Box$ \\
~\\
Conversely, assume that $\kQ_{a}(\LL) \cup \kQ_{b}(\LL) = \kQ_{a \vee
b}(\LL)$ holds for all $a, b \in \LL$. Then we get for all $a, b, c
\in \LL$: 
\begin{eqnarray*}
    \kQ_{(a \we b) \vee (a \we c)}(\LL)  & = & \kQ_{a \we b}(\LL) \cup
    \kQ_{a \we c}(\LL)  \\
     & = & (\kQ_{a}(\LL) \cap \kQ_{b}(\LL)) \cup (\kQ_{a}(\LL) \cap
     \kQ_{c}(\LL))  \\
     & = & \kQ_{a}(\LL) \cap (\kQ_{b}(\LL) \cup \kQ_{c}(\LL))  \\
     & = & \kQ_{a \we (b \vee c)}(\LL),
\end{eqnarray*}
i.e. if property \ref{eq:GP1} holds then also
\begin{equation}
    \kQ_{(a \we b) \vee (a \we c)}(\LL) = \kQ_{a \we (b \vee c)}(\LL).
    \label{eq:GP2}
\end{equation}
A lattice satisfying property \ref{eq:GP2} is called
\emph{quasidistributive}. Quasidistributivity does not imply
distributivity:

\begin{example}\label{SS14}
    Let $\kh$ be an infinite dimensional Hilbert space and 
    \begin{math}
	\llh_{cof} := \{ U \in \llh \ | \ \dim U^\pp < \∞ \} \cup \{0\}.
    \end{math}
    Then $\llh_{cof}$ is a non-distributive lattice. As the
    intersection of two subspaces of finite codimension is never zero,
    $\llh_{cof}$ contains only one quasipoint, namely 
    \begin{math}
	\llh_{cof} \smm \{0\}.
    \end{math}
    Therefore, this lattice is trivially quasidistributive but not
    distributive.
\end{example}
Such messy situations cannot occur for \emph{orthomodular lattices}:

\begin{lemma}\label{SS15}
    Let $\LL$ be an orthomodular lattice. Then $\qal = \qbl$ implies
    $a = b$.
\end{lemma}
\emph{Proof:} It suffices to prove that $\qal \tm \qbl$ implies $a ≤
b$. (In fact this is equivalent to the assertion.) If $a \nleq b$ then
$a \we b < a$ and therefore, as $a$ commutes
with $a \we b$, $a \we (a \we b)^\pp \ne 0$. Take a quasipoint $\frb$ 
that contains $a \we (a \we b)^\pp$. Then $a \in \frb$ and therefore
$b \in \frb$. Hence we get the contradiction $a \we b, a \we (a \we b)^\pp
\in \frb$. \ \ $\Box$ \\
~\\
Distributivity of an orthomodular lattice $\LL$ can now be characterized by 
properties of the topological space $\ql$:

\begin{proposition}\label{SS16}
    The following properties of an orthomodular lattice $\LL$ are
    equivalent:
    \begin{enumerate}
	\item  [(i)] $\LL$ is distributive.
    
	\item  [(ii)] $\kQ_{a}(\LL) \cup \kQ_{b}(\LL) = \kQ_{a \vee
	b}(\LL)$ for all $a, b \in \LL$.
	
	\item  [(iii)] $\qal \cup \kQ_{a^\pp}(\LL) = \ql$ for all $a
	\in \LL$.	
	
	\item  [(iv)] The only open closed subsets of $\ql$ are the
	sets $\qal \ \ (a \in \LL)$.  	
    \end{enumerate}  
\end{proposition}
\emph{Proof:} $(ii)$ follows from $(i)$ by remark \ref{SS13}
and $(iii)$ is a special case of $(ii)$. If $(iii)$ holds and if there
is a quasipoint $\frb$ that contains $a \vee b$ but neither $a$ nor
$b$, then $a^\pp, b^\pp \in \frb$ and therefore $(a \vee b)^\pp =
a^\pp \we b^\pp \in \frb$, contradicting $a \vee b \in \frb$. $(ii)$
implies that $\LL$ is quasidistributive, hence distributive by lemma
\ref{SS15}. If $(iv)$ holds, then for all $a, b \in \LL$ there is some 
$c \in \LL$ such that $\qal \cup \qbl = \kQ_{c}(\LL)$ holds. Hence, by
lemma \ref{SS15}, $a, b ≤ c ≤ a \vee b$, i.e. $c = a \vee b$. This
shows that $(iv)$ implies the distributivity of $\LL$. Conversely, if 
$\LL$ is distributive then, being orthomodular, it is a Boolean algebra.
Therefore $\ql$ is compact by Stone's theorem (\cite{birk}). Let $\kO 
\tm \ql$ be open and closed. Then $\kO$ can be be represented as a
\emph{finite} union of sets $\kQ_{a_{i}}(\LL)$ and therefore, by
$(ii)$, $\kO = \kQ_{\Ve_{i}a_{i}}(\LL)$. \ \ $\Box$ \\

\begin{definition}\label{SS17}
    A quasipoint $\frb$ in a lattice $\LL$ is called {\bf atomic} if
    $\frb$ is isolated in $\ql$. 
\end{definition}

\begin{proposition}\label{SS18}
    Let $\LL$ be an orthomodular lattice. Then $\frb \in \ql$ is atomic if
    and only if there is a (necessarily unique) atom $a_{0} \in \LL$ such that 
    \begin{equation}
	\frb = \{ a \in \LL \ | \ a_{0} ≤ a \}.
	\label{eq:GP3}
    \end{equation}
\end{proposition}
\emph{Proof:} If $a_{0} \in \LL$ is an atom such that \ref{eq:GP3} is 
satisfied, then $\{\frb \} = \kQ_{a_{0}}(\LL)$, so $\frb$ is atomic.
If, conversely, $\frb$ is atomic then there is some $a \in \LL$ such
that $\{\frb \} = \qal$. Assume that $a$ is not an atom, i.e. there is
some $b \in \LL$ such that $0 < b < a$. Then $\kQ_{a \we b^\pp}(\LL)$ 
is a proper nonempty subset of $\qal$, a contradiction. If $a_{0} \in 
\LL$ is an atom then clearly $\{ a \in \LL \ | \ a_{0} ≤ a \}$ is a
quasipoint in $\LL$. \ \ $\Box$ \\
~\\
Proposition \ref{SS18} is not valid for arbitrary lattices as is shown
by example \ref{SS14}. On the other hand, orthomodularity is not a
necessary assumption for \ref{SS18} because the proposition is true 
also for the lattice $\kT(M)$ of open subsets of a Hausdorff space $M$. 
\\

We will now show that the Stone spectrum $\ql$ of a \emph{completely
distributive lattice} $\LL$ is \emph{extremely disconnected}, i.e.
that the closure of every open subset of $\ql$ is open again. In order
to prove this we must characterize the closure of the union of an arbitrary
family of basic sets $\qal$. It is useful to do this for an arbitrary 
lattice. \\
Let $(a_{k})_{k \in \KK}$ be an arbitrary family in a lattice $\LL$.
Then $\overline{\bigcup_{k \in \KK}\kQ_{a_{k}}(\LL)}$ can be
characterized in the following way:
\begin{eqnarray*}
    \frb \in \overline{\bigcup_{k \in \KK}\kQ_{a_{k}}(\LL)} & \llra & 
    \all \ a \in \frb : \ \qal \cap (\bigcup_{k \in
    \KK}\kQ_{a_{k}}(\LL)) \neq \emptyset   \\
     & \llra & \all \ a \in \frb \ \ex \ k : \ \kQ_{a \we a_{k}}(\LL) 
     = \qal \cap \kQ_{a_{k}}(\LL) \neq \emptyset  \\
     & \llra & \all \ a \in \frb \ \ex \ k : \ a \we a_{k} \neq 0.
\end{eqnarray*}

\begin{proposition}\label{SS19}
    If $\LL$ is a completely distributive lattice then its Stone
    spectrum $\ql$ is extremely disconnected.
\end{proposition}
\emph{Proof:} We will prove that for an arbitrary family $(a_{k})_{k \in \KK}$
\begin{equation}
    \overline{\bigcup_{k \in \KK}\kQ_{a_{k}}(\LL)} = \kQ_{\Ve_{k \in
    \KK}a_{k}}(\LL)
    \label{eq:GP4}
\end{equation}
holds. Obviously, the left hand side of equation \ref{eq:GP4} is
contained in the right hand side. Conversely, let $\frb$ be a
quasipoint that contains $\Ve_{k}a_{k}$. Then we obtain for all $a \in
\frb$
\[
    0 \ne a \we (\Ve_{k \in \KK}a_{k}) = \Ve_{k \in \KK}(a \we a_{k}).
\]
Hence $a \we a_{k} \ne 0$ for some $k \in \KK$. This means, by the
foregoing characterization of the closure, that $\frb \in \overline{\bigcup_{k
\in \KK}\kQ_{a_{k}}(\LL)}$. \ \ $\Box$ \\
~\\
More general than completely distributive lattices are lattices of
finite type:

\begin{definition}\label{SS20}
    A lattice $\LL$ is called of {\bf finite type} if 
    \[
	\overline{\bigcup_{k \in \KK}\kQ_{a_{k}}(\LL)} = \kQ_{\Ve_{k \in
	\KK}a_{k}}(\LL)
    \]
    holds for all {\bf increasing} families $(a_{k})_{k \in \KK}$ in
    $\LL$.
\end{definition}

\begin{lemma}\label{SS21}
    An orthomodular lattice $\LL$ is of finite type if and only if
    \[
	a \we (\Ve_{k \ikk}a_{k}) = \Ve_{k \ikk}(a \we a_{k})
    \] 
    for all $a \in \LL$ and all \emph{increasing} families $(a_{k})_{k
    \ikk}$ in $\LL$.
\end{lemma}
\emph{Proof:} The proof rests on the following simple observation: Let
$M$ be a topological space and let $A, B \tm M$ be subsets such that
$A$ is closed and open. Then
\[
    \overline{A \cap B} = A \cap \overline{B}.
\]
Indeed, let $U$ be an open neighborhood of $x \in A \cap \overline{B}$.
Then $U \cap A$ is an open neighborhood of $x$ and therefore $U \cap A
\cap B \ne \emptyset$, i.e. $x \in \overline{A \cap B}$. The reverse
inclusion is obvious. \\
If $\LL$ is of finite type and if $a$ and $(a_{k})_{k \ikk}$ are as
above then $(a \we a_{k})_{k \ikk}$ is increasing and therefore
\begin{eqnarray*}
    \kQ_{\Ve_{k \ikk}(a \we a_{k})}(\LL) & = & \overline{\bigcup_{k
	\ikk}\kQ_{a \we a_{k}}(\LL)}  \\
     & = & \overline{\bigcup_{k \ikk}(\kQ_{a}(\LL) \cap
	 \kQ_{a_{k}}(\LL))}  \\
     & = & \overline{\qal \cap (\bigcup_{k \ikk}\kQ_{a_{k}}(\LL))}  \\
     & = & \qal \cap \overline{\bigcup_{k \ikk}\kQ_{a_{k}}(\LL)}  \\
     & = & \qal \cap \kQ_{\Ve_{k \ikk}a_{k}}(\LL)  \\
     & = & \kQ_{a \we (\Ve_{k \ikk}a_{k})}(\LL).
\end{eqnarray*}
Hence, by orthomodularity, $a \we (\Ve_{k \ikk}a_{k}) = \Ve_{k \ikk}
(a \we a_{k})$. 
The converse is shown by the same argument as in the proof of
proposition \ref{SS19}. \ \ $\Box$ \\

\begin{corollary}\label{SS22}
    A complete Boolean algebra is completely distributive if and only 
    if it is of finite type. 
\end{corollary}
\emph{Proof:} In a distributive lattice $\LL$ we have 
\begin{math}
    \kQ_{\Ve_{k \ikk}a_{k}}(\LL) = \bigcup_{k \ikk}\kQ_{a_{k}}(\LL)
\end{math}
for {\bf finite} $\KK$ and the join of an arbitrary family in $\LL$
can be written as the join of an increasing family of finite
subfamilies. \ \ $\Box$  \\

The term ``finite type'' is chosen because of the following

\begin{theorem}\label{SS23}
    The projection lattice $\pr$ of a von Neumann algebra $\rr$ is of 
    finite type if and only if $\rr$ is of finite type.
\end{theorem}
We need a simple lemma on tensor products:

\begin{lemma}\label{SS24}
    Let $\mm \tm \kL(\kK)$ be a von Neumann algebra acting on a
    Hilbert space $\kK$ with unity $I_{\mm} = id_{\kK}$ and let $\rr
    \tm \lh$ be a von Neumann algebra. Then for all $A, B \in \hr$
    and all $P, Q \in \pr$:
    \begin{enumerate}
	\item  [(i)] $I_{\mm} \otimes A ≤ I_{\mm} \otimes B$ if and
	only if $A ≤ B$. 
    
	\item  [(ii)] $I_{\mm} \ten (P \we Q) = (I_{\mm} \ten P) \we
	(I_{\mm} \ten Q)$.
    
	\item  [(iii)] $I_{\mm} \ten (P \vee Q) = (I_{\mm} \ten P) \vee
	(I_{\mm} \ten Q)$.
    
	\item  [(iv)] $I_{\mm} \ten (\Ve_{k \ikk}P_{k}) = \Ve_{k
	\ikk}(I_{\mm} \ten P_{k})$ for all families $(P_{k})_{k \ikk}$
	in $\ph$. An analogous property holds for arbitrary meets. 
    \end{enumerate}
\end{lemma}
\emph{Proof:} We use some results on tensor products that can be found
in \cite{kr1, kr2, tak1}.\\
Let $(e_{b})_{b \in \BB}$ be an orthonormal basis of $\kK$. Then 
\[
    U : \sum_{b \in \BB}x_{b} \mapsto \sum_{b \in \BB}(e_{b} \ten
    x_{b})
\]
is a surjective isometry from $\Ds_{b \in \BB}\kh_{b}$ (with
$\kh_{b} = \kh$ for all $b \in \BB$) onto $\kK \ten \kh$. Let $A \in
\rr$. Then $U$ intertwines $I_{\mm} \ten A$ and $A$ :
\[
    U^{-1}(I_{\mm} \ten A)U = \Ds_{b \in \BB}A_{b}
\] 
with $A_{b} = A$ for all $b \in \BB$. This immediately implies
$(i)$.\\
Note that $I_{\mm } \ten A$ is a projection if and only if $A$ is.
Then $(ii)$ and $(iii)$ follow from $(i)$ and the universal property
of minimum and maximum.\\
In order to prove $(iv)$ we use the fact that the 
mapping $A \mapsto I_{\mm} \ten A$ from $\rr$ to $\mm \bar{\ten} \rr$ 
is strongly continuous on bounded subsets of $\rr$:
\[
    \Ve_{k \ikk}(I_{\mm} \ten P_{k}) = I_{\mm} \ten (\Ve_{k
    \ikk}P_{k}).
\]
Hence also $(iv)$ follows. \ \ $\Box$ \\
~\\
\emph{Proof of theorem:} Due to lemma \ref{SS21} we have to show that $\rr$ is of
finite type if and only if for all $P \in \pr$ and every increasing
net $(P_{k})_{k \ikk}$ in $\pr$
\begin{equation}
	P \we (\Ve_{k \ikk}P_{k}) = \Ve_{k \ikk}(P \we P_{k})
    \label{eq:GP5}
\end{equation}
holds. Now the right hand side of \ref{eq:GP5} is the limit of the
increasing net $(P \we P_{k})_{k \ikk}$ in the strong operator
topology. If $\rr$ is of finite type, this limit is equal to the left 
hand side of \ref{eq:GP5} (see \cite{kr4}, p.412). \\
If $\rr$ is not of finite type we present an example for which \ref{eq:GP5}
does not hold. We use a construction which is quite similar to one
already used in \cite{deg2}. \\
Assume that $\rr$ is not 
finite. Then $\rr$ contains a direct summand of the form $\mm
\bar{\ten} \kL(\kh_{0})$, where $\mm \tm \kL(\kK)$ is a suitable von
Neumann algebra and $\kh_{0}$ a separable Hilbert space of infinite
dimension (see e.g. \cite{tak1}, Ch. V.1, essentially prop. 1.22: if
$\rr$ is not finite then $\rr$ has a direct summand with properly
infinite unity $I_{0}$. Use the halving lemma to construct a countable
infinite orthogonal sequence of pairwise equivalent projections with
sum $I_{0}$ (see the proof of theorem 6.3.4 in \cite{kr2})).

Now let $(e_{k})_{k \inn}$ be an orthonormal basis of $\kh_{0}$, 
$x := \sum_{k = 1}^{\∞}\frac{1}{k}e_{k}$, $P$ the projection onto
$\CC x$ and $P_{n}$  the projection onto
    \[
	U_{n} = \CC e_{1} + \ldots + \CC e_{n}.
    \]
Note that $x \notin U_{n}$ for all $n \in \NN$, hence $P \we P_{n} =
0$ for all $n \inn$ and therefore $\Ve_{n \inn}(P \we P_{n}) = 0$. On 
the other hand $\Ve_{n \inn}P_{n} = I$ and therefore $P \we (\Ve_{n
\inn}P_{n}) = P > 0$. Using lemma \ref{SS24}, we obtain
\begin{eqnarray*}
    (I_{\mm} \ten P) \we (\Ve_{n \inn}(I_{\mm} \ten P_{n})) & 
    = & (I_{\mm} \ten P) \we (I_{\mm} \ten (\Ve_{n \inn}P_{n}))  \\
     & = & I_{\mm} \ten P  \\
     & > & 0  \\
     & = & I_{\mm} \ten (\Ve_{n \inn}(P \we P_{n}))  \\
     & = & \Ve_{n \inn}((I_{\mm} \ten P) \we (I_{\mm} \ten P_{n})).
\end{eqnarray*}
Thus property \ref{eq:GP5} is not satisfied in $\mm \bar{\ten}
\kL(\kh_{0})$ and therefore also not in $\rr$. \ \ $\Box$ \\

\section{Stone Spectra of Some Distributive\\ Lattices}
\label{SSDL}  
\pagestyle{myheadings}
\markboth{The Stone Spectrum of a Lattice}{Stone Spectra of
Some Distributive Lattices}

In this section we discuss the structure of Stone spectra of two
classes of examples: $\gs$-algebras, the lattice $\kT(M)$ and the
sublattice $\kT_{r}(M)$ for topological spaces $M$.\\
~\\
We begin with the lattice $\kT(M)$ of open subsets of a locally
compact Hausdorff space $M$. \\ 
Let $\frb$ be a quasipoint in $\LL$. We distinguish two cases. In the 
first case we assume that $\frb$ has an element that is a relatively 
compact open subset of $M$. Let $U_{0} \in \frb$ be such an element. 
Then
\[  \bigcap_{U \in \frb}\bar{U} \ne \emptyset ,  \]
for otherwise  $\bigcap_{U \in \frb}\overline{U \cap U_{0}} = 
\emptyset$ and from the compactness of $\bar{U_{0}}$ we see that 
there are $U_{1},\ldots,U_{n} \in \frb$ such that $\bigcap_{i = 
1}^{n}\overline{U_{i} \cap U_{0}} = \emptyset$.\\ But then $U_{0} \cap 
U_{1} \cap \ldots \cap U_{n} = \emptyset$, contrary to the defining 
properties of a filter base. The maximality of $\frb$ implies that 
every open neighbourhood of $x \in \bigcap_{U \in \frb}\bar{U}$ 
belongs to $\frb$. Therefore, as $M$ is a Hausdorff space, $\bigcap_{U 
\in \frb}\bar{U}$ consists of precisely one element of $M$. We will 
denote this element by $pt(\frb)$ and call $\frb$ a \emph{quasipoint 
over $x = pt(\frb)$}.\\
Now consider the other case in which no element of the quasipoint 
$\frb$ is relatively compact. It can be easily shown, using the 
maximality of $\frb$ again, that in this case $M\setminus K \in \frb$ 
for every \emph{compact} subset $K$ of $M$. (See lemma \ref{DL18} for a
more general statement.) We summarize these facts in the following

\begin{proposition}\label{DL1}
Let $M$ be a locally compact Hausdorff space and $\frb$ a quasipoint 
in the lattice $\kT(M)$ of open subsets of $M$. Then either 
$M\setminus K \in \frb$ for all compact subsets $K$ of $M$ or there 
is a unique element $x \in M$ such that $\bigcap_{U \in 
\frb}\bar{U} = \{x\}$.\\
In the first case $\frb$ is called an unbounded quasipoint, in the 
second a bounded quasipoint over $x$.\\
For a non-compact space $M$ let $M_{\infty} := M \cup \{\infty\}$ be 
the one-point compactification of $M$. Then the unbounded quasipoints 
in $\kT(M)$ can be considered as quasipoints over $\infty$ in 
$\kT(M_{\infty})$.
\end{proposition}

~\\
Next we consider the \emph{Boolean $\sigma$-algebra $\kbb(M)$} of all
\emph{Borel subsets} of a Hausdorff topological space $M$. The
orthocomplement of $A \in \kbm$ is the ordinary set theoretic complement
which we denote by $A'$. For some of our results the topology of $M$ must
fulfill some countability conditions. We suppose here that
\begin{enumerate}
    \item  [(i)] $M$ satisfies the first axiom of countability, i.e. each $x \in M$
    has a countable base of neighbourhoods, and that

    \item  [(ii)] $M$ satisfies the \emph{Lindelöf condition}, i.e.
    every open covering of $M$ can be refined by an at most
    countable subcovering.     
\end{enumerate}
These conditions are satisfied if e.g. the topology of $M$ has a
countable base. 

\begin{proposition}\label{DL2}
    $\frp \tm \kbm$ is a point in the lattice $\kbm$ if and only if
    $\frp$ is an atomic quasipoint in $\kbm$. 
\end{proposition}
\emph{Proof:} An atomic quasipoint in $\kbm$ has the form $\{ A \in
\kbm \ | \ x \in A \}$ for some $x \in M$, so it is obviously a point 
in $\kbm$. Conversely, assume that $\frp \tm \kbm$ is a point and let 
$\frb$ be a quasipoint in $\kbm$ that contains $\frp$. Let $A \in \frb$.
Because of $A \cup A' = M \in \frp$ we have $A \in \frp$ or $A' \in \frp$.
As $A' \notin \frb$, this implies $A \in \frp$. Hence $\frp$ is a
quasipoint. Moreover $\bigcap_{n \in \NN}A_{n} \in \frp$ for all
sequences $(A_{n})_{n \in \NN}$ in $\frp$: $\bigcap_{n \in \NN}A_{n}
\notin \frp$ would imply $\bigcup_{n}A_{n}^{'} = (\bigcap_{n}A_{n})'
\in \frp$, hence $A_{n_{0}}' \in \frp$ for some $n_{0}$, contradicting 
$A_{n_{0}} \in \frp$. \\
Now assume that 
\begin{math}
    \bigcap_{A \in \frp}\overline{A} = \emptyset,
\end{math}
i.e.
\begin{math}
    \bigcup_{A \in \frp}\overline{A}' = M.
\end{math}
The Lindelöf property implies that there is a sequence
$(A_{n})_{n \in \NN}$ in $\frp$ such that 
\begin{math}
    \bigcup_{n \in \NN}\overline{A_{n}}' = M.
\end{math}
Hence $\overline{A_{n}}' \in \frp$ for some $n$, a contradiction. The 
maximality of $\frp$ implies that $\{y\} \in \frp$ for every $y \in
\bigcap_{A \in \frp}\overline{A}$. Hence there is a unique $x \in M$
such that 
\begin{math}
    \bigcap_{A \in \frp}\overline{A} = \{x\}.
\end{math}
This means that $\frp$ is an atomic quasipoint. \ \ $\Box$ \\
~\\
Using similar arguments, we easily obtain

\begin{proposition}\label{DL3}
    A quasipoint $\frb$ in a $\gs$-algebra $\kbb$ is a point in $\kbb$
    if and only if $\We_{n \in \NN}a_{n} \in \frb$ for every sequence 
    $(a_{n})_{n \in \NN}$ in $\frb$.
\end{proposition}
We will now present a sufficient condition for a $\gs$-algebra to have
no points. For the convenience of the reader we repeat some well known
notions.

\begin{definition}\label{DL4}
    A nonempty subset $\kI$ of an $\frm$-complete lattice $\LL$ is called
    an $\frm$-ideal if it has the following properties:
    \begin{enumerate}
        \item  [(i)] If $a \in \kI$, then $a \we b \in \kI$ for all $b \in \LL$.
    
        \item  [(ii)] $\Ve_{k \ikk}a_{k} \in \kI$ for every family
	$(a_{k})_{k \ikk}$ in $\kI$ such that $\#\KK  ≤ \frm$.
    \end{enumerate}
    An $\frm$-ideal is called proper if $1 \notin \kI$. 
\end{definition}
If $\kI$ is an $\frm$-ideal in $\LL$ then the quotient $\LL / \kI$ is
defined in the following way. We define an equivalence relation $\sim 
\ \tm \LL \times \LL$ by
\[
    a \sim b \ \ :\llra \ \ \ex \ p \in \kI : \ a \vee p = b \vee p,
\]
and we denote by $[a]$ the equivalence class of $a \in \LL$. We define
\[
    [a] \vee [b] := [a \vee b] 
\]
and
\[
    [a] ≤ [b] \ \ :\llra \ \ [a \vee b] = [b].   
\]
A routine calculation shows that these are well defined binary
relations on $\LL / \kI$ which turn the quotient into a
$\vee$-semilattice. The natural definition of a meet,
\[
    [a] \we [b] := [a \we b],
\]   
however, is only well defined if the lattice $\LL$ is \emph{distributive}.
If $\kbb$ is a Boolean algebra, $\kbb / \kI$ is also orthocomplemented by
\[
    [a]^\pp := [a^\pp].
\]
The well definedness of this operation is most easily proved by using 
the following characterization of the equivalence modulo $\kI$:
\[
    a \sim b \ \ \llra \ \ (a \we b^\pp) \vee (b \we a^\pp) \in \kI.
\]
We skip the essentially computational proof.\\
If $\LL$ is complete (with or without restrictions to the cardinal
defining the degree of completeness) then it is natural to define 
\[
    \Ve_{k \ikk}[a_{k}] := [\Ve_{k \ikk}a_{k}].
\]   
Obviously, this is well defined. If $\LL$ is orthocomplemented, we
define the infinite meet by
\[
    \We_{k \ikk}[a_{k}] := (\Ve_{k \ikk}[a]^\pp)^\pp.
\]
This definition avoids the assumption of complete distributivity of
$\LL$ which would be needed when defining the infinite meet by
$[\We_{k \ikk}a_{k}]$. Collecting these facts we obtain

\begin{proposition}\label{DL5}
    If $\kI$ is an $\frm$-ideal in an $\frm$-complete Boolean algebra 
    $\kbb$ then the quotient $\kbb / \kI$ is an $\frm$-complete Boolean
    algebra.
\end{proposition}
An $\frm$-complete Boolean algebra will be simply called an {\bf
$\frm$-algebra}. For $\frm = \aleph_{0}$ we use the traditional
notation ``$\gs$-algebra''.\\
~\\
The following result is merely a corollary to proposition \ref{DL2}:

\begin{proposition}\label{DL6}
    Let $M$ be a Hausdorff space that satisfies the first axiom of
    countability and the Lindelöf condition. If $\kI$ is a $\gs$-ideal
    in $\kbm$ that contains all atoms of $\kbm$ then the $\gs$-algebra
    $\kbb := \kbm / \kI$ has no points.
\end{proposition}
\emph{Proof:} Consider the canonical projection
\[
    \begin{array}{cccc}
	\pi : & \kbm & \to & \kbm / \kI  \\
	 & A & \tto & [A].
     \end{array}
\]
$\pi$ is a $\gs$-morphism of $\gs$-algebras. Assume that there exists
a point $\frp$ in $\kbb$. Then
\[
    \overset{-1}{\pi}(\frp) = \{ A \in \kbm \ | \ [A] \in \frp \}
\]
is a point in $\kbm$. By proposition \ref{DL2} there is a unique $x
\in M$ such that $\overset{-1}{\pi}(\frp) = \frb_{x}$ and therefore
$[\{x\}] \in \frp$. But $\{x\} \in \kI$, hence $[\{x\}] = [\emptyset] 
= 0$, a contradiction. \ \ $\Box$ \\
~\\
We recall that a subset $N$ of a Hausdorff topological space $M$ is said
to be \emph{nowhere dense}, if the interior of its closure is empty. $N$ is 
called a set of \emph{first category} (or \emph{meagre}) if it is a
countable union of nowhere dense subsets, otherwise a set of
\emph{second category}. $M$ is called a \emph{Baire space} if every
nonvoid open subset is of second category. By a theorem of Baire
(\cite{bou2}, p.193) every locally compact and every complete metric 
space is a Baire space. \\

For any topological space $M$ we signify by $\kI_{1}$ the $\gs$-ideal 
of all meagre Borel subsets of $M$. The following result, not difficult
to prove, can be found in \cite{Sik}:

\begin{proposition}\label{DL7}
    Let $M$ be a Baire space. Then every equivalence class $a \in \kbm
    / \kI_{1}$ contains a unique regular open set $U_{a}$. The
    mapping
    \[
    \begin{array}{cccc}
	\gF : & \kbm / \kI_{1} & \to & \kT_{r}(M)  \\
	 & a & \tto & U_{a}
    \end{array}
    \]
    is a $\gs$-isomorphism of $\gs$-algebras.
\end{proposition}
Together with proposition \ref{DL6} we obtain

\begin{corollary}\label{DL8}
    Let $M$ be a separable complete metric space. Then $\kT_{r}(M)$
    has no points.
\end{corollary}
							   
\begin{proposition}\label{DL9}
Let $\mathcal{I}$ be a $\sigma$-ideal in the $\sigma$-algebra $\kbb$ 
and let $\pi : \kbb \to \kbb/\mathcal{I}$ be the canonical projection 
onto the quotient $\sigma$-algebra $\kbb/\mathcal{I}$. Then $\frb 
\subseteq\kbb/\mathcal{I}$ is a quasipoint if and only if 
$\overset{-1}{\pi}(\frb)$ is a quasipoint in $\kbb$ such that 
\[ \overset{-1}{\pi}(\frb) \cap \mathcal{I} = \emptyset \].
\end{proposition}
\emph{Proof:} Let $\frb$ be a quasipoint in $\kbb / \kI$. Then
$\overset{-1}{\pi}(\frb)$ is a filter base in $\kbb$. If $a \in \kbb$
then $\pi(a)$ or $\pi(a)^\pp$ belongs to $\frb$, i.e. $a$ or $a^\pp$
belongs to $\overset{-1}{\pi}(\frb)$. Thus $\overset{-1}{\pi}(\frb)$
is a quasipoint in $\kbb$. Clearly $\kI \cap \overset{-1}{\pi}(\frb) =
\emptyset$ because $a \in \kI$ if and only if $\pi(a) = 0$. \\
Conversely let $\frb$ be a nonempty subset of $\kbb / \kI$ such that
$\overset{-1}{\pi}(\frb)$ is a quasipoint with $\kI \cap \overset{-1}
{\pi}(\frb) = \emptyset$. Then $0 \notin \frb$. If $a, b \in \frb$,
$A \in \overset{-1}{\pi}(a), B \in \overset{-1}{\pi}(b)$ then $A \we B
\in \overset{-1}{\pi}(\frb)$ and therefore $a \we b = \pi(A \we B) \in
\frb$. If $a$ is an arbitrary element of $\kbb / \kI$ and $A \in
\overset{-1}{\pi}(a)$ then $A$ or $A^\pp$ belongs to
$\overset{-1}{\pi}(\frb)$, hence $a \in \frb$ or $a^\pp \in \frb$.
This shows that $\frb$ is a quasipoint. \ \ $\Box$ \\

\begin{proposition}\label{DL10}
    Let $\kI$ be a $\gs$-ideal in a $\gs$-algebra $\kbb$ and let $\frb
    \tm \kbb$ be a quasipoint such that $\kI \cap \frb = \emptyset$.
    Then $\pi(\frb) \tm \kbb / \kI$ is a quasipoint and
    $\overset{-1}{\pi}(\pi(\frb)) = \frb$. 
\end{proposition}
\emph{Proof:} The same arguments as in the foregoing proof show that
$\pi(\frb)$ is a quasipoint. Then $\overset{-1}{\pi}(\pi(\frb))$ is a 
quasipoint in $\kbb$ that contains $\frb$ and therefore $\frb =
\overset{-1}{\pi}(\pi(\frb))$ by maximality. \ \ $\Box$ \\

\begin{corollary}\label{DL10a}
    Let $\kbb$ be a $\gs$-algebra and $\kI$ a $\gs$-ideal in $\kbb$.
    Then the set 
    \[
	\kQ^2 (\kbb) := \{ \frb \in \qb \ | \ \frb \cap \kI = \emptyset
	\}
    \]
    is compact and homeomorphic to the Stone spectrum of $\kbb / \kI$. 
\end{corollary}
\emph{Proof:} By proposition \ref{DL10} the canonical projection $\pi 
: \kbb \to \kbb / \kI$ induces a mapping
\[
    \begin{array}{cccc}
	\pi_{\ast} : & \kQ^2 (\kbb) & \to & \kQ(\kbb / \kI)  \\
	 & \frb & \tto & \pi(\frb)
    \end{array}
\] 
which is, according to propositions \ref{DL9} and \ref{DL10}, a
bijection $\pi_{\ast}^{-1} : \frc \tto \overset{-1}{\pi}(\frc)$. 
Let $A \in \kbb \smm \kI$. Then
$\pi_{\ast}(\kQ_{A}(\kbb) \cap \kQ^2 (\kbb)) \tm \kQ_{[A]}(\kbb /
\kI)$. If $\Tilde{\frb} \in \kQ_{[A]}(\kbb / \kI)$, then
$\pi_{\ast}^{-1}(\Tilde{\frb}) \in \kQ_{B}(\kbb) \cap \kQ^2 (\kbb)$
for some $B \in [A]$. This shows that
\[
    \pi_{\ast}^{-1}(\kQ_{[A]}(\kbb /\kI)) = \bigcup_{B \in
    [A]}(\kQ_{B}(\kbb) \cap \kQ^2 (\kbb)).
\]
Now observe that $A, B \in [A]$ if and only if $A \vee N = B \vee N$
for some $N \in \kI$. Therefore $A \in \frb$ if and only if $B \in
\frb$ for all $\frb \in \kQ^2 (\kbb)$. Thus
\[
\pi_{\ast}^{-1}(\kQ_{[A]}(\kbb /\kI)) = \kQ_{B}(\kbb) \cap \kQ^2 (\kbb).
\]
for all $B \in [A]$. Hence $\pi_{\ast}$ is a homeomorphism.\\
Now let $\frb \notin \kQ^2(\kbb)$. This means that there is some
$A \in \frb \cap \kI$ and therefore $\kQ_{A}(\kbb) \cap \kQ^2 (\kbb) =
\emptyset$. So $\kQ^2 (\kbb)$ is closed. As $\qb$ is compact,
$\kQ^2 (\kbb)$ is compact too. \ \ $\Box$ \\

\begin{remark}\label{DL10b}
    Since $\kbb$ is a distributive ortholattice, the condition $\frb
    \cap \kI = \emptyset$ is equivalent to $\kI^\pp \tm \frb$ where
    $\kI^\pp := \{ A^\pp \mid A \in \kI \}$. 
\end{remark}
By a theorem of Loomis and Sikorski (\cite{Sik}) every 
$\sigma$-algebra is the quotient of a $\sigma$-algebra of Borel sets of 
a compact space modulo a $\sigma$-ideal.\\
The construction is roughly as follows. Let $\kbb$ be a $\gs$-algebra 
and let 
\[
    \begin{array}{cccc}
	\eta : & \kbb & \to & \frA(\qb)  \\
	 & a & \tto & \qab
    \end{array}
\]
be the Stone isomorphism between the Boolean {\bf algebras} $\kbb$ and
$\frA(\qb)$. Note that $\frA(\qb)$ is in not a $\gs$-algebra, because 
$\bigcup_{\nin}\kQ_{a_{n}}(\qb)$ is open but, in general, not closed. 
Let $\frA_{\gs}(\qb)$ be the $\gs$-algebra generated by $\frA(\qb)$ in
$\kbb(\qb)$. Then $\eta$ can be considered as a homomorphism from
$\kbb$ to $\frA_{\gs}(\qb)$. It is not a $\gs$-homomorphism, because
\[
    \overline{\bigcup_{\nin}\kQ_{a_{n}}(\qb)} = \kQ_{\Ve_{\nin}a_{n}}(\qb)
\]
for every sequence $(a_{n})_{\nin}$ in $\kbb$. Thus the obstacles for 
$\eta$ to be a $\gs$-homomorphism are the \emph{boundaries} of open
sets of the form $\bigcup_{\nin}\kQ_{a_{n}}(\qb)$. It is easy to see
that cutting them away means to go over to the quotient $\frA_{\gs}(\qb)
/ \kI(\frN)$ where $\kI(\frN)$ is the $\gs$-ideal in
$\frA_{\gs}(\qb)$ generated by
\[
    \frN := \{ \bigcap_{\nin}\kQ_{a_{n}}(\qb) \mid \We_{\nin}a_{n} = 0
    \}. 
\]
Note that the elements of $\kI(\frN)$ are meagre subsets of $\qb$.
It is not difficult to prove that the composition $\eta_{\gr} := \gr \circ
\eta$ of $\eta$ with the canonical projection $\gr : \frA_{\gs}(\qb)
\to \frA_{\gs}(\qb) / \kI(\frN)$ is a \emph{surjective} $\gs$-homomorphism
\[
    \eta_{\gr} : \kbb \to \frA_{\gs}(\qb) / \kI(\frN).
\] 
The technically difficult part of Loomis's construction is the proof
that $\eta_{\gr}$ is injective. For this part we refer to Sikorski
(\cite{Sik}). \\
~\\
The theorem of Loomis and Sikorski represents an abstract $\gs$-algebra
$\kbb$ as a $\gs$-algebra of subsets of $\qb$ (with the standard set theoretic
operations) modulo a $\gs$-ideal. This is often not a very economic
representation because, as we mentioned already, the Stone spectrum is
typically very large. We will prove in the next section a $C^\ast$-algebraic
description of Stone spectra of $\gs$-algebras of the form $\frA(M) / \kI$ 
where $\frA(M)$ is an arbitrary $\gs$-algebra of sets\footnote{This
means in particular that the Boolean operations are the usual set
theoretic ones.} and $\kI$ is a $\gs$-ideal in $\frA(M)$.\\
~\\
~\\
If $\MM$ is a complete lattice, isomorphic to $\LL$ via a lattice 
isomorphism $\gF : \LL \to \MM$, then it is easy to see that $\gF$ 
induces a homeomorphism
\[  \gF_{*} : \kQ(\LL) \to \kQ(\MM) \]
of the corresponding Stone spectra:
\[  \gF_{*}(\frb) := \{ \gF(a) \mid a \in \frb \} . \]
The opposite conclusion, however, is not true.\\
In fact we can show that the Stone spectra $\kQ(\kT(M))$ and 
$\kQ(\kT_{r}(M))$ are homeomorphic for every topological space $M$. 
But in general the lattice $\kT(M)$ of open subsets of $M$ is not 
isomorphic to the lattice $\kT_{r}(M)$ of regular open subsets of 
$M$, because $\kT(M)$ possesses points whereas, in general, 
$\kT_{r}(M)$ does not.

In section \ref{latt} we have seen that $\kT_{r}(M)$ is a Boolean algebra with 
complement operation
\[  U \mapsto U^c \]
where $U^c := M\setminus \bar{U}$. Now it is easy to see that 
\[  U \cap V = \emptyset \quad \Longrightarrow \quad U^{cc} \cap 
V^{cc} = \emptyset \]
holds for all open sets $U,V \subseteq M$. From this fact we get

\begin{lemma}\label{DL11}
Let $M$ be a topological space and let $\frb$ be a quasipoint in 
$\kT(M)$. Then
\[  \frb^r := \{ U^{cc} \mid U \in \frb \} \]
is a quasipoint in $\kT_{r}(M)$.
\end{lemma}

\begin{proposition}\label{DL12}
The mapping
\begin{eqnarray*}
	\rho & : & \kQ(\kT(M)) \to \kQ(\kT_{r}(M))  \\
	 &  & \frb \longmapsto \frb^r
\end{eqnarray*}
is a homeomorphism of Stone spectra.
\end{proposition}
\emph{Sketch of proof:} The first thing to show is that every 
quasipoint $\frR$ in $\kT_{r}(M)$ is contained in exactly one 
quasipoint in $\kT(M)$. Thus $\rho$ is a bijection. Moreover
\[ U \in \frb \quad \Longleftrightarrow \quad U^{cc} \in \frb^r \]
for every quasipoint $\frb$ in $\kT(M)$. This implies
\[ \rho(\kQ_{U}(\kT(M))) = \kQ_{U^{cc}}(\kT_{r}(M)) \]
and
\[ \rho^{-1}(\kQ_{W}(\kT_{r}(M))) = \kQ_{W}(\kT(M)), \]
i.e. $\rho$ is a homeomorphism. $\Box$

\begin{corollary}\label{DL13}
The Stone spectrum $\kQ(\kT(M))$ is compact.
\end{corollary}

\begin{corollary}\label{DL14}
Let $M$ be a compact Hausdorff space and let
\[ pt : \kQ(\kT(M)) \to M \]
be the map that assigns to $\frb \in \kQ(\kT(M))$ the element 
$pt(\frb) \in M$ determined by $\bigcap_{U \in \frb}\bar{U}$. Then 
the quotient topology of $M$ induced by $pt$ coincides with the given 
topology of $M$.
\end{corollary}
This follows from the fact that $pt$ is a continuous mapping and 
therefore the quotient topology is finer than the given topology. It 
cannot be strictly finer because both topologies are compact and 
Hausdorff. \\
~\\ 
This result gives an extreme example for the fact that the projection 
onto the quotient by an equivalence relation need \emph{not} be an open 
mapping: let $M$ be a \emph{connected} compact Hausdorff space. The 
compactness of the Stonean space $\kQ(\kT(M)$ implies that $pt$ is a 
closed mapping. If it was also an open mapping the total 
disconnectedness of $\kQ(\kT(M))$ would imply that the image $M$ of 
$pt$ is totally disconnected, too. As $M$ is connected, this is only 
possible for the trivial case that $M$ consists of a single element.\\
~\\
We want to show up some relations between the Stone spectrum of $\kbm$
and the Stone spectrum of $\ktm$ for a Baire space $M$, in particular 
for a locally compact space $M$. \\
Let $M$ be a Baire space and $\kI_{1} \tm \kbm$ the $\gs$-ideal of meagre
Borel sets. The assumption that $M$ is a Baire space is expressed by
\[
    \ktm \cap \kI_{1} = \{\emptyset\}.
\]
In what follows, $\sim$ means equivalence modulo $\kI_{1}$. 

\begin{proposition}\label{DL15}
    Let $M$ be a Baire space and let $\frb \in \qm$. Then 
    \[
	 \frc(\frb) := \{ A \in \kbm \ | \ \ex \ U \in \frb : \ A \sim U \}
    \]
    is a quasipoint in $\kbm$ with $\frc(\frb) \cap \kI_{1} =
    \emptyset$.
\end{proposition}
\emph{Proof:} Let $\pi : \kbm \to \kbm / \kI_{1}$ be the canonical
projection. Then
\[
    \frc(\frb) = \overset{-1}{\pi}(\pi(\frb)).
\]
Since $\pi$ is a lattice homomorphism and $\frb \cap \kI_{1} =
\emptyset$, $\pi(\frb)$ is a filter base. Let $\Tilde{\frb}$ be a quasipoint
that contains $\pi(\frb)$. Then, by proposition \ref{DL9},
$\overset{-1}{\pi}(\Tilde{\frb})$ is a quasipoint in $\kbm$ such that 
$\overset{-1}{\pi}(\Tilde{\frb}) \cap \kI_{1} = \emptyset$. Clearly
$\frc(\frb) \tm \overset{-1}{\pi}(\Tilde{\frb})$. Assume that this
inclusion is proper and take an element $A \in \overset{-1}{\pi}(\Tilde{\frb})
\smm \frc(\frb)$. By proposition \ref{DL7} there is a $U \in \ktm$
such that $A \sim U$. According to our assumption we have $U \notin
\frb$. But then also $U \notin \overset{-1}{\pi}(\Tilde{\frb})$,
because $\frb \tm \overset{-1}{\pi}(\Tilde{\frb})$ and $U \in
\overset{-1}{\pi}(\Tilde{\frb})$ would imply $V \cap U \ne \emptyset$ 
for all $V \in \frb$ and therefore $U \in \frb$. Hence $M \smm U \in
\overset{-1}{\pi}(\Tilde{\frb})$, so $A \cap (M \smm U) \in
\overset{-1}{\pi}(\Tilde{\frb})$. But $A \sim U$ implies $A \cap (M
\smm U) \in \kI_{1}$ which contradicts $\overset{-1}{\pi}(\Tilde{\frb})
\cap \kI_{1} = \emptyset$. Therefore $\overset{-1}{\pi}(\Tilde{\frb}) 
= \frc(\frb)$. \ \ $\Box$

\begin{definition}\label{DL16}
    A quasipoint $\frc$ in $\kbm$ is called a quasipoint of second
    category if $\frc \cap \kI_{1} = \emptyset$. Otherwise it is
    called a quasipoint of first category. As in corollary \ref{DL10a} we denote
    the set of quasipoints of second category by $\kQ^2 (\kbm)$. 
\end{definition}

\begin{proposition}\label{DL17}
    Let $M$ be a Baire space. Then the mapping 
    \[
	\begin{array}{cccc}
	    \pi_{M} : & \qm & \to & \kQ^2 (\kbm)  \\
	     & \frb & \tto & \frc(\frb)
	\end{array}
    \]
    is a homeomorphism.
\end{proposition}
\emph{Proof:} From propositions \ref{DL7}, \ref{DL10a}, \ref{DL12} and
\ref{DL15} we have the following homeomorphisms:
\[
    \begin{array}{ccccccc}
	\qm & \to & \kQ(\kT_{r}(M)) & \to & \kQ(\kbm / \kI_{1}) & \to &
	\kQ^2 (\kbm)  \\
	\frb & \tto & \frb^r & \tto & \pi(\frb^r) & \tto &
	\overset{-1}{\pi}(\pi(\frb^r)).
    \end{array}
\]
$\pi_{M}$ is just the composition of these. \ \ $\Box$\\
~\\
Let $M$ be a Baire space. It is obvious that a quasipoint $\frc \in \qbm$
contains at most one quasipoint $\frb \in \qm$. If $\frc$ is of second 
category, then it is of the form $\frc(\frb)$ for exactly one $\frb \in \qm$ 
(proposition \ref{DL15}). If $\frc$ is an atomic quasipoint then it is of
second category if the defining atom is isolated in $M$. If $\frc \in 
\qbm$ is atomic and of first category then $\frc$ does not contain a
quasipoint $\frb \in \qm$: $\frc = \{ A \in \kbm \ | \ x \in A \}$ and
$\frb \tm \frc$ would imply $x \in U$ for all $U \in \frb$. But this
is a contradiction to the maximality of $\frb$ since $\{x\}$ is not open.
On the other hand, a quasipoint $\frc \in \qbm$ of first category can very
well contain a quasipoint $\frb \in \qm$. Let for example $M = \RR$ and
take any quasipoint $\frb$ in $\kQ(\kT(\RR))$. Then $\frb \cup \{\QQ\}$ is
contained in a quasipoint $\frc \in \kQ(\kbb(\RR))$. $\frc$ is of first
category because $\QQ$ is meagre in $\RR$. 

\begin{lemma}\label{DL18}
    Let $M$ be a topological space and let $\kj$ be a dual ideal in
    $\ktm$. Then $\kj$ is a quasipoint if and only if the following
    alternative is satisfied:
    \[
	\all \ U \in \ktm : \ U \in \kj \ \text{or} \ U^c \in \kj.
    \]
\end{lemma}
\emph{Proof:} We recall that the \emph{pseudocomplement} $U^c$ of $U
\in \ktm$ is defined as $U^c := M \smm \overline{U}$. If $\frb$ is a
quasipoint in $\ktm$ and $U \notin \frb$ then there is some $V \in
\frb$ such that $V \cap U = \emptyset$. But then also $V \cap
\overline{U} = \emptyset$, hence $V \tm U^c$ and therefore $U^c \in
\frb$.\\
Conversely, let $\kj$ be a dual ideal in $\ktm$ that satisfies the
alternative in the lemma. Let $\frb$ be a quasipoint that contains
$\kj$. If $W \in \frb$ then also $W \in \kj$, for otherwise $W^c \in
\kj$, hence also $W^c \in \frb$ and therefore $\emptyset = W \cap W^c 
\in \frb$, a contradiction. \ \ $\Box$ \\

We denote by $\partial A$ the \emph{boundary} of the subset $A$ of $M$,
i.e. $\partial A = \overline{A} \cap \overline{M \smm A}$.

\begin{proposition}\label{DL19}
    A quasipoint $\frc$ in $\kbm$ contains a quasipoint $\frb \in \qm$
    if and only if $\partial U \notin \frc$ for all $U \in \ktm$. 
\end{proposition}
\emph{Proof:} Let $U \in \ktm$ such that $\partial U \in \frc$. Then
$M \smm U, \ \overline{U} \in \frc$ and therefore $U, U^c \notin \frc$. 
If $\frc$ would contain a quasipoint $\frb \in \qm$ then, according to
lemma \ref{DL18}, $U$ or $U^c$ would belong to $\frb$, a contradiction.
\\ Conversely, if $\partial U \notin \frc$ for all $U \in \ktm$, then 
$U \cup U^c \in \frc$ and, therefore, $U \in \frc$ or $U^c \in \frc$. 
Hence $\frc \cap \ktm \in \qm$.  \ \ $\Box$ \\

\section{Stone Spectra of von Neumann algebras}
\label{SSVN}
\pagestyle{myheadings}
\markboth{The Stone Spectrum of a Lattice}{Stone Spectra of von Neumann
algebras}

Let $\rr$ be a von Neumann algebra, considered as a subalgebra of
$\lh$ for some Hilbert space $\kh$. The Stone spectrum of the
projection lattice $\pr$ is called the Stone spectrum of $\rr$ and we 
denote it by $\qr$.\\
~\\
The projection lattice $\pr$ reflects the structure of the von Neumann
algebra $\rr$. $\pr$ is an orthomodular lattice and therefore a
projection $P \in \pr$ can be identified with the open closed subset
$\qpr$ of $\qr$. This shows that the Stone spectrum $\qr$ of $\rr$ is 
in general a highly complicated object. So we cannot expect a simple
characterization of $\qr$ for a general von Neumann algebra $\rr$.
Indeed only partial results are known.\\
~\\
Putting aside von Neumann algebras of finite dimension, the most simple
case is the Stone spectrum of an \emph{abelian} von Neumann algebra
$\kaa$. We will prove now that the Stone spectrum $\qa$ of $\kaa$ is
homeomorphic to the Gelfand spectrum of $\kaa$ in a canonical manner. \\
~\\
We begin with some preparations.

\begin{definition}\label{def: 40}
    Let $A \in lin_{\CC}\poa$. 
    \[
	A = \sum_{j = 1}^{m}b_{j}P_{j}
    \]
    is called an \emph{orthogonal representation of $A$} if the
    projections $P_{j}$ are pairwise orthogonal.
\end{definition}
In analogy to standard measure theoretical methods one can easily see 
that each $A = \sum_{k}a_{k}E_{k} \in lin_{\CC}\poa$ has an orthogonal
representation:
{\begin{align}
   \sum_{k = 1}^{n}a_{k}E_{k} &= (a_{1}+\ldots+a_{n})E_{1} 
   \ldots E_{n} \notag \\
   &+ \sum_{i = 1}^{n}(a_{1}+\ldots 
   +\Hat{a}_{i}+\ldots+a_{n})E_{1}\ldots 
   (I - E_{i})\ldots E_{n} \notag \\
   &+\sum_{1\leq i < j\leq    n}^{n}(a_{1}+\ldots+\Hat{a}_{i}+\ldots 
   +\Hat{a}_{j}+\ldots+a_{n})
   E_{1}\ldots (I - E_{i})\ldots 
   (I - E_{j})\ldots E_{n} \notag \\
   &+\ldots+ \sum_{i =1}^{n}a_{i}(I - E_{1})\ldots 
   E_{i}\ldots (I - E_{n}). \notag
\end{align}}   
We call this the {\bf standard orthogonal representation of $A$}. 

\begin{lemma}\label{lem: 41}
    Let $\sum_{j = 1}^{m}b_{j}P_{j}$ and $\sum_{k = 1}^{n}c_{k}Q_{k}$ 
    be two orthogonal representations of $A \in lin_{\CC}\poa$. Then
    \[
	\sum_{j = 1}^{m}b_{j}\chi_{\qpja} = \sum_{k =
	1}^{n}c_{k}\chi_{\qqka}.
    \]
\end{lemma}
\emph{Proof:} Without loss of generality we may assume that the
occuring coefficients $b_{1}, \ldots, b_{m}, c_{1}, \ldots, c_{n}$ are
all different from zero. Then  $\sum_{j}b_{j}P_{j} =
\sum_{k}c_{k}Q_{k}$ implies that $P_{1} + \ldots + P_{m} = Q_{1} + \ldots
+ Q_{n}$ holds. Indeed let $V_{j} := im P_{j}, W_{k} := im Q_{k} \ \  
(j ≤ m, k ≤ n)$ and let $v_{j_{0}} \in V_{j_{0}}$. Then 
\[
    b_{j_{0}}v_{j_{0}} = (\sum_{j = 1}^{m}b_{j}P_{j})v_{j_{0}} =
    (\sum_{k = 1}^{n}c_{k}Q_{k})v_{j_{0}} \in W_{1} + \ldots + W_{n},
\]
hence $v_{j_{0}} \in W_{1} + \ldots + W_{n}$. This shows $V_{1} + \ldots
+ V_{m} \tm W_{1} + \ldots + W_{n}$ and by symmetry we obtain $V_{1} +
\ldots + V_{m} = W_{1} + \ldots + W_{n}$. \\
Let $\gb \in \qa$. If $P_{1} + \ldots + P_{m} \notin \gb$ (and consequently
$Q_{1} + \ldots + Q_{n} \notin \gb$) then both sides of the asserted
equality are zero. If $P_{1} + \ldots + P_{m} \in \gb$ then, due to
orthogonality, there are uniquely determined indices $j_{0}, k_{0}$
such that $P_{j_{0}}, Q_{k_{0}} \in \gb$. But then $P_{j_{0}}Q_{k_{0}}
\in \gb$, in particular $P_{j_{0}}Q_{k_{0}} \neq 0$. This shows
$b_{j_{0}} = c_{k_{0}}$ and because of $\sum_{j}b_{j}\chi_{\qpja}(\gb)
= b_{j_{0}}, \sum_{k}c_{k}\chi_{\qqka}(\gb) = c_{k_{0}}$ the assertion
follows also in this case. \ \ $\Box$ \\
~\\
The {\bf Gelfand spectrum of the abelian von Neumann algebra $\kaa$} is
the set of non-zero multiplicative linear functionals on $\kaa$. The
weak*-topology turns it into a compact Hausdorff space (\cite{kr1})
which is denoted by $\gO(\kaa)$.

\begin{theorem}\label{theo: 42}
    Let $\kaa$ be an abelian von Neumann algebra. Then the Gelfand spectrum
    $\gO(\kaa)$ is homeomorphic to the Stone  spectrum $\qa$ of $\kaa$.
\end{theorem}
\emph{Proof:} Let $\gt \in \gO(\kaa)$. Then
\[
    \gb_{\gt} := \{ P \in \poa \mid \gt(P) = 1 \}
\]
is a quasipoint of $\pa$: \\
For all $P \in \pa$ we have $\gt(P) \in \{0, 1\}$ because $\gt$ is
multiplicative. By definition $0 \notin \gb(\gt)$ and $P, Q \in
\gb(\gt)$ implies $\gt(PQ) = \gt(P)\gt(Q) = 1$, hence $PQ \in
\gb(\gt)$. Moreover $\gt(I - P) = 1 - \gt(P)$ and therefore
\[
    \all \ P \in \pa : \ P \in \gb_{\gt} \quad \text{or} \quad I - P
    \in \gb_{\gt}.
\]
As $\pa$ is distributive, this means that $\gb_{\gt}$ is a quasipoint.
\\
Let $\gs, \gt \in \gO(\kaa)$ such that $\gb_{\gs} = \gb_{\gt}$. Then
for all $P \in \pa$: $\gs(P) = 1$ if and only if $\gt(P) = 1$. Hence
$\gs$ and $\gt$ agree on $\pa$ and therefore also on $lin_{\CC}\pa$.
Using their continuity we conclude from the spectral theorem that $\gs
= \gt$. The mapping
\[
\begin{array}{ccc}
    \gO(\kaa) & \to & \qa  \\
    \gt & \tto & \gb_{\gt}
\end{array}
\]
is therefore injective. \\
Conversely, let $\gb \in \kQ(\kaa)$ be given. We define a mapping 
\[
    \gt_{\gb} : \pa \to \{0, 1\}
\] 
by
\[
    \gt_{\gb}(P) :=
    \begin{cases}
	1   &   \text{if} \ \ P \in \gb  \\
	0   &   \text{otherwise}. 
    \end{cases}
\]
Because of $P, Q \in \gb$ if and only if $PQ \in \gb$ we have
\[
    \all \ P, Q \in \pa : \ \gt_{\gb}(PQ) = \gt_{\gb}(P)\gt_{\gb}(Q).
\]
In the next step we extend $\gt_{\gb}$ to a linear functional on
$lin_{\CC}\pa$ (which will be denoted by $\gt_{\gb}$ too). Let $A =
\sum_{j = 1}^{m}b_{j}P_{j}$ be an orthogonal representation of $A \in 
lin_{\CC}\pa$. We may assume that $P_{1} + \ldots + P_{m} = I$ and
that $b_{1}, \ldots, b_{m - 1}$ are different from zero. We call such 
a representation a \emph{complete orthogonal representation}. Because 
of
\[
    \qa = \bigcup_{j ≤ m}\qpja
\]  
we have $\gb \in \qpja$ for exactly one $j ≤ m$. Denote this $j$ by
$j_{\gb}$. Of course we are forced to define
\[
    \gt_{\gb}(\sum_{j = 1}^{m}b_{j}P_{j}) := b_{j_{\gb}}.
\] 
We have to show that $\gt_{\gb}$ is well defined. Let $\sum_{j =
1}^{m}b_{j}P_{j}$ and $\sum_{k = 1}^{n}c_{k}Q_{k}$ be two (complete)
orthogonal representations of $A \in lin_{\CC}\pa$. Then, by lemma
\ref{lem: 41}, $b_{j_{\gb}} = c_{k_{\gb}}$. This proves that
$\gt_{\gb}$ is well defined on $lin_{\CC}\pa$. \\
Now let $\sum_{k = 1}^{n}a_{k}E_{k}$ be an arbitrary element of
$lin_{\CC}\pa$. Using the standard orthogonal representation we see
that
\[
    \gt_{\gb}(\sum_{k = 1}^{n}a_{k}E_{k}) = a_{j_{1}} + \ldots
    +a_{j_{s}}
\]
where $j_{1}, \ldots, j_{s}$ are precisely those indices for which
$E_{j_{1}}, \ldots, E_{j_{s}}$ are elements of $\gb$. This shows that 
$\gt_{\gb}$ is linear. Multiplicativity follows from linearity and the
fact that $\gt_{\gb}$ is multiplicative on projections. \\
$\gt_{\gb}$ is continuous in norm because for orthogonal
representations we have
\[
    |\sum_{k = 1}^{n}a_{k}P_{k}| = \max_{k ≤ n}|a_{k}|.
\]  
The spectral theorem assures that $\gt_{\gb}$ has a unique extension
to a multiplicative linear functional on $\kaa$ which we denote again 
by $\gt_{\gb}$. By construction we have
\[
    \gb_{\gt_{\gb}} = \gb
\] 
and therefore the mapping $\gt \tto \gb_{\gt}$ is a bijection from
$\gO(\kaa)$ onto $\qa$. In order to prove that this is a homeomorphism
we have only to show that it is continuous because $\gO(\kaa)$ and
$\qa$ are compact Hausdorff spaces. \\
Let $\gt_{0} \in \gO(\kaa), \ 0 < \eps < 1$ and let $P \in \pa$ such
that $\gt_{0}(P) = 1$. Then
\[
    N_{w}(\gt_{0}) := \{ \gt \in \gO(\kaa) | \ |\gt(P) - \gt_{0}(P)| <
    \eps \}
\] 
is an open neighborhood of $\gt_{0}$ and from $\eps < 1$ we conclude
\[
\begin{array}{ccc}
    \gt \in N_{w}(\gt_{0})  & \llra & \gt(P) = \gt_{0}(P)  \\
     & \llra & P \in \gb_{\gt}  \\
     & \llra & \gb_{\gt} \in \qpa.
\end{array}
\]
This means that $N_{w}(\gt_{0})$ is mapped bijectively onto the open
neighborhood $\qpa$ of $\gb_{\gt_{0}}$. The $\qpa$ with $P \in
\gb_{\gt_{0}}$ form a neighborhood base of $\gb_{\gt_{0}}$ in the
Stone topology of $\qa$. Hence $\gt \tto \gb_{\gt}$ is continuous. \ \
$\Box$ \\
~\\
~\\
We can use the proof of theorem \ref{theo: 42} (with some obvious
changings) to show that the Stone spectrum of a $\gs$-algebra of the
form $\frA(M) /\kI$, where $\frA(M)$ is a $\gs$-algebra of subsets of 
a non-empty set $M$ and $\kI \tm \frA(M)$ is a $\gs$-ideal, is the
Gelfand spectrum of an abelian $C^\ast$-algebra that is canonically
associated to the $\gs$-algebra $\frA(M) / \kI$. \\
~\\
Let $\kf := \kf_{\frA(M)}(M, \CC)$ be the algebra of all bounded
$\frA(M)$-measurable functions $M \to \CC$. $\kf$, equipped with the
norm of uniform convergence, is an abelian $C^\ast$-algebra. \\
For $f \in \kf$, let 
\[
    P(f) := \{ x \in M \mid f(x) \ne 0 \}.
\]
Because of 
\[
    P(fg) = P(f) \cap P(g) \ \ \text{and} \ \ P(f + g) \tm P(f) \cup
    P(g)
\]
for all $f, g \in \kf$, the set
\[
    \kf(\kI) := \{ h \in \kf \mid P(h) \in \kI \} 
\]
is an ideal in $\kf$. It is closed in the norm topology of $\kf$: let 
$(h_{n})_{\nin}$ be a sequence in $\kf(\kI)$ that converges uniformly 
to $h \in \kf$. If $x \in P(h)$, there is $n_{x} \inn$ such that
$|h_{n_{x}}(x) - h(x)| < |h(x)|/2$. Hence
\[
    P(h) \tm \bigcup_{\nin}P(h_{n}) \in \kI
\]
and, therefore, $h \in \kf(\kI)$. The quotient
\[
     \kaa_{\kbb} := \kf / \kf(\kI), 
\]
where $\kbb$ is an abbreviation for $\frA(M) / \kI$, is an abelian
$C^\ast$-algebra, which we call the \emph{$C^\ast$-algebra
associated to the $\gs$-algebra $\kbb$.} The norm of $\kaa_{\kbb}$ is 
the \emph{quotient norm}
\[
    |[f]| := \inf \{ |f + h| \mid h \in \ffi \}, 
\]
and it is easy to see that this norm is an \emph{essential supremum
norm}:
\[
    |[f]| = \inf \{ c > 0 \mid \{x \in M \mid |f(x)| > c \} \in \kI \}. 
\]
 In the following we will show
that the Stone spectrum $\qb$ of $\kbb$ is homeomorphic to the Gelfand
spectrum $\gO(\kaa_{\kbb})$ of $\kaa_{\kbb}$.\\
~\\
The Stone spectrum $\qb$ is homeomorphic to
\[
    \qAm_{\kI} := \{ \frb \in \qAm \mid \kI^\pp \tm \frb \}.   
\]
$h \in \ffi$ can be uniformly approximated by step functions of the
form
\[
    s_{\kI} = \sum_{k = 1}^{n}a_{k}\chi_{A_{k}}
\]
with $a_{1}, \ldots, a_{n} \in \CC$ and mutually disjoint $A_{1},
\ldots, A_{n} \in \kI$. Let $\gt_{\frb} \in \gO(\kf)$ be the
character induced by $\frb \in \qAm_{\kI}$. It is quite analogously
defined as in the case of abelian von Neumann algebras:
\[
    \gt_{\frb}(\chi_{A}) := 
    \begin{cases}
	1   & \ \ \text{if} \ \ A \in \frb  \\
	0   & \ \ \text{otherwise}
    \end{cases}
\]
for all $A \in \frA(M)$. It follows from the
definition of $\gt_{\frb}$ that $\gt_{\frb}(s_{\kI}) = 0$ for all step
functions $s_{\kI}$ of the above form and therefore, by continuity,
$\gt_{\frb}$ vanishes on $\ffi$. Hence $\gt_{\frb}$ induces a
character $\gtt_{\frb}$ of $\kaa_{\kbb}$. Conversely, if a character
$\gtt$ of $\kaa_{\kbb}$ is given, we obtain a character $\gt := \gtt \circ
\gr$ of $\kf$ by composition with the canonical projection $\gr : \kf \to
\kf / \ffi$. As in the case of von Neumann algebras, $\gt$ gives rise 
to a quasipoint
\[
    \frb_{\gtt} := \{ A \in \frA(M) \mid \gt(\chi_{A}) = 1 \}
\]
which, by construction, belongs to $\qAm_{\kI}$. Thus we get a
bijection $\frb \tto \gtt_{\frb}$ from $\qAm_{\kI}$ onto
$\gO(\kaa_{\kbb})$. The same argument as in the proof of theorem
\ref{theo: 42} shows that it is a homeomorphism. Therefore, we have
proved the following theorem which is a generalization of 5.7.20 in
\cite{kr3}:

\begin{theorem}\label{42a}
    Let $M$ be a non-empty set, $\frA(M)$ a $\gs$-algebra of subsets
    of $M$ and $\kI$ a $\gs$-ideal in $\frA(M)$. Furthermore, let
    $\kf_{\frA(M)}(M, \CC)$ be the abelian algebra of all bounded
    $\frA(M)$-measurable functions $M \to \CC$. $\kf_{\frA(M)}(M, \CC)$
    is a $C^\ast$-algebra with respect to the supremum-norm and the
    set $\ffi$ of all $f \in \kf_{\frA(M)}(M, \CC)$ that vanish outside
    some set $A \in \kI$ is a closed ideal in $\kf_{\frA(M)}(M, \CC)$.
    Then the Gelfand spectrum of the quotient $C^\ast$-algebra
    $\kf_{\frA(M)}(M, \CC) / \ffi$ is homeomorphic to the Stone spectrum
    of the $\gs$-algebra $\frA(M) / \kI$.
\end{theorem}
Note that $\kf_{\frA(M)}(M, \CC) / \ffi$ may fail to be a von Neumann 
algebra (\cite{kr3}, 5.7.21(iv)).
~\\
~\\
If the von Neumann algebra $\rr$ is a \emph{factor of type $I$}, i.e. if $\rr$
is isomorphic to $\lh$ for some Hilbert space $\kh$, we have the following
simple results for the Stone spectrum $\qr$. We formulate and prove these
results in the lattice $\llh$ rather than in the equivalent projection
lattice $\ph$.

\begin{proposition}\label{vn1}
Let $\frb$ be a quasipoint in $\LL(\kh)$. $\frb$ contains an element 
of finite dimension if and only if there is a unique line $\CC x_{0}$ in $\kh$ 
such that
\[ \frb = \{ U \in \LL(\kh) \ \mid \CC x_{0} \subseteq U \} . \]
$\frb$ does not contain an element of finite dimension if and only if $W \in 
\frb$ for all $W\in \LL(\kh)$ of finite codimension.
\end{proposition}
\emph{Proof:} Let $U_{0} \in \frb$ be finite dimensional. Then $U \cap 
U_{0} \ne 0$ for all $U \in \frb$ and therefore $\{ U \cap U_{0} \mid 
U \in \frb \}$ contains an element $V_{0}$ of minimal dimension. 
Hence $V_{0} \subseteq U$ for all $U \in \frb$ and by the maximality 
of $\frb$ $V_{0}$ must have dimension one.\\
Assume that a quasipoint $\frb$ in $\LL(\kh)$ contains every $W \in 
\LL(\kh)$ of finite codimension. Let $U$ be a finite dimensional 
subspace of $\kh$. Then $U^{\perp} \in \frb$ and therefore $U \notin 
\frb$ because of $U \cap U^{\perp} = 0$.\\
Let $V \in \LL(\kh)$ be of finite codimension and $V \notin \frb$. Then 
there is some $U \in \frb$ such that $U \cap V = 0$. Consider the 
orthogonal projection
\[ P_{V^{\perp}}: \kh \to V^{\perp} \] 
onto $V^{\perp}$. $U \cap V = 0$ means that the restriction of 
$P_{V^{\perp}}$ to $U$ is injective. As $V^{\perp}$ is finite 
dimensional, $U$ must be finite dimensional too. $\Box$ \\ 
~\\
Quasipoints in $\LL(\kh)$ that contain a line are \emph{atomic}. The
non-atomic quasipoints are called \emph{continuous}.\\
Whereas the structure of atomic quasipoints is trivial, the set of 
continuous quasipoints mirrors the whole complexity of spectral 
theory of linear operators in $\kh$. Therefore it is a real challenge 
to classify the continuous quasipoints in $\lh$.\\
~\\
In contrast to the case of Boolean algebras, Stone spectra are not 
compact in general. The situation can be even worse, as the following 
important example shows:

\begin{proposition}\label{vn2}
Let $\kh$ be a Hilbert space of dimension greater than one. Then the 
Stone spectrum $\qh := \kQ(\LL(\kh))$ is not compact and, if the 
dimension of $\kh$ is infinite, $\qh$ is not even locally compact. 
\end{proposition}
\emph{Proof:} This is an easy consequence of Baire's category theorem and the 
general fact that the Stonean space $\kQ(\LL_{U})$ of the principal 
ideal $\LL_{U} := \{ V \in \LL \mid V \leq U \}$ of an arbitrary 
lattice $\LL$ and $U \in \LL \setminus \{0\}$ is homeomorphic to 
$\kQ_{U}(\LL)$.\\
Indeed, assume that $\qh$ is compact. Then there are $U_{1}, \ldots,
U_{n} \in \LL(\kh)$ such that 
\[
    \bigcup_{k = 1}^n\kQ_{U_{k}}(\kh) = \qh. 
\]    
Let $x$ be an arbitrary nonzero element of $\kh$. Then the atomic
quasipoint $\frb_{\CC x}$ belongs to $\kQ_{U_{k}}(\kh)$ for at least
one $k$ and hence $\kh = \bigcup_{k = 1}^nU_{k}$, a contradiction. By 
Baire's category theorem, the same argument works not only for finite 
$n$ but also for $n = \aleph_{0}$. This shows that $\qh$ is not
Lindelöf compact. \ \   $\Box$ \\
~\\
The structure of Stone spectra of \emph{finite} von Neumann algebras
of type $I$ has been clarified by A. Döring in \cite{doe1a}.\\
The basis of Döring's result is the observation that the Stone
spectrum $\qr$ of an arbitrary von Neumann algebra $\rr$ with center
$\kcc$ can be mapped canonically onto the Stone spectrum $\qc$ of
$\kcc$. Later on we need a stronger and more general result:

\begin{proposition}\label{vn3}
    Let $\rr$ be a von Neumann algebra with center $\kcc$ and let
    $\kaa$ be a von Neumann subalgebra of $\kcc$. Then the mapping
    \[
	\gz_{\kaa} : \frb \tto \frb \cap \kaa
    \]
    is an open continuous, and therefore identifying, mapping from $\qr$
    onto $\qa$. Moreover
    \[
	\gz_{\kaa}(\frb) = \{ s_{\kaa}(P) \ | \ P \in \frb \}
    \]           
    for all $\frb \in \qr$, where 
    \[
	s_{\kaa}(P) := \We \{ Q \in \pa \ | \ P ≤ Q \}
    \]
     is the $\kaa$-support of $P \in \pr$.    
\end{proposition}
\emph{Proof:} $\frb \cap \kaa$ is clearly a filter base in $\pa$. Let 
$\gb \in \qa$ be a quasipoint that contains $\frb \cap \kaa$ and let
$C \in \gb$. If $C \notin \frb \cap \kaa$ then $C \notin \frb$. Hence 
there is some $P \in \frb$ such that $P \we C = 0$. Because $C$ is
central this means $PC = 0$. But then $P = PC + P(I - C) = P(I - C)$, 
i.e. $P ≤ I - C$. This implies $I - C \in \frb \cap \kaa \tm \gb$, a
contradiction to $C \in \gb$. Hence $\frb \cap \kaa$ is a quasipoint
in $\kaa$.\\
It follows immediately from the definition of the $\kaa$-support that
\[
    \all \ P, Q \in \pr : \ P ≤ s_{\kaa}(P) \ \ \text{and} \ \
    s_{\kaa}(P \we Q) ≤ s_{\kaa}(P) \we s_{\kaa}(Q)
\]
holds. This implies that $\{ s_{\kaa}(P) \ | \ P \in \frb \}$ is a
filter base contained in $\frb \cap \kaa$. Because of $s_{\kaa}(P) =
P$ for all $P \in \pa$, we must have equality. \\
Now we prove that
\begin{enumerate}
    \item  [(i)] $\all \ P \in \pr : \ \gz_{\kaa}(\qpr) =
    \kQ_{s_{\kaa}(P)}(\kaa)$ and

    \item  [(ii)] $\all \ Q \in \pa : \ \overset{-1}{\gz_{\kaa}}(\kQ_{Q}
    (\kaa)) = \kQ_{Q}(\rr)$
\end{enumerate}
hold: It is obvious that $\gz_{\kaa}(\qpr)$ is contained in
$\kQ_{s_{\kaa}(P)}(\kaa)$. Let $\gga \in \kQ_{s_{\kaa}(P)}(\kaa)$.
Then $P \in \urb{s_{\kaa}}(\gga)$, and we shall show that this implies
that $\{P\} \cup \gga$ is a filter base in $\pr$. $P$ being a central 
projection, $\{P\} \cup \gga$ is a filter base if and only if
\[
    \all \ Q \in \gga : \ PQ \ne 0.
\]
Assume that $PQ = 0$ for some $Q \in \gga$. Then $P ≤ I - Q$, hence
also $s_{\kaa}(P) ≤ I - Q$, contradicting $s_{\kaa}(P) \in \gga$.
Let $\frb$ be a quasipoint in $\pr$ that contains $\{P\} \cup \gga$.
Because of $s_{\kaa}(Q) = Q$ for all $Q \in \gga$ we obtain
\[
    \gga = s_{\kaa}(\{P\} \cup \gga) \tm s_{\kaa}(\frb) =
    \gz_{\kaa}(\frb).
\]
Hence $\gga = \gz_{\kaa}(\frb)$ since $\gz_{\kaa}(\frb)$ and $\gga$
are quasipoints in $\pa$. This proves $(i)$. $(ii)$ follows from the
fact that each quasipoint in $\pa$ is contained in a quasipoint in
$\pr$. 
Properties $(i)$ and $(ii)$ imply that $\gz_{\kaa}$ is open, continuous
and surjective. \ \ $\Box$ \\
~\\
In \cite{doe1a} a quasipoint $\frb \in \qr$ is called \emph{abelian}
if it contains an abelian projection. The term ``abelian quasipoint'' 
is motivated by the following fact: If $E \in \frb$ is an abelian
projection then the ``\emph{$E$-socle}''
\[
    \frb_{E} := \{ P \in \frb \ | \ P ≤ E \},
\]
which determines $\frb$ uniquely, consists entirely of abelian
projections. Moreover every subprojection of an abelian projection $E$
is of the form $CE$ with a suitable central projection $C$. Hence
\[
    \frb_{E} = \{ CE \ | \ C \in \frb \cap \kcc \}
\]
if $E$ is abelian.\\
~\\
Let $\gtt \in \rr$ be a partial isometry, i.e. $E := \gtt^\ast \gtt$ and
$F := \gtt \gtt^\ast$ are projections. $\gtt$ has kernel $E(\kh)^\pp$ and
maps $E(\kh)$ \emph{isometrically} onto $F(\kh)$. Now it is easy to
see that for any projection $P_{U} ≤ E$ we have
\begin{equation}
    \gtt P_{U} \gtt^\ast = P_{\gtt(U)}.
    \label{eq:vn1}
\end{equation}
A consequence of this relation is
\begin{equation}
    \all \ P, Q ≤ E : \ \gtt (P \we Q) \gtt^\ast = (\gtt P \gtt^\ast) \we 
	(\gtt Q \gtt^\ast). 
    \label{eq:vn2}
\end{equation}
If $\frb \in \kQ_{E}(\rr)$ then
\begin{equation}
    \gtt_{\ast}(\frb_{E}) := \{ \gtt P \gtt^\ast \ | \ P \in \frb_{E} \}
    \label{eq:vn3}
\end{equation}               
is the $F$-socle of a (uniquely determined) quasipoint $\gtt_{\ast}(\frb)
\in \kQ_{F}(\rr)$: Equation \ref{eq:vn2} guarantees that
$\gtt_{\ast}(\frb_{E})$ is a filter base. Let $\Tilde{\frb}$ be a
quasipoint that contains $\gtt_{\ast}(\frb_{E})$. Then $\gtt_{\ast}(\frb_{E})
\tm \Tilde{\frb}_{F}$. Assume that this inclusion is proper. If $Q \in
\Tilde{\frb}_{F} \smm \gtt_{\ast}(\frb_{E})$ then $\gtt^\ast Q \gtt
\notin \frb_{E}$ and therefore there is some $P \in \frb_{E}$ such that
$P \we \gtt^\ast Q \gtt = 0$. But then $\gtt P \gtt^\ast \we Q = 0$, a
contradiction. This shows that we obtain a mapping
\[
    \begin{array}{cccc}
	\gtt_{\ast} : & \kQ_{\gtt^\ast \gtt}(\rr) & \to & \kQ_{\gtt
	\gtt^\ast}(\rr)  \\
	 & \frb & \tto & \gtt_{\ast}(\frb).
     \end{array}
\]
It is easy to see that $\gtt_{\ast}$ is a homeomorphism with inverse
$(\gtt^\ast) _{\ast}$. Note that $\gtt_{\ast}$ is globally defined if 
$\gtt$ is a unitary operator. The following result is fundamental:

\begin{proposition}(\cite{doe1a})\label{vn4}
    Let $\rr$ be a von Neumann algebra with center $\kcc$ and let
    $\frb_{1}, \frb_{2} \in \qr$. Then $\gz_{\kcc}(\frb_{1}) =
    \gz_{\kcc}(\frb_{2})$ if and only if there is a partial isometry
    $\gtt \in \rr$ such that $\gtt_{\ast}(\frb_{1}) = \frb_{2}$. If
    $\rr$ is finite then $\gtt$ can be chosen as a unitary operator in
    $\rr$.
\end{proposition}
It is well known (\cite{kr2}) that a von Neumann algebra $\rr$ of type
$I_{n}$, $n < \∞$, is isomorphic to the algebra $\MM_{n}(\kcc)$ of
all $(n, n)$-matrices with entries from the center $\kcc$ of $\rr$.                           
Hence $\rr$ can be considered as the endomorphism algebra of the free 
$\kcc$-module $\kcc^n$. Using methods from the theory of
$C^\ast$-modules one can prove

\begin{theorem}( \cite{doe1a})\label{vn5}
    Let $\rr$ be a finite von Neumann algebra of type $I$. Then all
    quasipoints in $\rr$ are abelian.
\end{theorem}

Applying proposition \ref{vn4} we obtain the structure of Stone
spectra of finite von Neumann algebras of type $I$:

\begin{theorem}(\cite{doe1a})\label{vn6}
    Let $\rr$ be a finite von Neumann algebra  of type $I$ and let
    $\kcc$ be the center of $\rr$. Then the orbits of the action of
    the unitary group $\frU_{\rr}$ of $\rr$ on $\qr$ are the fibres
    of $\gz_{\kcc}$.
\end{theorem}

\section{Boolean Quasipoints}
\label{BQ}
\pagestyle{myheadings}
\markboth{The Stone Spectrum of a Lattice}{Boolean Quasipoints}

\emph{Boolean filter bases} in an \emph{orthomodular lattice} $\LL$ are
defined as filter bases in $\LL$ whose elements \emph{commute} with
each other. Analogous to (ordinary) quasipoints we define:

\begin{definition}\label{bq1}
    Let $\LL$ be an orthomodular lattice. A subset $\gb$ of $\LL$ is
    called a Boolean quasipoint if it satisfies the following
    requirements:
    \begin{enumerate}
	\item  [(i)] $0 \notin \gb$,     
    
	\item  [(ii)] $\all \ a, b \in \gb \ \ex \ c \in \gb : \ c ≤ a
	\we b$,
    
	\item  [(iii)] $\all \ a, b \in \gb : \ a \kcc b$, 
    
	\item  [(iv)] $\gb$ is a maximal subset that satisfies $(i),
	(ii), (iii)$. 
    \end{enumerate}
    The set of all Boolean quasipoints in $\LL$ is denoted by $\bpl$.
\end{definition}
A Boolean quasipoint is therefore a \emph{maximal Boolean filter base}.
As for ordinary quasipoints it is easy to see that
\begin{enumerate}
    \item  [(v)] Let $\gb$ be a Boolean quasipoint in an orthomodular
    lattice $\LL$. If $c \in \LL$ has the property that $a \kcc c$ for
    all $a \in \gb$ and $b ≤ c$ for some $b \in \gb$ then $c \in \gb$.  
\end{enumerate}
holds.

\begin{lemma}\label{bq2}
    A Boolean sector $\BB$ of an orthomodular lattice $\LL$ is
    generated by every Boolean quasipoint contained in $\BB$. 
\end{lemma}
\emph{Proof:} Let $\gb \in \bpl$. Then $\gb$ is contained in some
Boolean sector $\BB$ because the lattice $\LL(\gb)$ generated by
$\gb$ is distributive. Note that $\gb$ is a quasipoint in $\BB$. Let
$a \in \BB$. Then, by remark \ref{SS13}, $a \in \gb$ or $a^\pp \in \gb$.
Hence $a \in \LL(\gb)$ in any case, i.e. $\LL(\gb) = \BB$. \ \ $\Box$ \\ 

\begin{corollary}\label{bq3}
    A Boolean quasipoint in an orthomodular lattice $\LL$ is contained
    in exactly one Boolean sector of $\LL$.
\end{corollary}
This means that the maximal distributive sublattices of $\LL$ induce a
partition of $\bpl$. Therefore the name ``Boolean sector''. 

\begin{proposition}\label{bq4}
    The Stone spectrum $\kQ(\BB)$ of a Boolean sector of $\LL$
    coincides with the set of Boolean quasipoints contained in $\BB$.
\end{proposition}
\emph{Proof:} A Boolean quasipoint contained in $\BB$ is obviously a
quasipoint of the lattice $\BB$. Conversely, let $\frb \in \kQ(\BB)$
and assume that $\frb$ is not a Boolean quasipoint. Then $\frb$ is
properly contained in a Boolean quasipoint $\gb$ in $\LL$. Let $b \in 
\gb \smm \frb$. Then there is some $a \in \BB$ not commuting with $b$.
But $a$ or $a^\pp$ belongs to $\frb$ hence also to $\gb$. This gives a
contradiction. \ \ $\Box$ 

\begin{proposition}\label{bq5}
    Let $\rr$ be a von Neumann algebra. Then there is a canonical one to
    one correspondence between the maximal abelian subalgebras of $\rr$
    and the Boolean sectors of the projection lattice $\pr$.
\end{proposition}
\emph{Proof:} Let $\BB \tm \pr$ be a Boolean sector and denote by
$\mm(\BB)$ the von Neumann algebra generated by $\BB$, i.e. the double
commutant of $\BB$: 
\[
    \mm(\BB) = \BB^{\kcc\kcc}.
\]
$\mm(\BB)$ is an abelian von Neumann subalgebra of $\rr$. It is
contained in a maximal abelian subalgebra $\mm$ of $\rr$. Let $A \in
\mm_{sa}$. By the spectral theorem the spectral projections of $A$
commute with every operator that commutes with $A$. Hence all spectral
projections of $A$ commute with all elements of $\BB$. It follows from
the maximality of $\BB$ that all spectral projections of $A$ belong to
$\BB$. Therefore $A \in \mm(\BB)$, i.e. $\mm(\BB)$ is maximal.\\
Conversely, let $\mm$ is a maximal abelian subalgebra of $\rr$ and let
$P \in \pr$ commute with all elements of $\pmm$. Then, again by the
spectral theorem, $P$ commutes with all elements of $\mm$ and
therefore $P \in \mm$ since $\mm$ is maximal. This shows that $\pmm$
is a Boolean sector of $\pr$ and that the mapping $\BB \tto
\BB^{\kcc\kcc}$ is a bijection from the set of Boolean sectors of $\pr$
onto the set of maximal abelian subalgebras of $\rr$. \ \ $\Box$ \\
~\\
For a general von Neumann algebra it is an intricate problem to
determine its maximal abelian subalgebras. In the case $\rr = \lh$
however, they can be determined up to unitary equivalence. Moreover
every abelian von Neumann algebra is isomorphic to a maximal abelian
subalgebra of $\lh$ for a suitable Hilbert space $\kh$. These are
well known results (\cite{kr2, tak1}). We like to reprove them here in
order to demonstrate the use of Stone spectra. \\
~\\
We recall some well known notions and facts. A Hausdorff space $\gO$
is called a {\bf Stone space} if it is compact and \emph{extremely
disconnected}. The Stone spectrum of an abelian von Neumann algebra is
a Stone space, but the converse is not true (\cite{kr3}, p.228). If
$\gO$ is a Stone space, then $C(\gO)$ is a von Neumann algebra if and
only if $\gO$ carries a family of {\bf normal} probability measures
that separates the continuous functions on $\gO$. In this case $\gO$
is called a {\bf hyperstonean space} (\cite{tak1}, III.1). \\
~\\
Let $\kaa$ be an abelian von Neumann algebra acting on a
\emph{separable} Hilbert space $\kh$. By the Gelfand representation
theorem $\kaa$ is isomorphic to $C(\gO)$, where $\gO$ is the Gelfand
spectrum of $\kaa$, and by theorem \ref{theo: 42} $\gO$ is
homeomorphic to the Stone spectrum $\qa$ of $\kaa$.\\
Let $D$ be the set of isolated points of $\gO$. Since $\kh$ is
separable, $D$ is at most countably infinite. $D$ is an open subset of
the extremely disconnected space $\gO$. So its closure
\[
    \gO_{d} := \overline{D}
\]  
is open and closed. This leads to a partition
\[
    \gO = \gO_{c} \sqcup \gO_{d}
\]
of $\gO$ into open and closed sets. By the way, if $D$ is infinite
then $\gO_{d}$ is a very large set: it can be shown that $\gO_{d}$ is 
the Stone-\v{C}ech compactification of $D$. Because $\gO_{c}$ and
$\gO_{d}$ are open and closed, this partition induces an isometric
isomorphism
\begin{eqnarray*}
    C(\gO) & \to & C(\gO_{c}) \times C(\gO_{d})  \\
    f & \tto & (f|_{\gO_{c}}, f|_{\gO_{d}}).
\end{eqnarray*} 
Using the embedding $C(\gO_{c}) \to C(\gO), \ f \tto f_{c}$, defined
by
\[
    f_{c} := 
    \bg{cases}
     f   &   \ \text{on} \ \gO_{c}  \\
     0   &   \ \text{on} \ \gO_{d}
     \end{cases}
\]
and the similar defined embedding $C(\gO_{d}) \to C(\gO)$, we see that
$C(\gO)$ can be written as a direct sum
\[
    C(\gO) = C(\gO_{c}) \oplus C(\gO_{d}).
\]
From this representation it is immediate that $\gO$ is hyperstonean if
and only $\gO_{c}$ and $\gO_{d}$ are. \\
~\\
Since $\gO_{c}$ has no isolated points, $C(\gO_{c})$ is isomorphic to
$L^\∞ (]0, 1[, \gl)$ where $\gl$ denotes the Lebesgue measure on $] 0,
1[$ (\cite{tak1}, III.1). It remains to find a canonical Hilbert space
representation for $C(\gO_{d})$.\\
~\\
If $\gO$ is a Stonean space we denote by $\kO\kcc(\gO)$ the set of
open and closed subsets of $\gO$. Since the interior of a closed
subset of $\gO$ is open and closed, $\oco$ is a \emph{complete}
distributive lattice with respect to the operations
\begin{eqnarray*}
    \Ve_{k \ikk}U_{k} & := & \overline{\bigcup_{k \ikk}U_{k}},  \\
    \We_{k \ikk}U_{k} & := & int (\bigcap_{k \ikk}U_{k}).
\end{eqnarray*}
It is easy to see that $\oco$ is even \emph{completely distributive}.

\begin{lemma}\label{bq6}
    Let $\gO$ be a Stonean space. Then the Stone spectrum of $\oco$ is
    canonically homeomorphic to $\gO$.
\end{lemma}
\emph{Proof:} Let $\go \in \gO$ and let $\frb$ be a quasipoint over
$\go$. Then $\go \in \overline{U}$ for all $U \in \frb$. But $U$ is
open and closed, so $\frb$ consists of all open closed neighborhoods
of $\go$. This means that there is exactly one quasipoint over $\go$. 
Hence the mapping
\[
      pt : \kQ(\oco) \to \gO, 
\]                   
defined by $\{pt(\frb)\} := \bigcap_{U \in \frb}U$, is a bijection.
Obviously
\[
    pt(\kQ_{U}(\oco)) = U
\]
for all $U \in \oco$, so $pt$ is a homeomorphism. \ \ $\Box$ 

\begin{proposition}\label{bq7}
    Let $\kh$ be a Hilbert space (of arbitrary dimension) and let
    $(e_{k})_{k \ikk}$ be an orthonormal basis for $\kh$. Then there
    is a unique Boolean sector $\BB$ of $\llh$ that contains all $\CC 
    e_{k} \ (k \ikk)$. $\BB$ consists of all $U \in \llh$ with the
    property
    \[
	\all \ k \ikk : \ e_{k} \in U \cup U^\pp.
    \]
\end{proposition}
\emph{Proof:} It is obvious that $\{ \CC e_{k} \ | \ k \ikk \}$ is
contained in a Boolean sector $\BB$. Let $U \in \BB$. $\CC e_{k} \kcc 
U$ means
\[
    \CC e_{k} = (\CC e_{k} \cap U) + (\CC e_{k} \cap U^\pp) 
\]
and therefore, according to the minimality of $\CC e_{k}$,
\[
    e_{k} \in U \cup U^\pp.
\]
Hence it suffices to show that the elements of $\{ U \in \llh \ | \ 
\all \ k \ikk : \ e_{k} \in U \cup U^\pp \}$ commute with each other. 
But this follows easily from remark \ref{ol8}. \ \ $\Box$ \\
~\\
The Boolean sector $\BB$ occuring in proposition \ref{bq7} is called
{\bf maximal atomic} because it contains a maximal number of atoms of 
$\llh$.\\
~\\
We return to the study of $C(\gO_{d})$. Because $\gO_{d} =
\overline{D}$ and the elements of $D$ are isolated, each open and
closed subset of $\gO_{d}$ is the closure of its intersection with
$D$:
\[
    \all \ O \in \ocod : \ O = \overline{O \cap D},             
\]
that is
\[
    \all \ O \in \ocod : \ O = \Ve_{\gd \in O \cap D}\{\gd\}.
\]
The projections of $C(\gO_{d})$ are the characteristic functions of
the elements $O \in \ocod$. Therefore they can be written as
\[
    \chi_{O} = \Ve_{\gd \in O \cap D}\chi_{\gd}.
\]
Let $\kh_{D}$ be a Hilbert space of dimension $\# D$ and take an
orthonormal basis $(e_{\gd})_{\gd \in D}$, labeled with the elements 
of $D$. Moreover, let $\BB$ be the Boolean sector of $\LL(\kh_{D})$
that is defined according to proposition \ref{bq7} by this orthonormal
basis and let
\[
    \kP(\BB) := \{ P_{U} \ | \ U \in \BB \}.
\]
Since each $U \in \BB$ is the join of the lines $\CC e_{\gd}$
that are contained in $U$, we obtain a mapping
\[
    \begin{array}{cccc}
	\gt : & \kP(\BB) & \to & \kP(C(\gO_{d}))  \\
	 & P_{U} & \tto & \Ve \{ \chi_{\gd} \ | \ e_{\gd} \in U \}.
    \end{array}
\]
It is easy to see that $\gt$ is a lattice isomorphism, so it induces a
homeomorphism from the Stone spectrum $\kQ(\kP(\BB))$ of $\kP(\BB)$
onto the Stone spectrum $\kQ(\kP(C(\gO_{d})))$. According to the
foregoing considerations and to lemma \ref{bq6} the latter is
homeomorphic to $\gO_{d}$. This shows that the von Neumann algebra
$C(\gO_{d})$ is isomorphic to the maximal abelian subalgebra
$\mm(\kP(\BB))$ of $\kL(\kh_{D})$. Altogether we have proved:

\begin{proposition}\label{bq8}
    Let $\kaa$ be an abelian von Neumann algebra acting on a
    separable Hilbert space $\kh$. Then there is a separable Hilbert
    space $\kh_{D}$ and a maximal atomic Boolean sector $\BB$ of
    $\kP(\kh_{D})$ such that $\kaa$ is isomorphic to the maximal
    abelian subalgebra $L^\∞ (]0, 1[, \gl) \oplus \mm(\BB)$ of
    $\kL(L^2 (]0, 1[, \gl) \oplus \kh_{D})$, where $\gl$ denotes the
    Lebesgue measure on $]0, 1[$. The isomorphy class of $\kaa$ is
    determined by the cardinality of the set $D$ of isolated points of
    the Gelfand spectrum $\gO$ of $\kaa$ and fact whether $\gO \smm
    \overline{D}$ is empty or not.
\end{proposition}
Applying this result we can describe the isomorphy classes of Boolean 
sectors of $\ph$ for a separable Hilbert space $\kh$. If $\mm_{1},
\mm_{2}$ are maximal abelian von Neumann subalgebras of $\lh$ and if
$\gf : \mm_{1} \to \mm_{2}$ is an isomorphism (recall that
``isomorphism between von Neumann algebras'' always means
``$\ast$-isomorphism between von Neumann algebras'') then there is a
unitary $U \in \lh$ such that $\gf(A) = UAU^\ast$ for all $A \in
\mm_{1}$ (\cite{kr2}, p.661). We know that $\mm_{k} = \mm(\BB_{k}) =
\BB_{k}^{\kcc\kcc} \quad (k = 1, 2)$ for suitable Boolean sectors $\BB_{1},
\BB_{2} \tm \ph$. Therefore the isomorphy of maximal abelian subalgebras
of $\lh$ is the same as the unitary equivalence of the corresponding
Boolean sectors of $\ph$. \\
To each maximal abelian subalgebra $\mm(\BB)$ of $\lh$ we assign a
pair of numbers $(p, n) \in \{0, 1\} \times \{ n \ | \ 0 ≤ n ≤
dim \kh \}$ defined in the following way: Let $D_{\BB}$ be the set 
of isolated points in the Stone spectrum $\kQ(\BB)$ and let
\[
    n := \# D_{\BB} 
\]
and
\[
    p := 
    \begin{cases}
	0  &  \text{if} \quad \overline{D_{\BB}} = \kQ(\BB)  \\
	1  &  \text{otherwise}.
    \end{cases}
\]
Then we obtain from proposition \ref{bq8}:

\begin{corollary}\label{bq9}
    Let $\kh$ be a separable Hilbert space. Then the unitary
    equivalence classes of Boolean sectors of $\ph$ correspond to the 
    elements of $\{0, 1\} \times \{ n \ | \ 0 ≤ n ≤ dim \kh \}$.
\end{corollary}

\end{document}